\pdfoutput=1
\newcommand*{\ATLASLATEXPATH}{./}
\newcommand*{\trfmj}{\ensuremath{\text{TRF}_{\text{MJ}}}\xspace}
\newcommand*{\mjeps}{\ensuremath{\varepsilon_{\text{MJ}}}\xspace}
\documentclass[cernpreprint,texlive=2011,txfonts,UKenglish]{\ATLASLATEXPATH atlasdoc}

\PreprintIdNumber{CERN-EP-2016-058}
\newcommand{\rmt}[1]{\textcolor{black}{ #1}}

\usepackage{\ATLASLATEXPATH atlaspackage}
\usepackage{\ATLASLATEXPATH atlasbiblatex}
\usepackage{\ATLASLATEXPATH atlascontribute}
\usepackage{\ATLASLATEXPATH atlasphysics}
\usepackage{multirow}
\usepackage{lscape} 
\graphicspath{{logos/}{figures/}}
\usepackage{mydocument-defs}
\AtlasTitle{Search for the Standard Model Higgs boson decaying into $b\bar{b}$
produced in association with top quarks decaying hadronically in $pp$
collisions at $\sqrt{s}$=8~\TeV with the ATLAS detector}

\author{The ATLAS Collaboration}
\date{\today}
\AtlasRefCode{HIGG-2015-05}
\PreprintIdNumber{CERN-EP-2016-058}
 \AtlasJournal{JHEP}
\AtlasAbstract{%
A search for Higgs boson production in association with a pair of top
quarks ($t\bar{t}H$) is performed, 
where the Higgs boson decays to $b\bar{b}$, and both top quarks decay hadronically. 
The data used correspond to an integrated luminosity of 20.3~\ifb~of
$pp$ collisions at $\sqrt{s}=8$~\TeV\ collected with the ATLAS detector at the Large Hadron Collider. 
The search selects events with at least six energetic jets and uses
a boosted decision tree algorithm to discriminate between signal and Standard Model background. 
The dominant multijet background is estimated using a dedicated data-driven technique. 
For a Higgs boson mass of 125~\GeV, an upper limit of 6.4 (5.4) times the
Standard Model cross section is observed (expected) at
95\% confidence level.
The best-fit value for the signal strength is $\mu = 1.6 \pm 2.6$ times the
Standard Model expectation for $m_{\text{H}} = 125$~\GeV.  
Combining all \tth\ searches carried out by ATLAS at $\sqrt{s}=8$ and $7$~\TeV, an
observed (expected) upper limit of 3.1 (1.4) times the Standard
Model expectation is obtained at 95\% confidence level, with a
signal strength
$\mu = 1.7 \pm 0.8$. 
}

\hypersetup{pdftitle={ATLAS draft},pdfauthor={The ATLAS Collaboration}}
\begin{document}

\maketitle

\tableofcontents
\newpage

\section{Introduction} 
\label{sec:intro}
%-------------------------------------------------------------------------------
 
After the discovery of a new boson with a mass of around 125~\GeV\ in July 2012
by the ATLAS~\cite{ATLASHiggsDiscovery} and CMS~\cite{CMSHiggsDiscovery}
collaborations, the focus has now shifted to confirming 
whether this particle is the Standard Model (SM) Higgs boson~\cite{Englert:1964aa,Higgs:1964aa,Higgs:1964ab,Guralnik:1964aa} or another boson.
While any deviation from SM predictions would indicate the presence 
of new physics, all measurements of the properties of this new boson thus far performed at the Large Hadron Collider (LHC), 
including spin, parity, total width, and coupling to SM particles, 
are consistent with the SM
prediction~\cite{Aad:2015gba,Aad:2015mxa,Khachatryan:2014jba,Khachatryan:2014kca,Aad:2015xua,Khachatryan:2014iha}.  

Because of its large mass, the top quark is the fermion with the
largest Yukawa coupling ($y_t$) to the Higgs field in the SM, with a
value close to unity. 
The coupling  $y_t$ is experimentally accessible by measuring the gluon
fusion (ggF) production process or the $H \rightarrow \gamma \gamma$
decay, where a sizeable contribution derives from a top-quark loop. 
This case requires the assumption that no new physics contributes with
additional induced loops in order to measure $y_t$. Currently, the only
process where  $y_t$ can be accessed directly is the production of
a top-quark pair in association with a Higgs boson (\tth).   

The results of searches for the Higgs boson are usually expressed in terms of the
 signal-strength parameter $\mu$, which is defined as the
ratio of the observed to the expected number of signal events. The latter 
is calculated using the SM cross section times branching ratio~\cite{lhcxs}.  
The combined \tth~signal strength measured by the CMS Collaboration~\cite{Khachatryan:2014qaa}, obtained by merging searches in several final
    states, is $\mu = 2.8\pm 1.0$. 
The~ATLAS Collaboration has searched
for a \tth~signal~in events enriched in Higgs boson decays to
two massive vector bosons or $\tau$ leptons in the
multilepton channel~\cite{Aad:2015iha}, finding $\mu = 2.1^{+1.4}_{-1.2}$, 
for \tth($H \rightarrow b\bar{b}$)~\cite{Aad:2015gra}~in final states with at least one lepton obtaining $\mu = 1.5 \pm 1.1$,
and for \tth($H \rightarrow \gamma\gamma$)~\cite{ttHyy} measuring $\mu = 1.3^{+2.6}_{-1.7}$.

Among all \tth\ final states, the one where both $W$ bosons from $t\rightarrow Wb$ decay hadronically
and the Higgs boson decays into a $b\bar{b}$ pair has the largest branching
ratio, but also the least signal purity. This paper describes a search for this all-hadronic
$t\bar{t}H(H \rightarrow b\bar{b})$ decay mode. The analysis uses proton--proton collision data corresponding to an integrated luminosity of 20.3~\ifb~at center-of-mass energy $\sqrt{s}=8$~\TeV\ recorded with the ATLAS detector at the LHC.  

%-------------------------------------------------------------------------------

At Born level, the signal signature is eight jets, four of which are $b$-quark jets.
The dominant background is the non-resonant production of multijet events.
For this analysis, a data-driven method is applied to estimate the multijet
background by extrapolating its contribution from a control region with the same jet
multiplicity, but a lower multiplicity of jets containing $b$-hadrons than the
signal process. The parameters used for the extrapolation are measured from a control region and checked using Monte Carlo (MC) simulations. Other subdominant background processes are estimated
using MC simulations. To maximise the signal sensitivity, the events
are categorised according to their number of jets and jets identified as containing $b$-hadrons ($b$-tagged). A
boosted decision tree (BDT) algorithm, based on event shape and kinematic variables, is
used to discriminate the signal from the background.
The extraction of $\mu$ is performed through a fit to the
BDT discriminant distribution.
After the fit the dominant uncertainty is the  $t\bar{t}$\ +\ $b\bar{b}$ production cross section. The sensitivity is also limited by systematic uncertainties from
  the data-driven method used for the modelling of the large non-resonant multijet production. 

\newcommand{\AtlasCoordFootnote}{
ATLAS uses a right-handed coordinate system with its origin at the nominal interaction point (IP)
in the centre of the detector and the $z$-axis along the beam pipe.
The $x$-axis points from the IP to the centre of the LHC ring,
and the $y$-axis points upwards.
Cylindrical coordinates $(r,\phi)$ are used in the transverse plane, 
$\phi$ being the azimuthal angle around the $z$-axis.
The pseudorapidity is defined in terms of the polar angle $\theta$ as $\eta = -\ln \tan(\theta/2)$.
Angular distance is measured in units of $\Delta R \equiv \sqrt{(\Delta\eta)^{2} + (\Delta\phi)^{2}}$.}

%-------------------------------------------------------------------------------
\section{The ATLAS detector}
\label{sec:atlas}
%-------------------------------------------------------------------------------

The ATLAS detector~\cite{ATLASTechnicalPaper}
consists of an inner tracking detector surrounded by a thin superconducting
solenoid magnet providing a \SI{2}{\tesla} axial magnetic field,
electromagnetic and hadron calorimeters, and a muon spectrometer incorporating
three large superconducting toroid magnets. The inner detector (ID) comprises the high-granularity silicon pixel detector and the silicon microstrip
tracker covering the pseudorapidity\footnote{ATLAS uses a right-handed
  coordinate system with its origin at the nominal interaction point (IP) in the centre of the 
  detector and the $z$-axis coinciding with the axis of the beam pipe. The
  $x$-axis points from the IP to the centre of the LHC ring, and the $y$-axis
  points upward. Cylindrical coordinates ($r$,$\phi$) are used in the
  transverse plane, $\phi$ being the azimuthal angle around the beam pipe. The
  pseudorapidity is defined in terms of the polar angle $\theta$ as $\eta = -
  \ln \tan(\theta/2)$. Transverse momentum and energy are defined as $\pt =
  p \sin \theta$ and $\et = E \sin \theta$ respectively.}
range $|\eta| < 2.5$, and the straw-tube transition radiation tracker
covering $|\eta| < 2.0$. The electromagnetic calorimeter covers $|\eta| < 3.2$
and consists of a barrel and two endcap high-granularity 
lead/liquid-argon (LAr) calorimeters. An additional thin LAr presampler covers
$|\eta| < 1.8$. 
Hadron calorimetry is provided by a steel/scintillator-tile calorimeter, which 
covers the region $|\eta| < 1.7$, and two copper/LAr hadron endcap calorimeters.
To complete the pseudorapidity coverage, copper/LAr and tungsten/LAr forward calorimeters cover up to $|\eta|=4.9$.
Muon tracking chambers precisely measure the deflection of muons in the magnetic field generated by superconducting air-core toroids in the region $|\eta| < 2.7$. 
A three-level trigger system selects events for offline analysis~\cite{PERF-2011-02}. The hardware-based Level-1 trigger 
is used to reduce the event rate to a maximum of \SI{75}{\kHz},
 while the two software-based trigger levels, Level-2
 and Event Filter (EF), 
reduce the event rate to about \SI{400}{\Hz}.

\section{Object reconstruction}
\label{sec:reconstruction}

The all-hadronic  $t\bar{t}H$ final state is composed of jets
originating from ($u$, $d$, $s$)-quarks or gluons (light jets) and jets from
  $c$- or $b$-quarks (heavy-flavour jets). Electrons and muons, selected in the same way
  as in Ref.~\cite{Aad:2015gra}, are used only to veto events that would overlap 
with the $t\bar{t}H$ searches in final states with leptons.

At least one reconstructed primary vertex is required, with at least five associated 
tracks with $\pt~\geq~400$~\MeV, and a position consistent
with the luminous region of the beams in the transverse plane.
If more than one vertex is found, the primary vertex is taken
to be the one which has the largest sum of the squared transverse momenta
of its associated tracks.

Jets are reconstructed with the anti-$k_t$ 
algorithm~\cite{ref:Cacciari2008,ref:Cacciari2006,ref:fastjet}, 
with a radius parameter $R = 0.4$ in the $(\eta,\phi)$ plane. They are built from
calibrated topological clusters of energy deposits 
in the calorimeters~\cite{ATLASTechnicalPaper}. 
Prior to jet finding, a local cluster calibration scheme \cite{LCW1,LCW2} 
is applied to correct the topological cluster energies for the effects of non-compensating calorimeter response, dead material, and out-of-cluster leakage. 
 After energy calibration based on in-situ measurements~\cite{ATLASJetEnergyMeasurement}, jets are required to have transverse momentum $\pt
>25$~\GeV\ and $|\eta|<2.5$. During jet reconstruction, no distinction is made
between identified electrons and jet energy deposits.
To avoid double counting electrons as jets, any jet within a cone of size $\Delta R = \sqrt{(\Delta\phi)^2 + (\Delta\eta)^2} = 0.2$
around a reconstructed electron is discarded. After this, electrons within
a $\Delta R = 0.4$ of a remaining jet are removed. 

To avoid selecting jets from additional $pp$ interactions in the same
event (pile-up), a loose selection is applied to the jet vertex fraction (JVF),
defined as the ratio of the scalar sum of the $\pt$ of tracks 
matched to the jet and originating from the primary vertex to that of all tracks
matched to the jet. This criterion, JVF $\geq 0.5$,  is only applied to jets with $\pt< 50$~\GeV\ and $|\eta|<2.4$.

Jets are $b$-tagged by means of the MV1 algorithm~\cite{btagpaper}. It combines information from track impact parameters and topological
properties of secondary and tertiary decay vertices which are
reconstructed within the jet. 
The working point used for this search corresponds to a 60\% efficiency to tag
a $b$-quark jet, a light-jet rejection factor of approximately 700 and a charm-jet rejection factor of $8$, as determined for jets with $\pt > 25$~\GeV\ and
$|\eta| < 2.5$ in simulated $t\bar{t}$ events~\cite{btagpaper}. 
The tagging efficiencies obtained in simulation are adjusted to match the
results of the calibrations performed in data~\cite{btagpaper}.

%-------------------------------------------------------------------------------

\section{Event selection}
\label{sec:selection}

This search is based on data collected using a multijet trigger, which
requires at least five jets passing the EF stage,
each having $\pt>55$~\GeV\ and $|\eta|<2.5$. Events are discarded if any jet 
with $\pt > 20$~\GeV\ is identified as out-of-time activity 
from a previous $pp$ collision or as calorimeter noise~\cite{ref:ATLAS-CONF-2012-020}.

The five leading jets in $\pt$ are required to have $\pt > 55$~\GeV\ with $|\eta|<2.5$ and all other jets are
required to have $\pt > 25$~\GeV\ and  $|\eta|<2.5$. Events are required to have at least six jets, of which at least two must be $b$-tagged.
Events with well-identified isolated muons or electrons with  $\pt > 25$~\GeV\ are discarded in order to avoid overlap with other \tth\ analyses. 

To enhance the sensitivity, the selected events are categorised into various distinct regions, according to their jet and $b$-tag multiplicities: 
the region with $m$ jets, of which $n$ are $b$-jets, is referred to as ``($m {\rm j}, n {\rm b}$)''.

\newcommand{\ShOL}{{\sc Sherpa+OpenLoops}}

\section{Signal and background modelling}
\label{sec:sigback}

%-------------------------------------------------------------------------------
\subsection{Signal model}
\label{sec:sigmodel}

The \tth\ signal process is modelled
using matrix elements calculations obtained from the HELAC-Oneloop package~\cite{Helac}
with next-to-leading order (NLO) accuracy in $\alpha_{\rm s}$. {\sc Powheg-box}~\cite{powheg,powbox1,powbox2} 
serves as an interface to the MC programs used to simulate the parton shower and hadronisation. The samples 
created using this approach are referred 
to as {\sc PowHel} samples~\cite{ttH_NLO}. They include all {SM Higgs boson 
and top-quark decays and 
use the {\sc CT10NLO}~\cite{ct10} parton 
distribution function (PDF) sets with the factorisation ($\mu_{\rm F}$) and
renormalisation ($\mu_{\rm R}$) scales set to 
$\mu_{\rm F} = \mu_{\rm R} = \mt+\mH/2$. The {\sc PowHel} \ttH\ samples
  use {\sc Pythia} 8.1~\cite{PythiaManual8} to simulate the parton
  shower with the {\sc CTEQ6L1}~\cite{cteq6} PDF and the AU2 underlying-event
  set of generator parameters (tune)~\cite{ATLAS:2012uec}, while {\sc
    Herwig}~\cite{herwig} is used to estimate systematic uncertainties due to
  the  fragmentation modelling. 

For these $t\bar{t}H$ samples the cross-section normalisations and the Higgs
boson decay branching fractions are taken from the NLO QCD and from the NLO
QCD + EW theoretical calculations~\cite{lhcxs} respectively.  
The masses of the Higgs boson and the top quark are set to
$125$~\GeV\ and to $172.5$~\GeV\ respectively.

\subsection{Simulated backgrounds}
\label{sec:bgmodel}

The dominant background to the all-hadronic \tth\ signal is multijet production,
followed by $t\bar{t}$\ +\ jets production. 
Small background contributions come from
the production of a single top quark and from the associated production of
a vector boson and a $t\bar{t}$ pair, $t\bar{t}V$ $(V=$ \textit{W}, \textit{Z}$)$.
The multijet background is determined from 
data using a dedicated method described in Section~\ref{sec:multijet}.
The other background contributions are estimated using MC simulations.

The multijet events, which
are used for jet trigger studies and for the validation of the data-driven multijet background estimation, are simulated 
with {\sc Pythia} 8.1 using the NNPDF2.3 LO~\cite{nnpdf} PDFs.

The main $t\bar{t}$ sample is generated using the {\sc Powheg} NLO generator
with the {\sc CT10NLO} PDF set, assuming a value of the top-quark mass of $172.5$~\GeV. It is 
interfaced to {\sc Pythia} 6.425~\cite{PythiaManual} with the {\sc CTEQ6L1} PDF set and the Perugia2011C~\cite{Skands:2010ak} underlying-event tune; this
combination of generator and showering programs is hereafter referred to  as {\sc Powheg}+{\sc Pythia}. 
The sample is normalised to the top++2.0 theoretical calculation performed at 
next-to-next-to leading order (NNLO) in QCD and includes resummation of next-to-next-to leading logarithmic (NNLL) soft gluon terms~\cite{ref:xs1,ref:xs2,ref:xs3,ref:xs4,ref:xs5,ref:xs6}. 
A second $t\bar{t}$ sample is generated using fully matched NLO predictions with massive $b$-quarks~\cite{sherpa_nlo_ttbb} 
within the {\sc Sherpa} with {\sc OpenLoops} framework~\cite{Gleisberg:2008ta,open_loops} 
henceforth referred to as \ShOL. The \ShOL\ NLO sample is generated
following the four-flavour scheme using the {\sc Sherpa}2.0 pre-release and 
the {\sc CT10NLO} PDF set.   
The renormalisation scale is set to   
$\mu_{\rm R}=\prod_{i=t,\bar{t},b,\bar{b}}\;  E_{\mathrm{T},i}^{1/4} $,
where $E_{\mathrm{T},i}$ 
is the transverse energy of parton $i$, and the 
factorisation and resummation scales are both set
to $(E_{\mathrm{T}, t }+E_{\mathrm{T}, \bar{t} })/2$.

The prediction from \ShOL\ is expected to model the \ttbb\
contribution more accurately than {\sc Powheg}+{\sc Pythia}, since the
  latter MC produces \ttbb\ exclusively via the parton shower. The
  \ShOL\ sample is not passed through full detector simulation. Thus,
  $t\bar{t}$\ +\ jets events from  {\sc Powheg}+{\sc Pythia} are categorised
  into three non-overlapping samples, $t\bar{t}$\ +\ $b\bar{b}$,
  $t\bar{t}$\ +\ $c\bar{c}$, and $t\bar{t}$\ +\ light-jets, hereafter called  $t\bar{t}$\ +\ light, using a labelling based
  on an algorithm that matches hadrons to particle jets. Then,
  \ttbar+\ \bbbar\ events from {\sc Powheg}+ {\sc Pythia}  are
  reweighted to reproduce the \ShOL\ NLO \ttbar+\ \bbbar prediction.
  The reweighting is done at generator level using a finer categorisation to distinguish  
 events where one particle jet is matched to two $b$-hadrons, or
  where only one $b$-hadron is matched. The reweighting is
  applied using several kinematic variables such as the top-quark \pt,
 the \ttbar\ system \pt, and, where this can be defined, $\Delta R$ and \pt\ of the
 dijet system not originating from the top-quark decay~\cite{Aad:2015gra}.

Unlike \ttbar+\ \bbbar, no fully matched NLO predictions exist for \ttbar+\ \ccbar\ and  $t\bar{t}$\ +\ light events.
A dedicated reweighting is therefore applied to the top-quark $\pt$
  spectra as well as to the \pt\ spectra of the $\ttbar$ system of
  \ttbar+\ light and \ttbar+\ \ccbar\ events in {\sc Powheg}+{\sc Pythia},
  based on the ratio  of data to simulation of the measured
  differential cross sections at $\sqrt{s}=7$~\TeV~\cite{topdiff_7TEV}. 
No such reweighting is applied to the \ttbar+\ \bbbar\ sample, which is
already corrected to match the best available theory calculation. 

Samples of single-top-quark events produced in the $s$- and $Wt$-channels are generated with 
{\sc Powheg-box} 2.0 using the {\sc CT10NLO} PDF set. The samples are interfaced 
to {\sc Pythia} 6.425 with the {\sc CTEQ6L1} set of parton 
distribution functions and Perugia2011C underlying-event tune.  
The $t$-channel production mode is generated with {\sc AcerMC}~\cite{acermc}
interfaced to {\sc Pythia} 6.425 with the {\sc CTEQ6L1} PDF set and the
Perugia2011C underlying-event tune. 
Overlaps between the \ttbar\ and $Wt$ final states are removed~\cite{mcatnlo_2}.
The single-top-quark samples are normalised to
the approximate NNLO theoretical cross sections~\cite{stopxs,stopxs_2}
using the {\sc MSTW2008} NNLO PDF set~\cite{mstw1,mstw2}. 

The samples of $t\bar{t}V$ ($V=W,Z$) events are generated with the {\sc MadGraph v5} LO
generator~\cite{madgraph} and the {\sc CTEQ6L1} PDF set. {\sc Pythia} 6.425 with the AUET2B tune is used to generate the parton shower. The $t\bar{t}V$ samples
are normalised to NLO cross-sections~\cite{ttbarVxs1,ttbarVxs2}.

Finally, event samples for single top quark plus Higgs boson production, $tHqb$ and $tHW$, are generated. The cross sections are computed using the MG5\_{\sc a}MC@NLO
  generator~\cite{Alwall:2014hca} at NLO in QCD. For \textit{tHqb}, samples are
  generated with {\sc MadGraph} in the four-flavour scheme and $\mu_{\rm F} =
  \mu_{\rm R} = 75$~\GeV\ then showered with  {\sc Pythia} 8.1 with the {\sc
    CTEQ6L1} PDF and the AU2 underlying-event tune. For \textit{tHW}, computed
  with the five-flavour scheme, dynamic $\mu_{\rm F}$ and $\mu_{\rm R}$ scales
  are used and events are generated at NLO with MG5\_{\sc a}MC@NLO+{\sc
    Herwig}++~\cite{Bahr:2008pv,Bellm:2013lba}.  
These two processes together are referred to as $tH$.  

A summary of the cross-section values and their uncertainties for the signal as well as for the MC simulated background processes is given in Table~\ref{tab:SigBkgXSec}.

\begin{table*}  
  \caption{\label{tab:SigBkgXSec} Production cross sections for signal \tth , at $m_H$ = 125~\GeV, and various simulated background processes. The quoted errors arise from variations of the renormalisation and factorisation scales and uncertainties in the parton distribution functions.}

  \small
\begin{center}
  \begin{tabular}{l|c}
\hline    
Process &  $\sigma$ [pb] \\
\hline 
\rule{0pt}{3ex} \rule[-2.0ex]{0pt}{0pt} \ttH & $0.129^{+0.012}_{-0.016}$ \\
\hline
\rule{0pt}{3ex}  \ttbar & $253^{+13}_{-15}$ \\ [0.1cm]
Single top $Wt$-channel & $ 22.4 \pm 1.5$ \\ [0.1cm]
Single top $t$-channel & $ 87.7^{+3.4}_{-1.9}$ \\ [0.1cm]
Single top $s$-channel & $5.61 \pm 0.22$  \\ [0.1cm]
$\ttbar$ + $W$ &  $0.232 \pm 0.070$ \\[0.1cm]
$\ttbar$ + $Z$ &  $0.205 \pm 0.061$  \\[0.1cm]
$tHqb$ &  $0.0172^{+0.0012}_{-0.0011}$  \\[0.1cm]
\rule[-2.0ex]{0pt}{0pt}$WtH$ &  $0.0047^{+0.0010}_{-0.0009}$  \\[0.1cm]
\hline
\end{tabular}
\end{center}  
\end{table*}

\subsection{Common treatment of MC samples}
\label{sec:MC}

All samples using {\sc Herwig} are also interfaced to {\sc Jimmy}
v4.31~\cite{jimmy} to simulate the underlying event.
With the exception of {\sc Sherpa}, all MC samples use {\sc Photos 2.15}~\cite{PhotosPaper} to simulate
photon radiation and {\sc Tauola 1.20}~\cite{TauolaPaper} to simulate
$\tau$ decays.  The samples are then processed through a simulation~\cite{atlas_sim}
of the detector geometry and response using {\sc Geant4}~\cite{geant}.
The single-top-quark sample produced in the $t$-channel is simulated with a parameterised
calorimeter response~\cite{ATLFAST2}.

All simulated events are processed through the same reconstruction 
software as the data. Simulated events are
corrected so that the lepton and jet identification efficiencies, energy
scales and energy resolutions match those in data.

When selecting based on the output value of the $b$-tagging algorithm, the number of selected simulated events is significantly reduced, leading to large statistical fluctuations in the resulting distributions for samples with a high $b$-tag multiplicity. Therefore, rather than tagging the jets individually, the normalisation and the shape of these distributions are predicted by calculating the probability that a jet with a given flavour, \pt, and \eta is $b$-tagged~\cite{D0btagPRD}. The method is validated by verifying that the predictions reproduce the normalisation and shape obtained for a given working point of the $b$-tagging algorithm. The method is applied to all simulated signal and background samples.

\subsection{Multijet background estimation using data: the \trfmj method}
\label{sec:multijet}

A data-driven technique, the tag rate function for multijet events
(\trfmj) method, is used to estimate the multijet background. After measuring
\mjeps, the probability of $b$-tagging a third jet in a sample of events
with at least two $b$-tagged jets, the \trfmj method uses \mjeps to
extrapolate the multijet background from the regions with lower $b$-tag
multiplicity to the search regions with higher $b$-tag multiplicity but
otherwise identical event selection. 

In the first step, the $b$-tagging rate is measured in data samples selected
with various single-jet triggers, which are enriched in multijet events and have limited
($\approx$10\%) overlap with the search region.
 The events in this \trfmj extraction region are required to have at
least three jets with $\pt > $ 25~\GeV\ and $|\eta|< 2.5$, with at least two
$b$-tagged jets.
Excluding the two jets with the highest $b$-tagging weight in the event,
\mjeps is defined as the rate of $b$-tagging any other 
jet in the event. It is parameterised as a function of the jet \pt and
$\eta$, and also of the average $\Delta R$ between this jet and the two jets in the event with
highest $b$-tagging weight, $\langle \Delta R_{(j,\text{hMV1})} \rangle$.
The \pt and \eta dependence of \mjeps reflects the corresponding sensitivity of the $b$-tagging efficiency to these variables.
In multijet events, the $\Delta R$ dependence of \mjeps is correlated with the multi-$b$-jet production mechanism. 
This affects \mjeps, shown in Figure~\ref{fig:trfqcd}, which decreases by up to a factor two as  $\Delta R$ increases for fixed \pt and \eta.

\begin{figure}[ht!]
\centering
\includegraphics[width=1.02\textwidth]{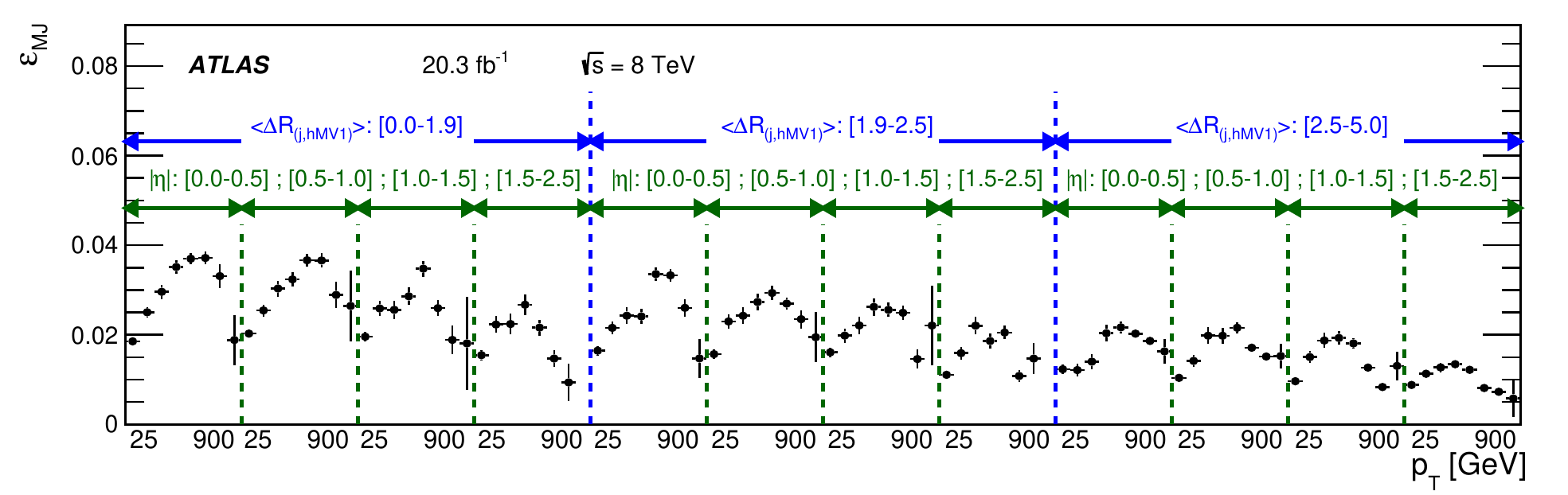}
\caption{Dependence of \mjeps on the jet transverse momentum \pt, in regions of jet pseudorapidity $\eta$ and average $\Delta R$ between this jet and the two jets in the event with
highest $b$-tagging weight, $\langle \Delta R_{(j,\text{hMV1})} \rangle$.
The \pt bin boundaries are 25 (lowest), 40, 55, 70, 100, 200, 400, 600, 900~\GeV\ (highest), chosen such as to have uniform number of events across bins of $\langle \Delta R_{(j,\text{hMV1})} \rangle$.}
\label{fig:trfqcd}
\end{figure}
 
In the search region the \trfmj method starts from the data sample with exactly two $b$-tagged jets
subtracting the contributions from all other backgrounds obtained from
MC simulation. Multijet background samples containing $m$ jets ($m \geq 6$),
out of which $n$ are $b$-tagged ($n \geq 3$) are then constructed, 
using an event weight $w$($m {\rm j}, n {\rm b}$), which is calculated from
$\mjeps$ analogously to the method described in Ref.~\cite{D0btagPRD}, accounting for the fact that the starting sample contains two $b$-tagged jets. In each multijet event emulated using \trfmj by means of $\mjeps$, $(m-2)$ jets not originally $b$-tagged can be used for the emulation of the properties of additional $b$-tagged jets. This procedure allows to emulate observables that depend on the number of $b$-tagged jets.

\subsection{Validation of the \trfmj method in data and simulation}

Validation of the \trfmj method is performed by a `closure test', separately in data and simulation. This is performed using the same data samples that were employed to
estimate \mjeps.
In these low jet multiplicity samples, the \trfmj method, which is applied to the events with exactly two $b$-tagged jets, is used to predict distributions in events with at least three $b$-tagged jets.  
Using \mjeps derived independently in data and simulation, the predicted distributions are  compared to those resulting when directly applying $b$-tagging. This is done for a number of variables, such as $b$-tagged jet \pt, angular distance between $b$-tagged jets, and event shapes. As an example, 
for events with at least three jets and at least three $b$-tagged jets
($\geq$3j, $\geq$3b), Figure~\ref{fig:data_33} shows the closure test in data for the third-leading-jet \pt, \HT (the scalar sum of the $\pt$ of all jets), and Centrality$_{\text{Mass}}$ (defined as \HT divided by the invariant
  mass of the jets).
Figure~\ref{fig:mc_34} shows the results of the closure test in simulated multijet events for distributions of the leading-jet \pt, the minimum mass of all jet pairs 
in the event ($m_{jj}^\text{min}$),
 and the third-leading $b$-tagged jet \pt. The definitions of these variables can be found in
 Table~\ref{tab:BDTVariables}.
 In both data and simulated multijet events with at least three $b$-tagged jets, the predicted and observed number of events agree within 5\%. In events with a higher $b$-tagged jet
  multiplicity the numbers agree within the large statistical uncertainty. For
  this reason the systematic uncertainties related to the \trfmj method are
  not estimated in the validation regions.

\begin{figure*}[!ht]
  \begin{center}   
    \subfloat[]{\includegraphics[width=0.32\textwidth]{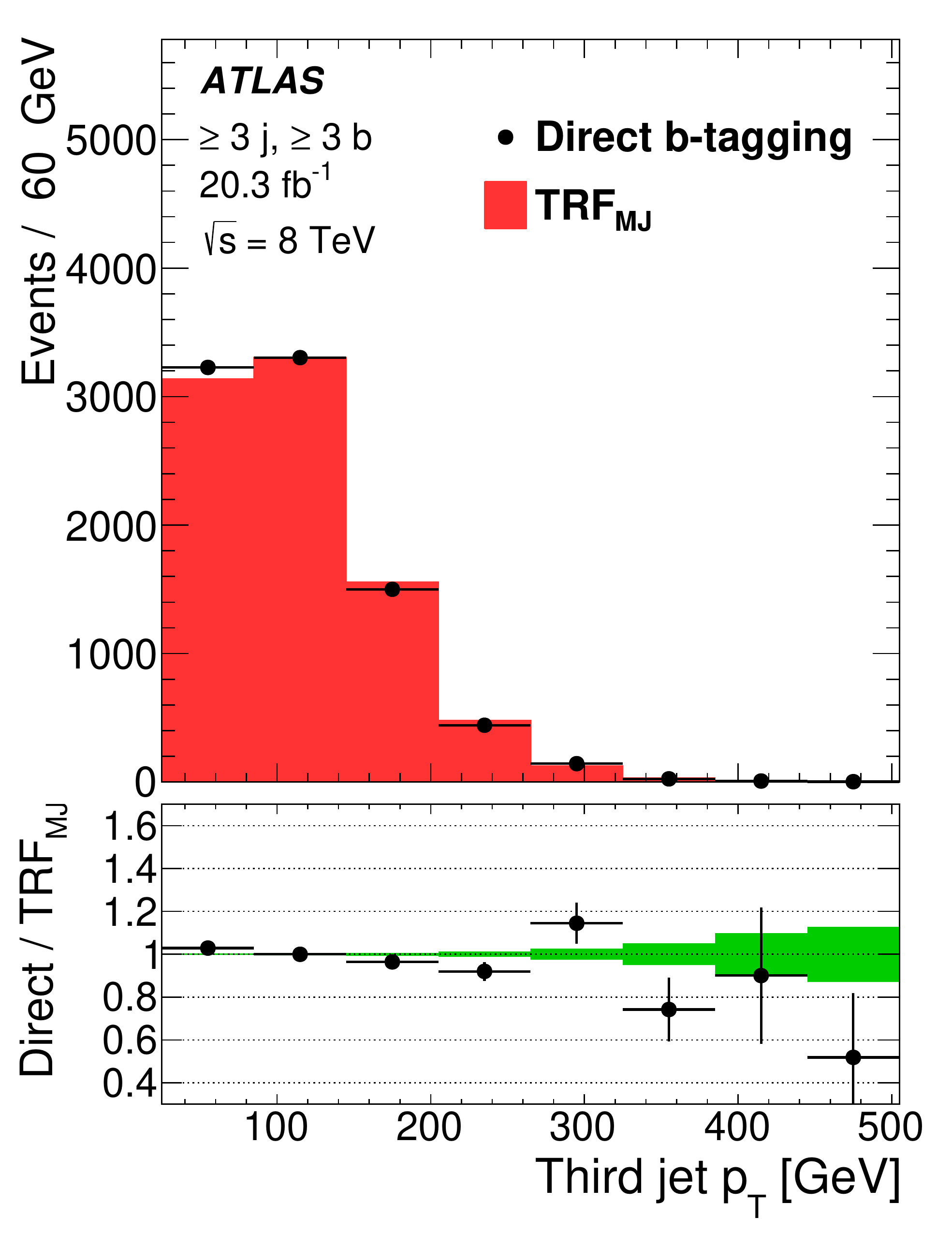}}\label{fig:TRFQCD_Validation_a}
    \subfloat[]{\includegraphics[width=0.32\textwidth]{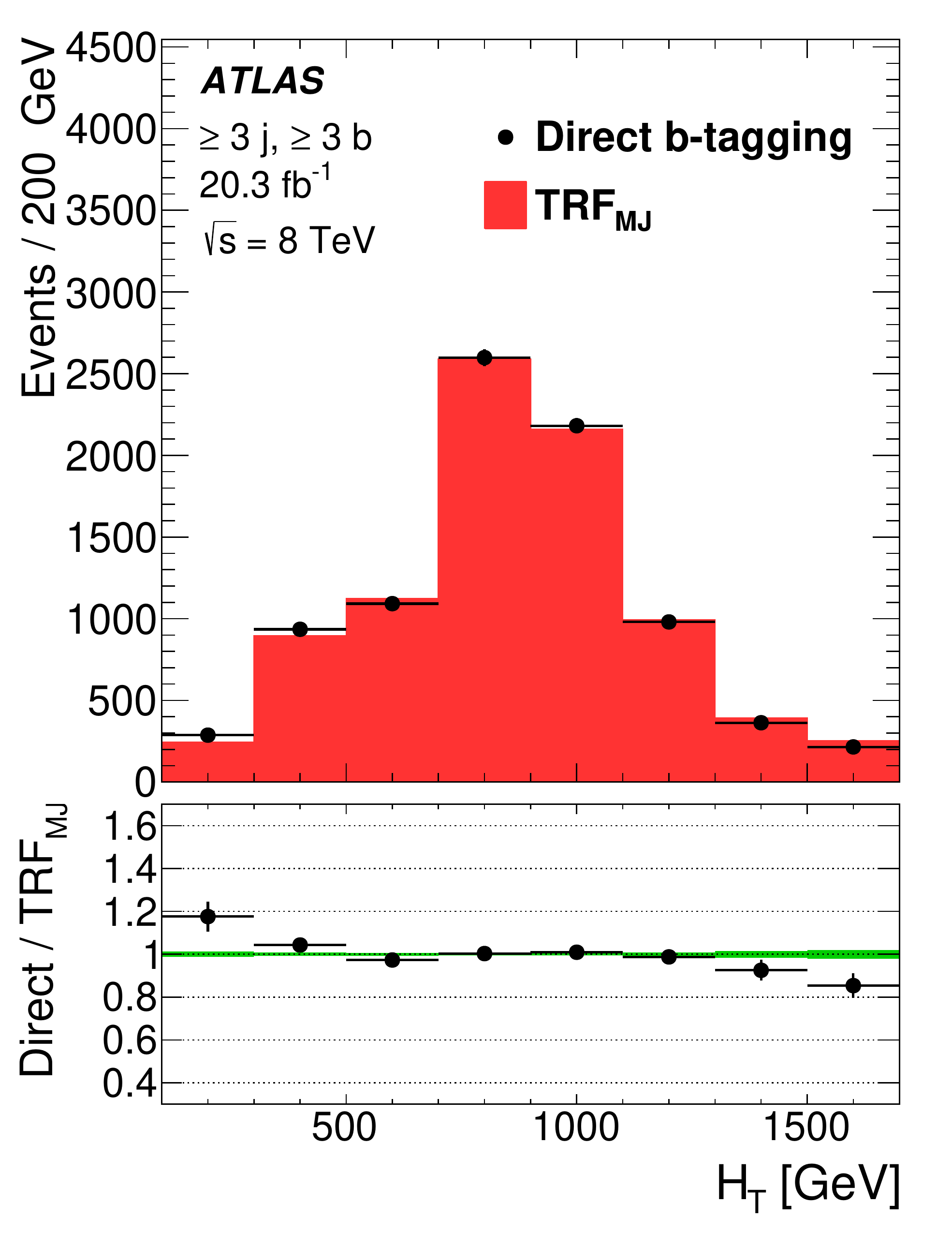}}\label{fig:TRFQCD_Validation_b}    
    \subfloat[]{\includegraphics[width=0.32\textwidth]{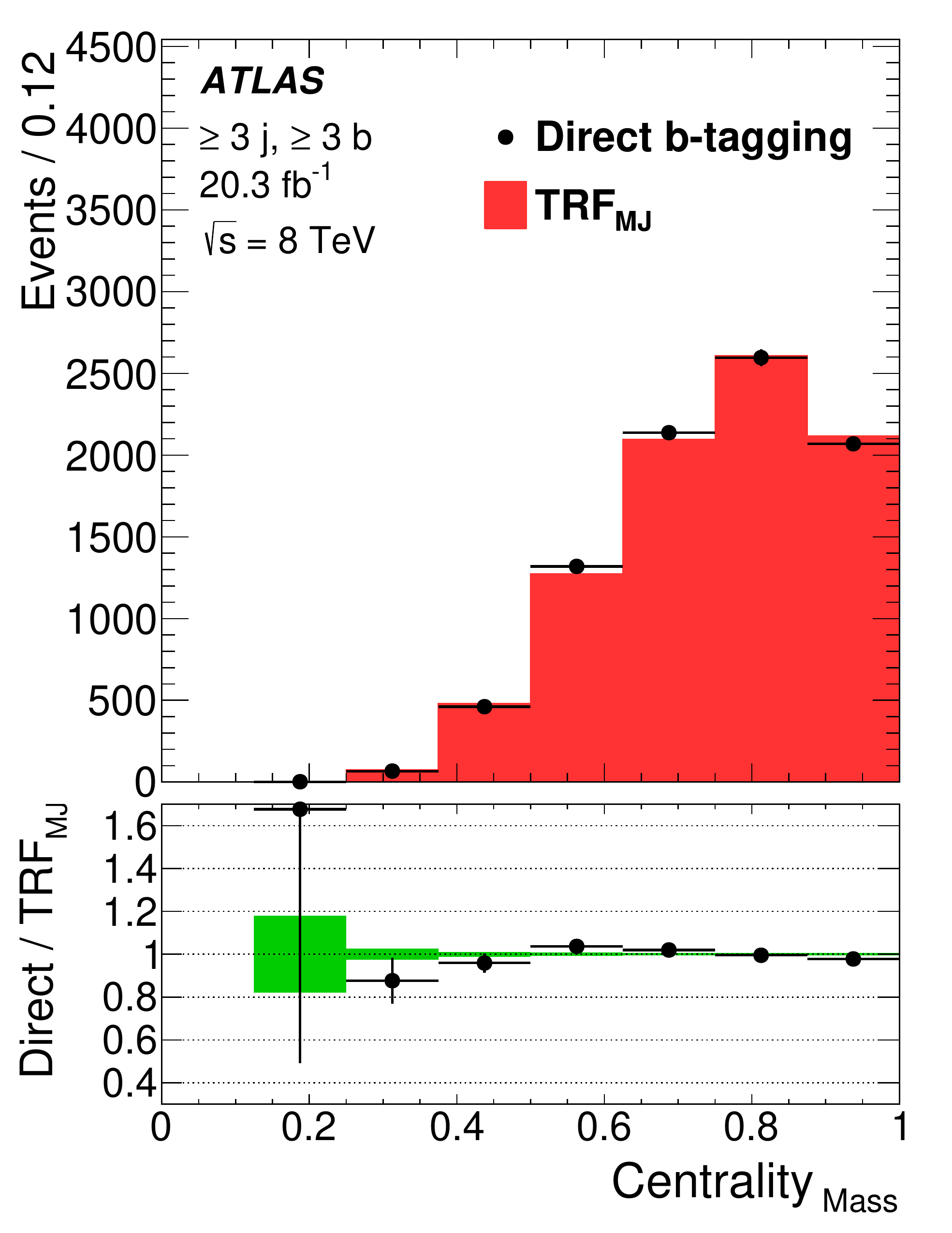}}\label{fig:TRFQCD_Validation_c}
    \caption{Comparison of the shapes predicted by the $\text{TRF}_{\text{MJ}}$
      method (red histograms) and direct $b$-tagging (black circles) in data events with at least three 
      jets and at least three $b$-tagged jets for (a) the third-leading $b$-tagged jet
      \pt, (b) \HT, and (c) Centrality$_{\text{Mass}}$. The definitions of the variables are listed in Table~\ref{tab:BDTVariables}. Events were selected with
      various single-jet triggers. The $\text{TRF}_{\text{MJ}}$ prediction is
      normalised to the same number of events as the data. The
        uncertainty band for the \trfmj predictions shown in the ratio plot
        represents statistical uncertainties only.}
    \label{fig:data_33} 
  \end{center}
\end{figure*}

\begin{figure*}[!ht]
  \begin{center}   

    \subfloat[]{\includegraphics[width=0.32\textwidth]{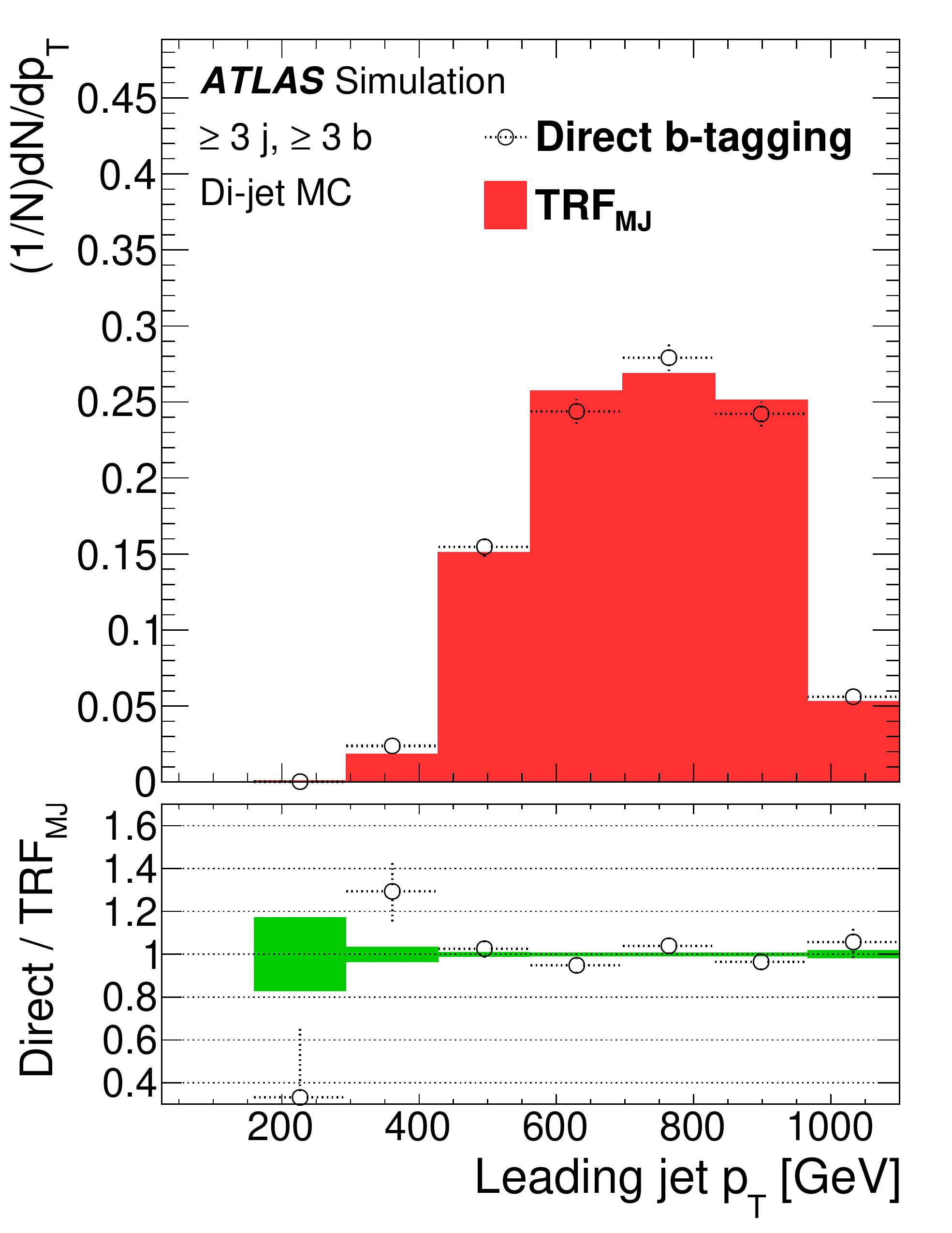}}\label{fig:TRFQCD_Validation_d}
    \subfloat[]{\includegraphics[width=0.32\textwidth]{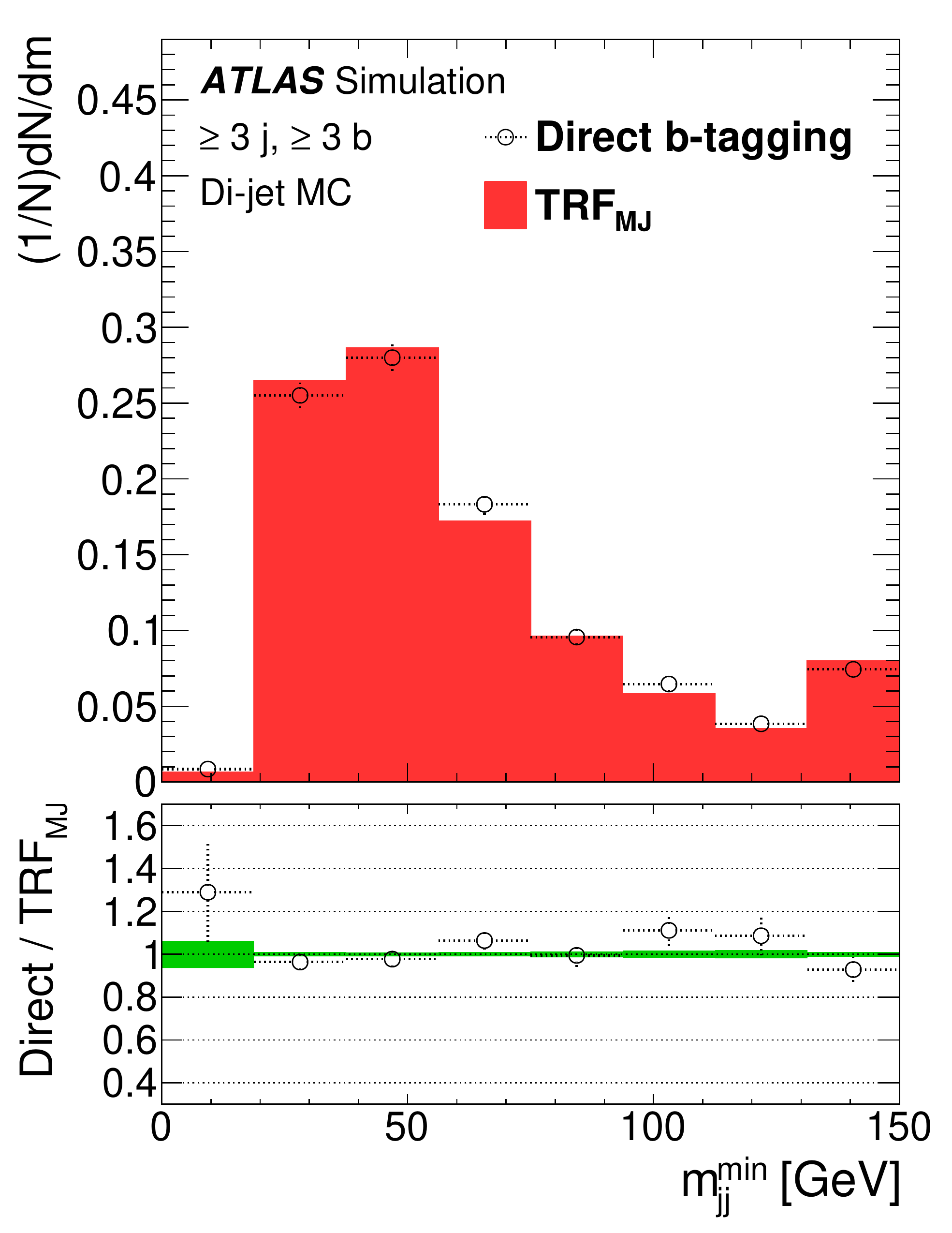}}\label{fig:TRFQCD_Validation_e}
    \subfloat[]{\includegraphics[width=0.32\textwidth]{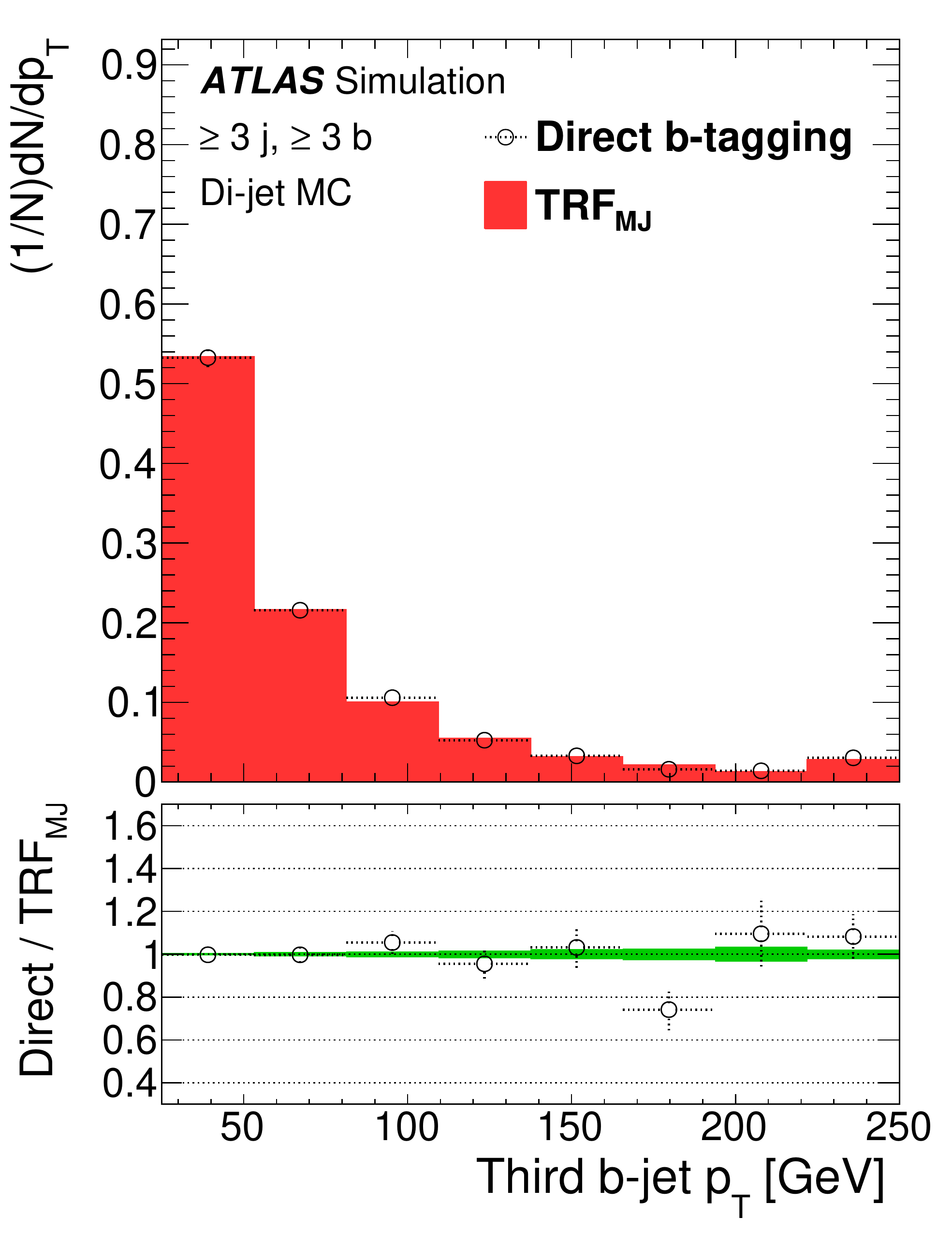}}\label{fig:TRFQCD_Validation_f}
    \caption{Comparison of the shapes predicted for the $\text{TRF}_{\text{MJ}}$
      method (red histograms) and direct $b$-tagging (black
      circles) in {\sc Pythia 8.1} multijet events with at least three jets and at
      least three $b$-tagged jets for (a) leading-jet \pt, (b) $m_{jj}^\text{min}$ 
      and (c) the third-leading $b$-tagged jet \pt in the event. The definitions of the variables are listed in Table~\ref{tab:BDTVariables}. Distributions are
      normalised to the same area. The uncertainty band for the \trfmj
        predictions shown in the ratio plot represents statistical
        uncertainties only.}
    \label{fig:mc_34} 
  \end{center}
\end{figure*}

\section{Multijet trigger efficiency}
\label{sec:trigger}

Not all jets are reconstructed at the trigger level, mainly due to the Level-1 sliding window algorithm and the Level-1 resolution~\cite{1748-0221-3-03-P03001}. %As a result, not all multijet events pass the trigger requirement.
The multijet trigger efficiency with respect to the offline selection is derived in terms of the efficiency for a single
jet to be associated with a complete jet trigger chain, i.e., a complete
  sequence of jets reconstructed at Level-1, Level-2 and EF satisfying the
  requirements described in Section~\ref{sec:selection}. This single-jet
trigger efficiency, $\epsilon_{\mathrm{trig}}$, is evaluated in intervals of offline reconstructed \pT\ and $\eta$:  
\begin{equation}
  \label{eq:trig:trigjetprob}
  \epsilon_{\mathrm{trig}} ( \pt, \eta ) = \frac{N_{\mathrm{trig}}  ( \pt, \eta )}{N ( \pt, \eta )},
\end{equation}
where $N_{\mathrm{trig}} ( \pt , \eta )$ is the number of jets matched with a
trigger chain and $N ( \pt, \eta)$ is the total number of jets within a given
 offline reconstructed $\pt$ and $\eta$ interval. 
Figure~\ref{fig:trig_eff} shows that for large jet $\pt$, $\epsilon_{\mathrm{trig}}$ reaches a plateau close to unity. 

For both data and simulation, $ \epsilon_{\mathrm{trig}} ( \pt, \eta )$ is derived using events triggered by a single-jet trigger with a \pt\ threshold of 110~\GeV, and only the
offline jets which are in the hemisphere opposite to the trigger jet are used.  To avoid additional trigger bias, events are
discarded if more than one jet with $\pt \geq 110 $~\GeV\ is reconstructed.
The ratio of $ \epsilon^{\mathrm{data}}_{\mathrm{trig}} ( \pt, \eta )$ to 
$ \epsilon^{\mathrm{MC,dijet}}_{\mathrm{trig}}$, where the latter is estimated in simulated dijet events, is referred to as $\mathrm{SF}_{\mathrm{trig}} ( \pt, \eta )$.
 In the analysis,
for each MC sample $\alpha$ considered, the final number of events passing the multijet
trigger is estimated by weighting each jet by the product of $ \epsilon^{\mathrm{MC},\alpha}_{\mathrm{trig}}( \pt, \eta )$
and $\mathrm{SF}_{\mathrm{trig}}( \pt, \eta )$. The parameters
$\epsilon_{\mathrm{trig}} ( \pt, \eta )$ and $\mathrm{SF}_{\mathrm{trig}}(
\pt, \eta )$ are estimated for jet $\pt$ up to 100~\GeV. Figure~\ref{fig:trig_eff} shows the \pt dependence of
$\epsilon^{\mathrm{data}}_{\mathrm{trig}} ( \pt, \eta )$,
$\epsilon^{\mathrm{MC},\tth}_{\mathrm{trig}}( \pt, \eta )$,
$\epsilon^{\mathrm{MC,dijet}}_{\mathrm{trig}}( \pt, \eta )$ and
$\mathrm{SF}_{\mathrm{trig}}( \pt, \eta )$ for jets within $|\eta| <
2.5$, 
together with the uncertainties from the difference between
$\epsilon^{\mathrm{MC},\tth}_{\mathrm{trig}}( \pt, \eta )$ and
$\epsilon^{\mathrm{MC,dijet}}_{\mathrm{trig}}( \pt, \eta )$, which is taken as
the systematic uncertainty of the method.  

\begin{figure*}[!ht]
  \begin{center}
    \includegraphics[width=0.48\textwidth]{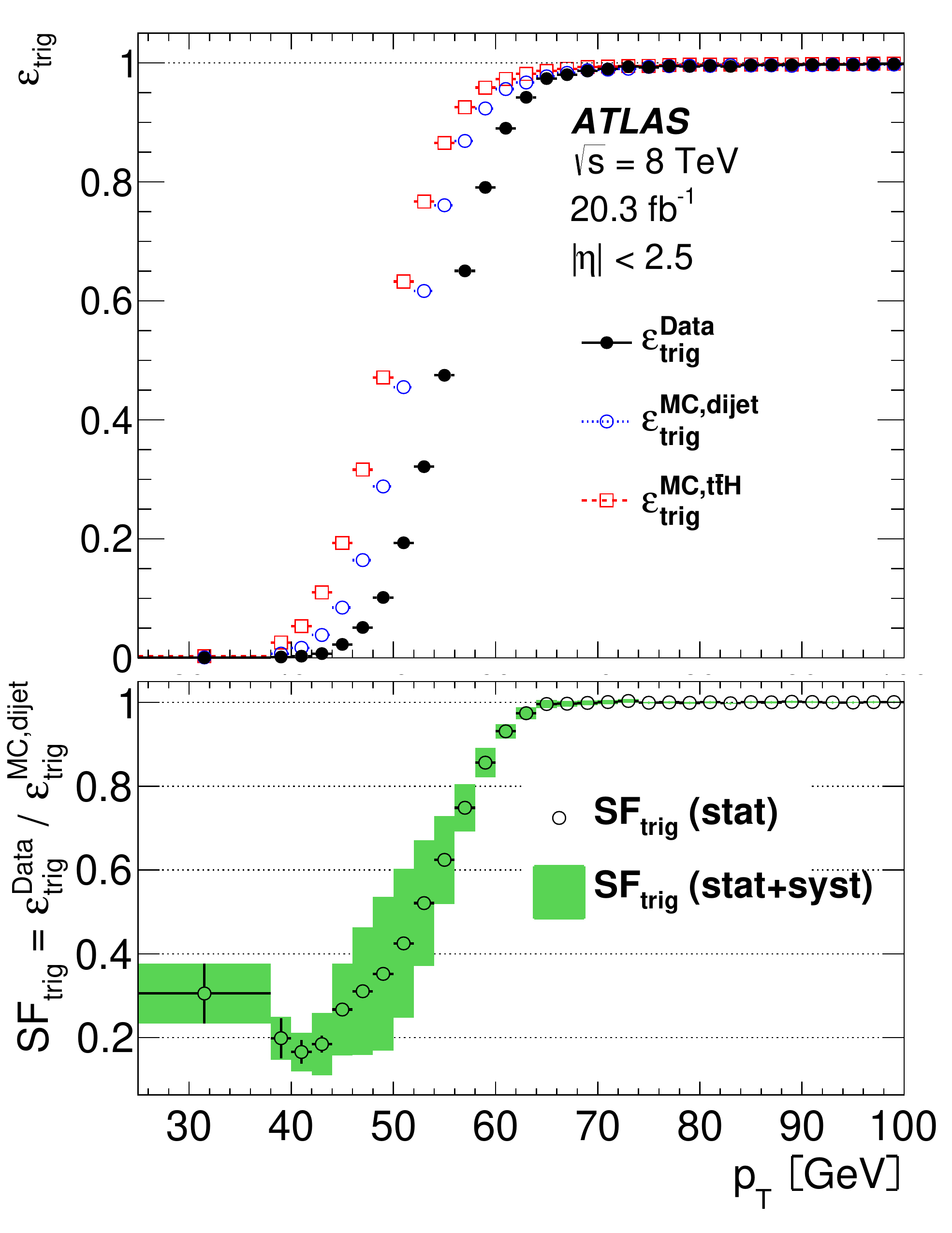}\label{fig:trig_eff_b}\\
    \caption{Single-jet trigger efficiencies, $\epsilon_{\mathrm{trig}}$, (top) for data, 
      simulated dijet events, and \tth\ events,
      as a function of jet $\pt$ for jets with $|\eta| < 2.5$;
      (bottom) $\mathrm{SF}_{\mathrm{trig}}( \pt, \eta )=
      \epsilon^{\mathrm{data}}_{\mathrm{trig}} ( \pt, \eta )/
      \epsilon^{\mathrm{MC,dijet}}_{\mathrm{trig}}( \pt, \eta )$. The
      uncertainty on  $\mathrm{SF}_{\mathrm{trig}}$, shown as the green shaded
      area, is estimated from the difference between the efficiencies in dijet and \tth\ simulated events in the denominator of $\mathrm{SF}_{\mathrm{trig}}$.}
    \label{fig:trig_eff} 
  \end{center}
\end{figure*}

\section{Event classification}
\label{sec:eventclass}

Six independent analysis regions are considered for the fit used in the
analysis: two control regions (6j, 3b), (6j, $\geq$4b) and four signal regions (7j, 3b), (7j, $\geq$4b), ($\geq$8j, 3b) and 
($\geq$8j, $\geq$4b). In addition, the three regions with exactly two $b$-tagged jets, (6j, 2b), (7j, 2b) and ($\geq$8j, 2b), are used to predict the multijet contribution to higher $b$-tagging multiplicity
regions, using the $\text{TRF}_{\text{MJ}}$ method, as described above.
The event yields in the different analysis regions prior to the fit are summarised in Table~\ref{tab:prefityields}.  

\begin{table*}  
  \caption{\label{tab:prefityields} Event yields from simulated
    backgrounds and the signal as well as data in each of the analysis regions prior to the fit (pre-fit). The quoted
uncertainties are the sum in quadrature of the statistical and
systematic uncertainties in the yields for all samples but the multijet
background. The multijet normalisation and its systematic uncertainty are
determined by the fit, so only its statistical uncertainty is quoted
here. Since the numbers are rounded, the sum of all contributions may not equal the
total value. The signal-to-background ratio, $S/B$, and the significance,
$S/\sqrt{\text{B}}$, are also given. The $tH$ background is not shown as it amounts to fewer than 1.5 events in each region. }

  \small
\begin{center}
\begin{tabular}{l|ccccccc}
\hline
                      & 6j, 3b           &  6j, $\geq$4b  & 7j, 3b           &7j, $\geq$4b  & $\geq$8j, 3b             &  $\geq$8j, $\geq$4b \\ \hline  
Multijet	      & \phantom{.}16380 $\pm$ 130\phantom{.} & \phantom{.}1112 $\pm$ 33\phantom{.} & \phantom{.}12530 $\pm$ 110\phantom{.} & \phantom{.}1123 $\pm$ 34\phantom{.} & \phantom{.}10670 $\pm$ 100\phantom{.} & \phantom{.}1324 $\pm$ 36\phantom{.} \\
$t\bar{t}$+light      &	\phantom{0.}1530 $\pm$ 390\phantom{.}  & \phantom{0.0}48 $\pm$ 18\phantom{.}   & \phantom{0.}1370 $\pm$ 430\phantom{.}  & \phantom{0.0}45 $\pm$ 18\phantom{.}   & \phantom{0.}1200 $\pm$ 520\phantom{.}  & \phantom{0.0}40 $\pm$ 23\phantom{.0} \\
$t\bar{t}+c\bar{c}$   & \phantom{0.0}280 $\pm$ 180\phantom{.}   & \phantom{0.0}17 $\pm$ 12\phantom{.}   & \phantom{0.0}390 $\pm$  240\phantom{.}   & \phantom{0.0}21 $\pm$  15\phantom{.}   & \phantom{00.}560 $\pm$ 350\phantom{.}   & \phantom{0.0}48 $\pm$ 33\phantom{.0} \\	 
$t\bar{t}+b\bar{b}$   & \phantom{0.0}330 $\pm$ 180\phantom{.}   & \phantom{0.0}44 $\pm$ 26\phantom{.}   & \phantom{0.0}490 $\pm$ 270\phantom{.}   & \phantom{0.0}87 $\pm$ 51\phantom{.}   & \phantom{00.}760 $\pm$ 450\phantom{.}   & \phantom{.0}190 $\pm$ 110\phantom{.} \\
$t\bar{t}+V$	      & \phantom{00}14.2 $\pm$ 6.3\phantom{0}    & \phantom{00}1.8 $\pm$ 1.5 & \phantom{00}22.0 $\pm$ 9.0\phantom{0}  & \phantom{00}3.5 $\pm$ 2.3 & \phantom{00.0}40 $\pm$ 15\phantom{0.} & \phantom{00}8.0 $\pm$ 4.2\phantom{0}\\
Single top	      & \phantom{0.0}168 $\pm$ 63\phantom{.0}    & \phantom{00}6.0 $\pm$ 3.7 & \phantom{0.0}139 $\pm$ 55\phantom{0.}    & \phantom{00}8.3 $\pm$ 4.6 & \phantom{00.}110 $\pm$ 49\phantom{.0}    & \phantom{0}10.6 $\pm$ 5.9\phantom{0}\\ \hline
Total background      & \phantom{.}18700 $\pm$ 480\phantom{.} & \phantom{.}1229 $\pm$ 48\phantom{.} & \phantom{.}14940 $\pm$ 580\phantom{.} & \phantom{.}1288 $\pm$ 66\phantom{.} & \phantom{.}13330 $\pm$ 780\phantom{.}  & \phantom{.}1620 $\pm$ 130\phantom{.} \\ \hline
$t\bar{t}H$ ($m_H$=125~\GeV)     & \phantom{00}14.3 $\pm$ 4.6\phantom{0}  & \phantom{00}3.3 $\pm$ 2.1 & \phantom{00}23.7 $\pm$ 6.4\phantom{0}  & \phantom{00}7.2 $\pm$ 3.3 & \phantom{000.}48 $\pm$ 11\phantom{0.}     & \phantom{0}16.8 $\pm$ 6.1\phantom{0} \\ \hline
Data events                  & 18508           & 1545          & 14741           & 1402          & 13131           & 1587 \\ \hline
$S/B$                   & < 0.001         & 0.003         & 0.002           & 0.006         & 0.004           & 0.010 \\
$S/\sqrt{\text{B}}$    & 0.10           & 0.095         & 0.194           & 0.20          & 0.415           & 0.417 \\ \hline
 
\end{tabular}
\end{center} 
\end{table*} 

The regions are analysed separately and combined statistically
to maximise the overall sensitivity. The most sensitive regions, ($\geq$8j,
3b) and ($\geq$8j, $\geq$4b), are expected to contribute more than
50\% of the total significance.

\section{Analysis method}
\label{sec:analysis}

The Toolkit for Multivariate Data Analysis (TMVA)~\cite{ref:TMVA} is
used to train a BDT to separate the~\tth\ signal
from the background. A dedicated BDT is defined and optimised in each of the six analysis regions. 
The variables entering the BDT and their definitions are listed in
Table~\ref{tab:BDTVariables}.

\begin{landscape} 
\null\vfill
\begin{table}[ht!]
\caption{List of variables used in the BDT in the six analysis regions. The
numbers indicate the ranking of the corresponding variables, ordered by
  decreasing discriminating power. Variables not used in the BDT of a specific
  region are marked by a dash.}
  \centering     % 1 2 3 4 5 
\resizebox{22 cm}{!}{%
\begin{tabular}{|l|l|c|c|c|c|c|c|}
\hline        
\multirow{ 2}{*}{Variable} & \multirow{ 2}{*}{Definition}  & \multicolumn{6}{c|}{BDT rank}
\\ %\hline 
\cline{3-8}
& &  $\,6 {\rm j},\,3 {\rm b}$ 
& $\,6 {\rm j},\ge\,4 {\rm b}$ 
& $\,7 {\rm j},\,3 {\rm b}$ 
& $\,7 {\rm j},\ge\,4 {\rm b}$ 
& $\ge\,8 {\rm j},\,3 {\rm b}$ 
& $\ge\,8 {\rm j},\ge\,4 {\rm b}$  \\\hline
$\text{Centrality}_\text{Mass}$                            & Scalar sum of the jet $\pt$ divided by the invariant mass of the jets       &  1  &  1  &  1  &  1  &  9  &  6 \\[0.1cm]
\multirow{2}{*} {Aplanarity}              &  $1.5 \lambda_2$, where $\lambda_2$ is the second eigenvalue of the momentum & \multirow{2}{*} {--}   &  \multirow{2}{*} {11}  &  \multirow{2}{*} {--}  &  \multirow{2}{*} {--}  &  \multirow{2}{*} {6}  &  \multirow{2}{*} {--} \\ [-0.1cm]
                                          & tensor built with all jets   & & & & & &  \\
$S_\text{T}$                              & The modulus of the vector sum of jet \pt       & 2  &  2  & 2  &  4 &  2  &  2 \\[0.1cm]
$H_{\text{T}\,5}$                  & Scalar sum of jet $\pt$  starting from the fifth jet        &  8  &  --  &  --  &  7  &  --  &  -- \\[0.1cm]
$m_{jj}^\text{min}$                  & Smallest invariant mass of any combination of two jets      &  9  &  --  &  6   &  10  &  11  &  12 \\[0.1cm]
$\Delta R^\text{min}$                        & Minimum $\Delta R$ between two jets      & 6  &  5  &  9  &  --  & 8  &  4 \\[0.1cm]
$p_\text{T}^\text{softest jet}$                      & $p_\text{T}$ of the softest jet       &  --  &  6  &  10  &  --  &  --  &  10 \\[0.1cm]
$\Delta R(b,b)^{p_\text{T}^\text{max}}$          & $\Delta R$ between two $b$-tagged jets with the largest vector sum $p_\text{T}$         &  11  &  --  &  7  &  5  &  5  &  3 \\[0.1cm]
$m_{bb}^{\Delta R(b,b)^\text{min}}$                 & Invariant mass of the combination of two $b$-tagged jets with the smallest $\Delta R$         &  3  &  3  &  8  &  9  &  3  &  9 \\[0.1cm]
$\frac{E_\text{T\,1} + E_\text{T\,2}}{\sum E_\text{T}^\text{jets}}$   & Sum of
the $E_{\text{T}}$ of the two jets with leading $E_\mathrm{T}$ divided by the sum of the $E_{\text{T}}$ of all jets        &  5  &  8   &  4   &  2  &  7  &  5 \\[0.1cm]
\multirow{2}{*} {$m_{2\,\text{jets}}$}                & The mass of the dijet pair,
  which, when combined with any $b$-tagged jet, & \multirow{2}{*} {10}   &  \multirow{2}{*} {--}  &  \multirow{2}{*} {--}  &  \multirow{2}{*} {8}  &  \multirow{2}{*} {--}  &  \multirow{2}{*} {--} \\ [-0.1cm]
                                          &  maximises the
  magnitude of the vector sum of the \pt of the three-jet system & & & &  & & \\[0.1cm]
\multirow{2}{*} {$m_{2\,\text{b--jets}}$}  & The invariant mass of the two $b$-tagged jets which are selected by requiring & \multirow{2}{*} {12}   &  \multirow{2}{*} {7}  &  \multirow{2}{*} {--}  &  \multirow{2}{*} {6}  &  \multirow{2}{*} {--}  &  \multirow{2}{*} {8} \\ [-0.1cm]
                                          & that the invariant mass of all the remaining jets is maximal & & & &  & & \\[0.1cm]
$m_{\text{top},1}$                 & Mass of the reconstructed top quark      &  13  &  10  &   --   & -- &  4  &  11 \\[0.1cm]
$m_{\text{top},2}$                 & Mass of the reconstructed top quark calculated from the jets not entering $m_{\text{top},1}$      &  7  &  9  &  5  &  --  &  10  &  7 \\[0.1cm]
\multirow{2}{*} {$\Lambda$}          & The logarithm of the ratio of event probabilities under the signal and    & 4  &  4  &  3  &  3   &  1  &  1 \\
                                 & background hypotheses   & & & & & &  \\ [0.1cm]
\hline
\end{tabular}
}
\label{tab:BDTVariables}
\end{table}
\vfill\null
\end{landscape}

The input variables include event-shape variables such as
$\text{Centrality}_\text{Mass}$ and aplanarity, global event variables,
such as $S_\text{T}$ (the modulus of the vector sum of the jet \pt), $H_{\mathrm{T}\,5}$ (the scalar sum of
the jet \pt starting from the fifth jet in \pt order), $m_{jj}^\text{min}$ (the smallest invariant mass of all dijet combinations), and
the minimum $\Delta R$ between jets. The \pt of the softest jet in the event is the only individual
kinematic variable that enters the BDT directly. Other variables are
calculated from pairs of objects:
$\Delta R(b,b)^{p_\text{T}^\text{max}}$ (the $\Delta R$ between the two $b$-tagged jets with highest vector sum \pt),
$m_{bb}^{\Delta R(b,b)^\text{min}}$ (the invariant mass of the two $b$-tagged
jets with the smallest $\Delta R$), $(E_\text{T\,1} + E_\text{T\,2}) / \sum
E_\text{T}^\text{jets}$ (the sum of the transverse energies of the two leading jets divided by the sum of
the transverse energies of all jets), $m_{2\,\text{jets}}$ (the mass of the dijet pair, which, when 
combined with any $b$-tagged jet, maximises the magnitude of the vector sum of the \pt of the
three-jet system) and $m_{2\,\text{b--jets}}$ (the invariant mass of the two
  $b$-tagged jets which are selected by requiring that the invariant mass of
  all the remaining jets is maximal). Two variables are calculated as the invariant mass of three
jets: $m_{\text{top},1}$ is computed from the three jets whose invariant
  mass is nearest to the top quark mass, taking into account the jet energy resolutions;
 the $m_{\text{top},2}$ calculation uses the same algorithm but
excludes the jets which enter $m_{\text{top},1}$.  
Finally, a log-likelihood ratio variable, $\Lambda$, is used; it
is related to the probability of an event to be a signal candidate, compared to the
probability of being a background candidate. \\

The $\Lambda$ variable is the sum of the logarithms of ratios of relative
probability densities for $W$ boson, top quark and Higgs boson resonances to be 
reconstructed in the event. For a given resonance $X$ decaying to two
  jets, the $\Lambda$ component is built as $\Lambda_X(m_{jj}) =
\ln\frac{P_{\mathrm{sig}}(m_{jj})}{P_{\mathrm{bkg}}(m_{jj})}$ within a mass
window $\rmt{w_X}=\pm 30$~\GeV\ around the given particle mass: \\
\begin{equation}  
P_{\mathrm{sig}}(m_{jj})=
\begin{cases}
s \cdot G(m_{jj} | \rmt{m}_X,\sigma_X), & \text{for}~|m_{jj} - \rmt{m}_X|
\leq \rmt{w}_X, \\
1 - s,  & \text{for}~|m_{jj} - \rmt{m}_X| > \rmt{w}_X.
\end{cases}
\end{equation}
\begin{equation}
P_{\mathrm{bkg}}(m_{jj})=
\begin{cases}
b\cdot \text{Rect}(\rmt{m}_X,{\rmt w}_X), & \text{for}~|m_{jj} - \rmt{m}_X|
\leq {\rmt w}_X, \\
1 - b,  & \text{for}~|m_{jj} - \rmt{m}_X| > \rmt{w}_X.
\end{cases}
\end{equation}
Here $s$ and $b$ are the probabilities to find a jet pair with an invariant
mass within $\pm \rmt{w}_X$ of $ \rmt{m}_X$. They are calculated from the signal simulation and from the multijet background respectively.
The signal mass distribution is modelled with a Gaussian
$G(m_{jj}|\rmt{m}_X,\sigma_X)$, while the background is modelled with a uniform distribution 
$\text{Rect}(\rmt{m}_X,\rmt{w}_X)$ between $\rmt{m}_X-\rmt{w}_X$ and
  $\rmt{m}_X+\rmt{w}_X$. Both functions $P_{\mathrm{sig}}(m_{jj})$ and $P_{\mathrm{bkg}}(m_{jj})$ are
normalised to unity.
For the top quark resonance the three-particle mass, $m_{jjb}$, is
used. The width of the Gaussian is set to $\sigma_X = 18$~\GeV\
for all resonances; this value corresponds to the expected experimental width
of a Higgs boson with no combinatoric background. 

The expression for the complete event $\Lambda$ is:
\begin{equation}
\begin{split}
  \Lambda (m_{jj},m_{jjb},m_{bb}) = \Lambda_{W}(m_{jj}|\rmt{m}_W,\sigma_X) +
  \Lambda_{\text{top}}(p_{\text{T},jjb},m_{jjb}|\rmt{m}_{\text{top}},\sigma_{X})\; +    \\
  + \Lambda_{H}(p_{\text{T},bb},m_{bb}|\rmt{m}_H,\sigma_X). 
\end{split}
\label{eq:llm}
\end{equation}
                                                                       
The three terms refer to \textit{W}, top, and Higgs resonances respectively.
For the top quark and Higgs boson resonances the masses, $m_{jjb}$ and $m_{bb}$, as
well as the $p_{\text{T}}$, defined as the magnitude of the vector sum of
the \pt of the jets used to reconstruct the top quark, $p_{\text{T},jjb}$, and
to reconstruct the Higgs boson, $p_{\text{T},bb}$, are used. The value of $\Lambda$ is
calculated for all possible jet combinations and the maximum $\Lambda$ of
the event is chosen.   

%=========

The variables entering the BDT are selected and ranked according to their
separation power with an iterative procedure, which stops when adding more
variables does not significantly improve the separation between signal and background. The cut-off corresponds to the point
  when adding a variable increases the significance, defined as $\sqrt{ \sum_i S_i^2/B_i^2 }$ where $S_i$ and $B_i$ are the expected signal and background yields in the $i^{th}$ bin of the BDT discriminant, by less than 1\%.\\
Signal and background samples are classified as described in Section~\ref{sec:eventclass},
and then each subsample is further subdivided
 randomly into two subsamples of equal size for training and for testing.
 
The ranking of the input variables in terms of separation power for each analysis
region is shown in Table~\ref{tab:BDTVariables}.
The distributions of the BDT outputs for simulated signal and background
events are shown in Figure~\ref{fig:BDTallregions} for each
 analysis region. The Figure shows a better separation between signal
  and background for low jet multiplicities than for high jet multiplicities. This is
  explained by the number of possible jet permutations. The number of jet
  permutations increases giving the background more configurations to mimic the signal.

\begin{figure}
\begin{center}
  \centering
  \subfloat[]{\includegraphics[width=0.31\textwidth]{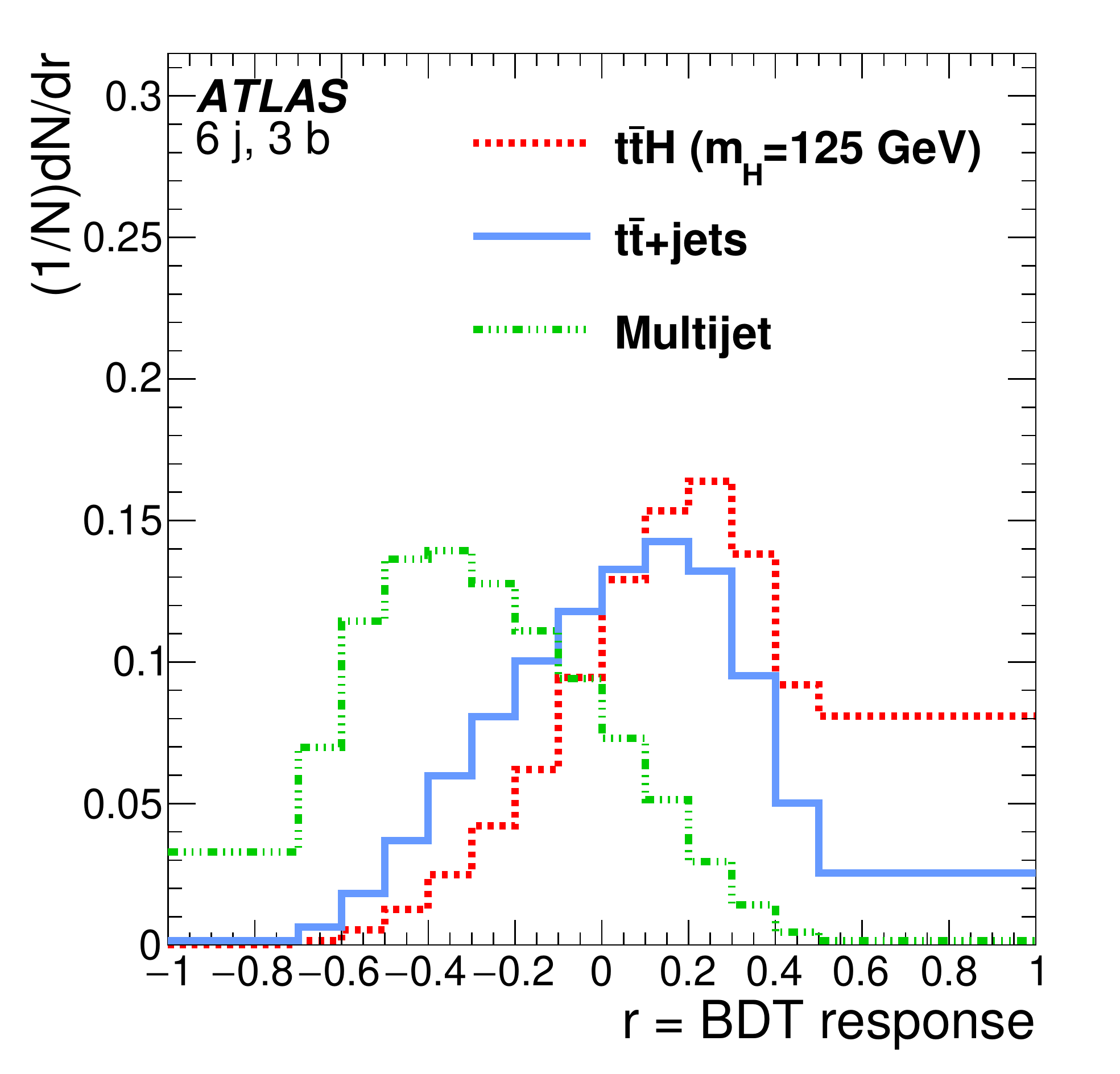}}
  \subfloat[]{\includegraphics[width=0.31\textwidth]{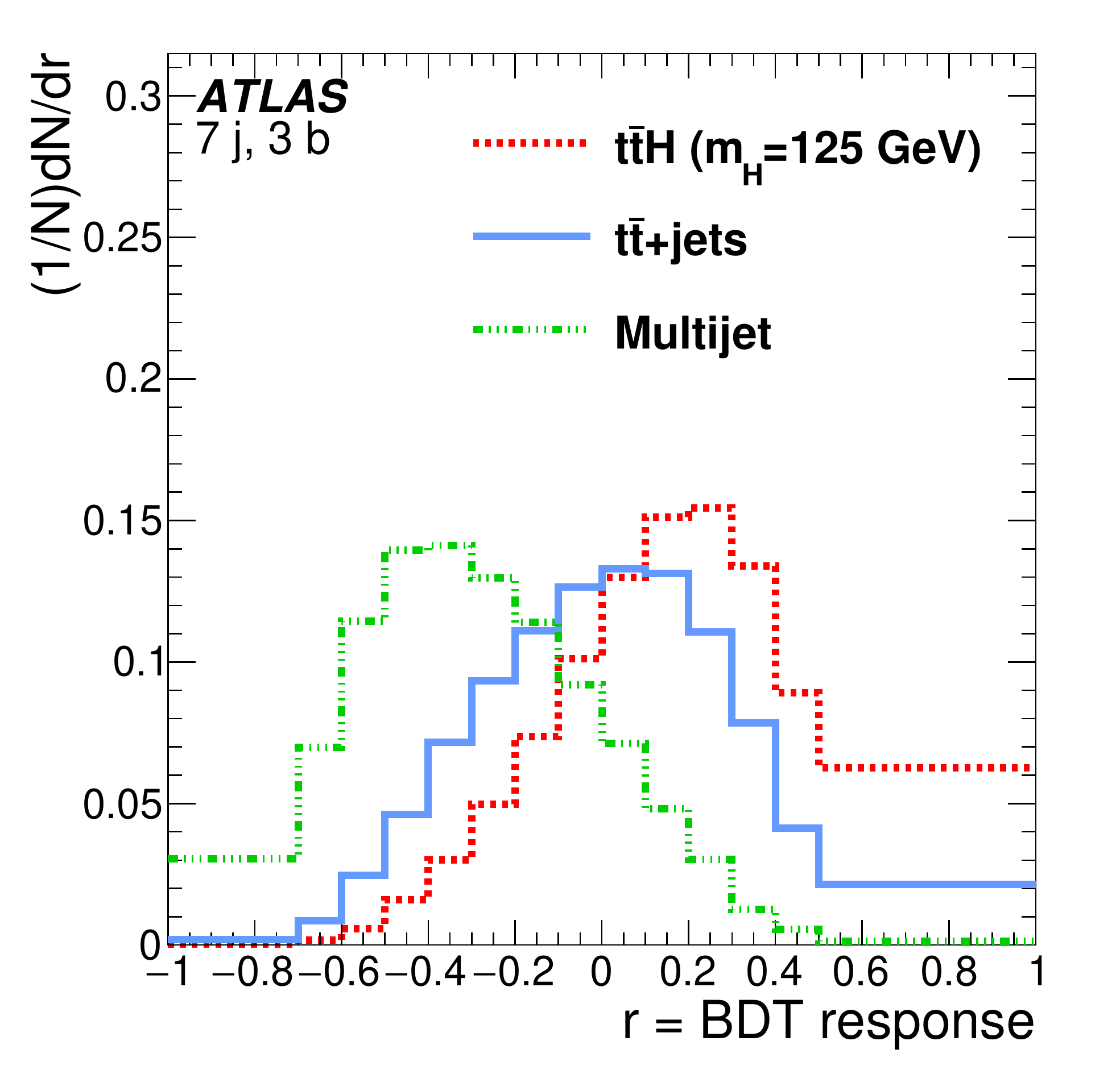}}
  \subfloat[]{\includegraphics[width=0.31\textwidth]{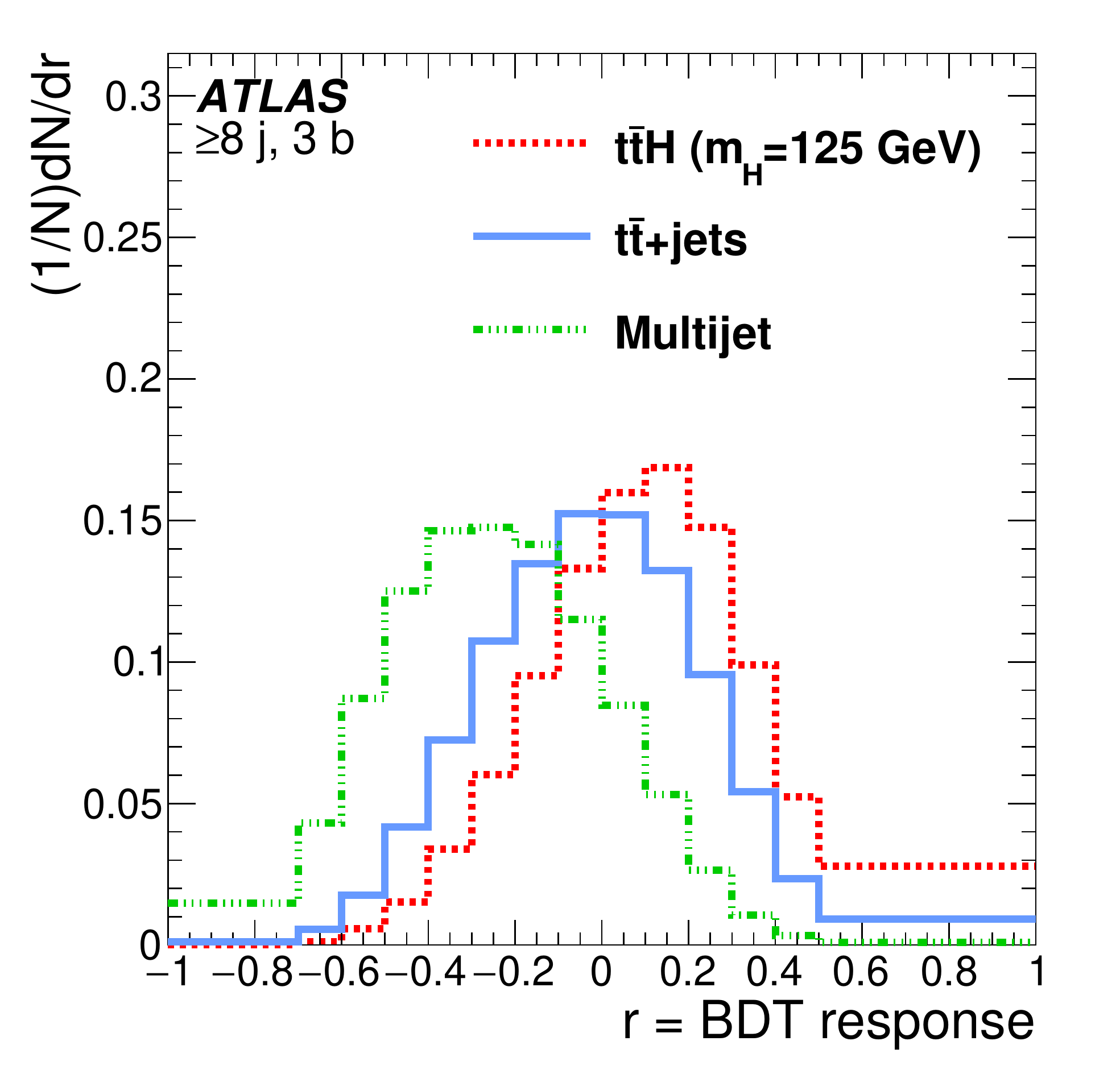}} \\
  \subfloat[]{\includegraphics[width=0.31\textwidth]{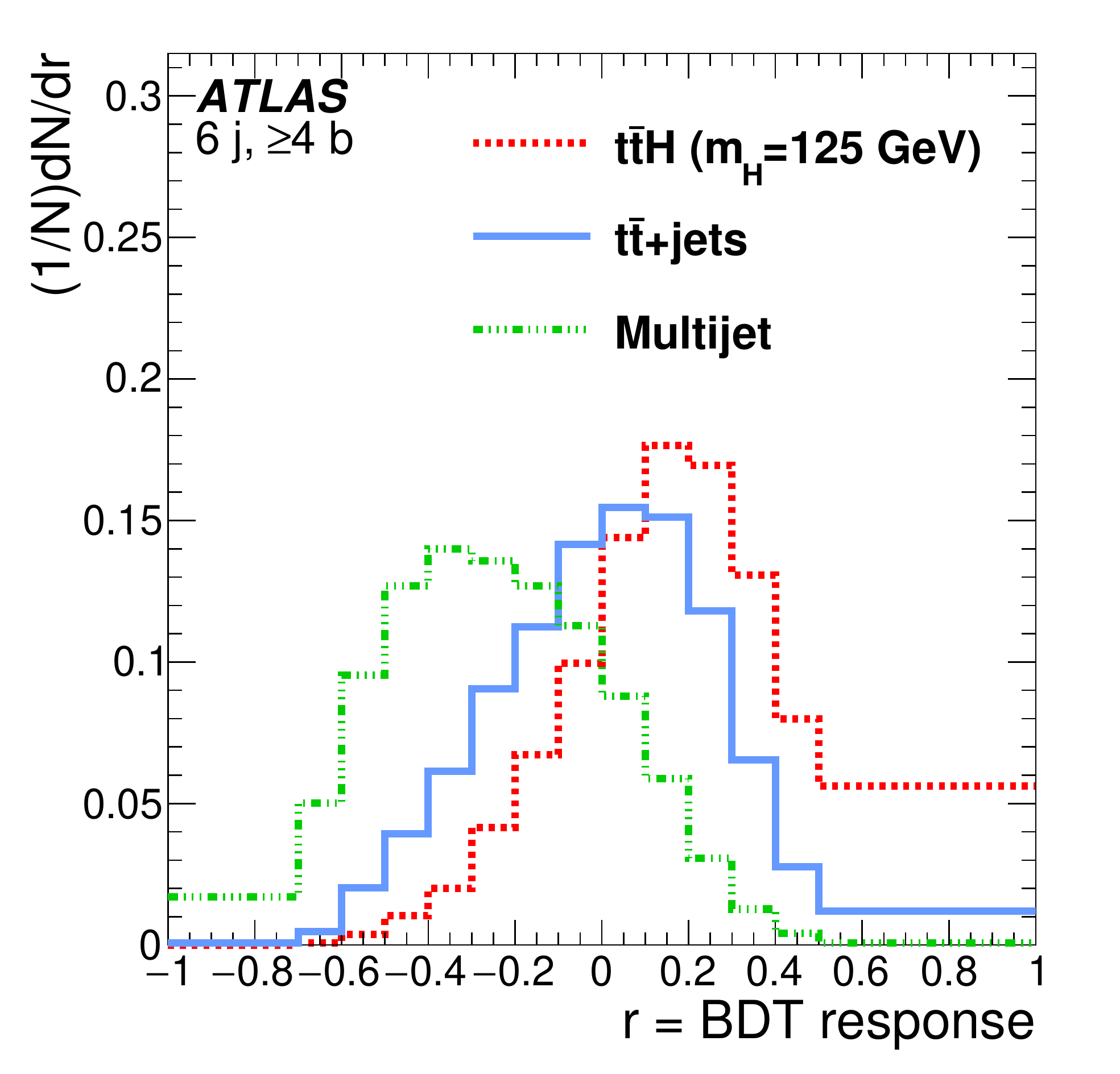}}
  \subfloat[]{\includegraphics[width=0.31\textwidth]{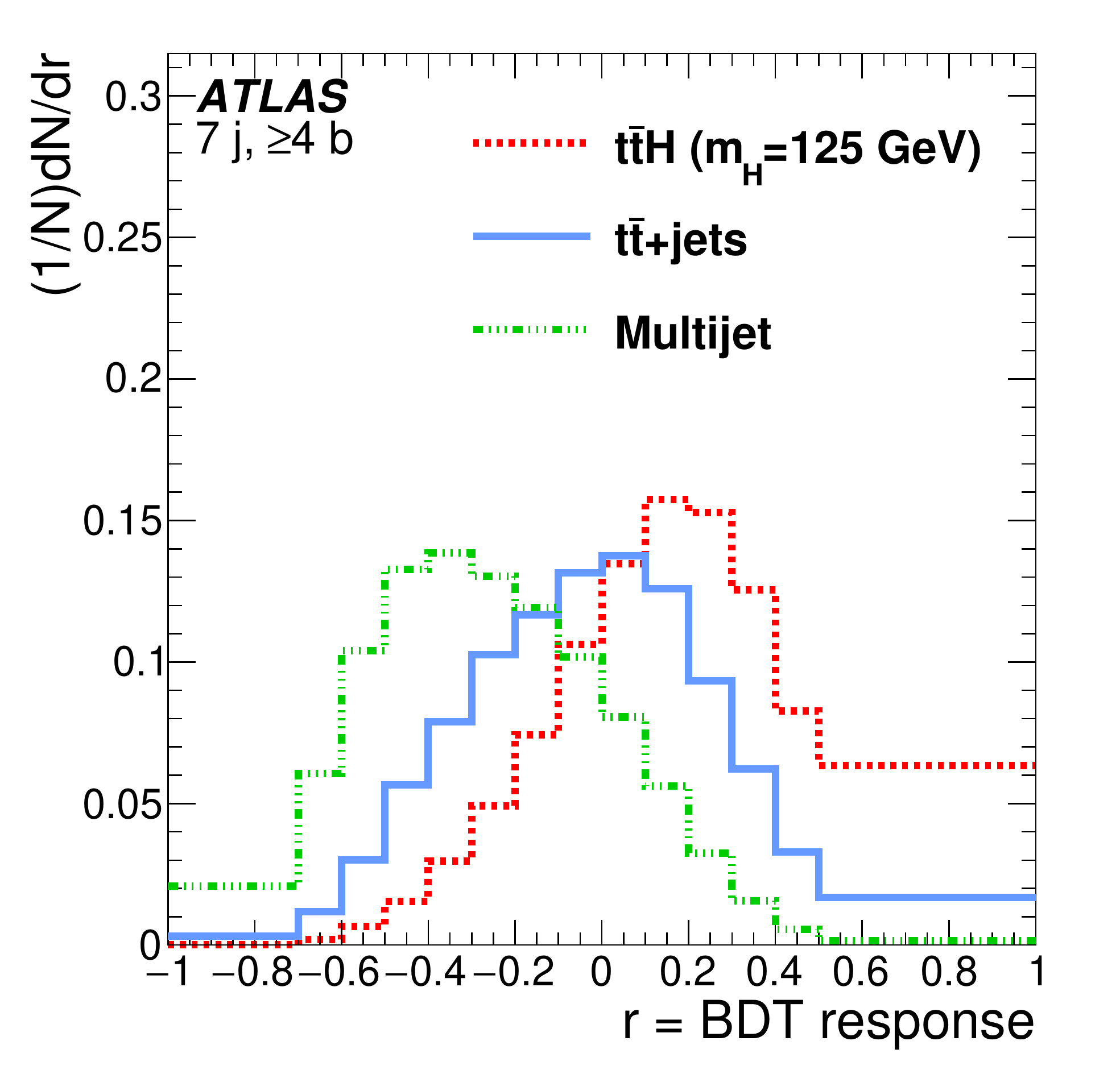}}
  \subfloat[]{\includegraphics[width=0.31\textwidth]{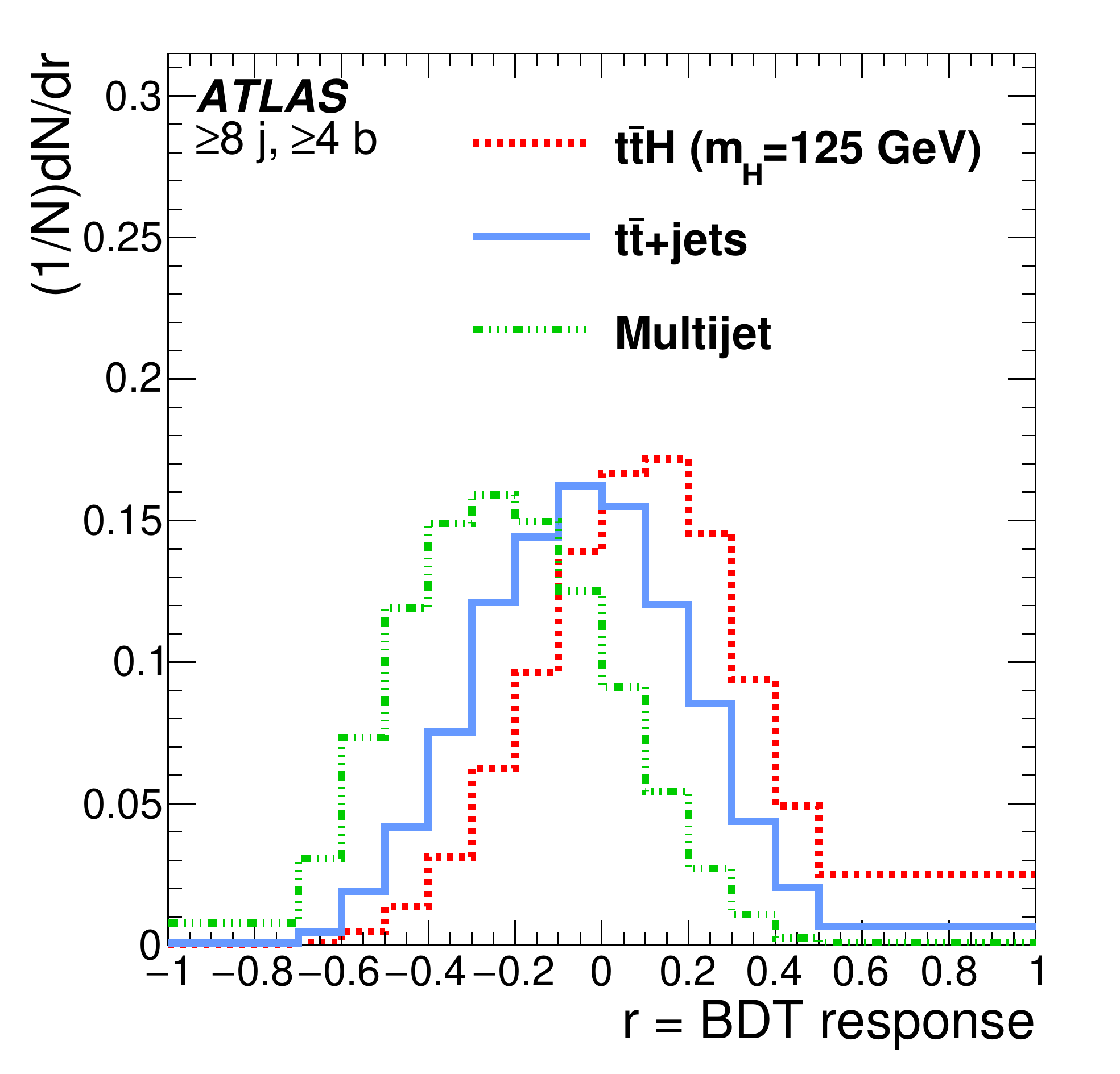}}
  \caption{Response of the BDT algorithm for simulated signal
    (dashed red),
    $t\bar{t}$+jets background (solid blue) and
    multijet background (dotted green) events in the (top)
    regions with ~$3$ $b$-tags ((a) $6$, (b) $7$ and (c) $\ge 8$ jets) and in the
    (bottom) regions with $\ge 4$ $b$-tags ((d) $6$, (e) $7$ and (f)
      $\ge 8$ jets). The binning is the same as that used in the fit.}
  \label{fig:BDTallregions}
\end{center}
\end{figure}

\section{Systematic uncertainties}
\label{sec:systunc}
The sources of systematic uncertainty considered in this analysis can
be grouped into six main categories as summarised in
Table~\ref{tab:SystSummary}. Each systematic uncertainty is represented
  by an independent parameter, referred to as a nuisance parameter, and is
  parameterised with a Gaussian function for the shape uncertainties and a log-normal
  distribution for the normalisations~\cite{cls_3}. They are centred around zero and one, respectively, with a
  width that corresponds to the given uncertainty.
The uncertainties in the integrated luminosity, reconstruction of the 
physics objects, and the signal and background MC models
are treated as in Ref.~\cite{Aad:2015gra}. The uncertainties related
to the jet trigger as well as those related to the data-driven method to estimate the
multijet background are discussed below. In total, 99
 fit parameters are considered. The determination and treatment of the
  systematic uncertainties are detailed in this section. Their impact on the
  fitted signal strength is summarised in Table~\ref{tab:elemer} in Section~\ref{sec:results}.

%%%%%%%%%%%%%%
\begin{table}[h!]
\caption{\label{tab:SystSummary} Sources of systematic uncertainty
  considered in the analysis grouped in six categories. 
``N'' denotes uncertainties affecting only the normalisation for the
  relevant processes and channels, whereas ``S'' denotes uncertainties
  which are considered to affect only the shape of normalised distributions.   
``SN'' denotes uncertainties affecting both shape and normalisation.
Some sources of systematic uncertainty are split into several
components. The number of components is also reported.}
\centering
\begin{tabular}{lcr}
\\
  \hline
Systematic uncertainty source & Type  & Number of components\\
\hline
Luminosity                  &  N & 1\\ \hline
Trigger                     & SN & 1\\ \hline
{\em Physics Objects}                 &   & \\
Jet energy scale            & SN & 21\\ 
Jet vertex fraction         & SN & 1\\
Jet energy resolution       & SN & 1\\
$b$-tagging efficiency      & SN & 7\\ 
$c$-tagging efficiency      & SN & 4\\
Light-jet tagging efficiency & SN & 12\\ \hline
{\em Background MC Model}                 &   & \\
$t\bar{t}$ cross section    &  N & 1\\
$t\bar{t}$ modelling: $p_{\mathrm{T}}$ reweighting   & SN & 9\\
$t\bar{t}$ modelling: parton shower & SN & 3\\
$t\bar{t}$+heavy-flavour: normalisation & N & 2 \\
$t\bar{t}$+$c\bar{c}$: heavy-flavour reweighting  & SN & 2 \\
$t\bar{t}$+$c\bar{c}$: generator & SN & 4 \\
$t\bar{t}$+$b\bar{b}$: NLO Shape & SN & 8 \\
$t\bar{t}V$ cross section   &  N & 1\\
$t\bar{t}V$ modelling   &  SN & 1\\
Single top cross section    &  N & 1\\ \hline
{\em Data driven background} & & \\ 
Multijet normalisation      &  N & 6\\
Multijet $\text{TRF}_{\text{MJ}}$ parameterisation              &  S & 6\\
Multijet $H_\text{T}$ correction             &  S & 1\\ 
Multijet $S_\text{T}$ correction             &  S & 1\\ \hline
{\em Signal Model}          &   & \\
$t\bar{t}H$ scale    & SN & 2 \\
$t\bar{t}H$ generator       & SN & 1 \\
$t\bar{t}H$ hadronisation   & SN & 1 \\
$t\bar{t}H$ parton shower   & SN & 1 \\
\hline
\end{tabular}
\end{table}

\begin{figure*}[!ht]
  \begin{center}
    \subfloat[]{\includegraphics[width=0.45\textwidth]{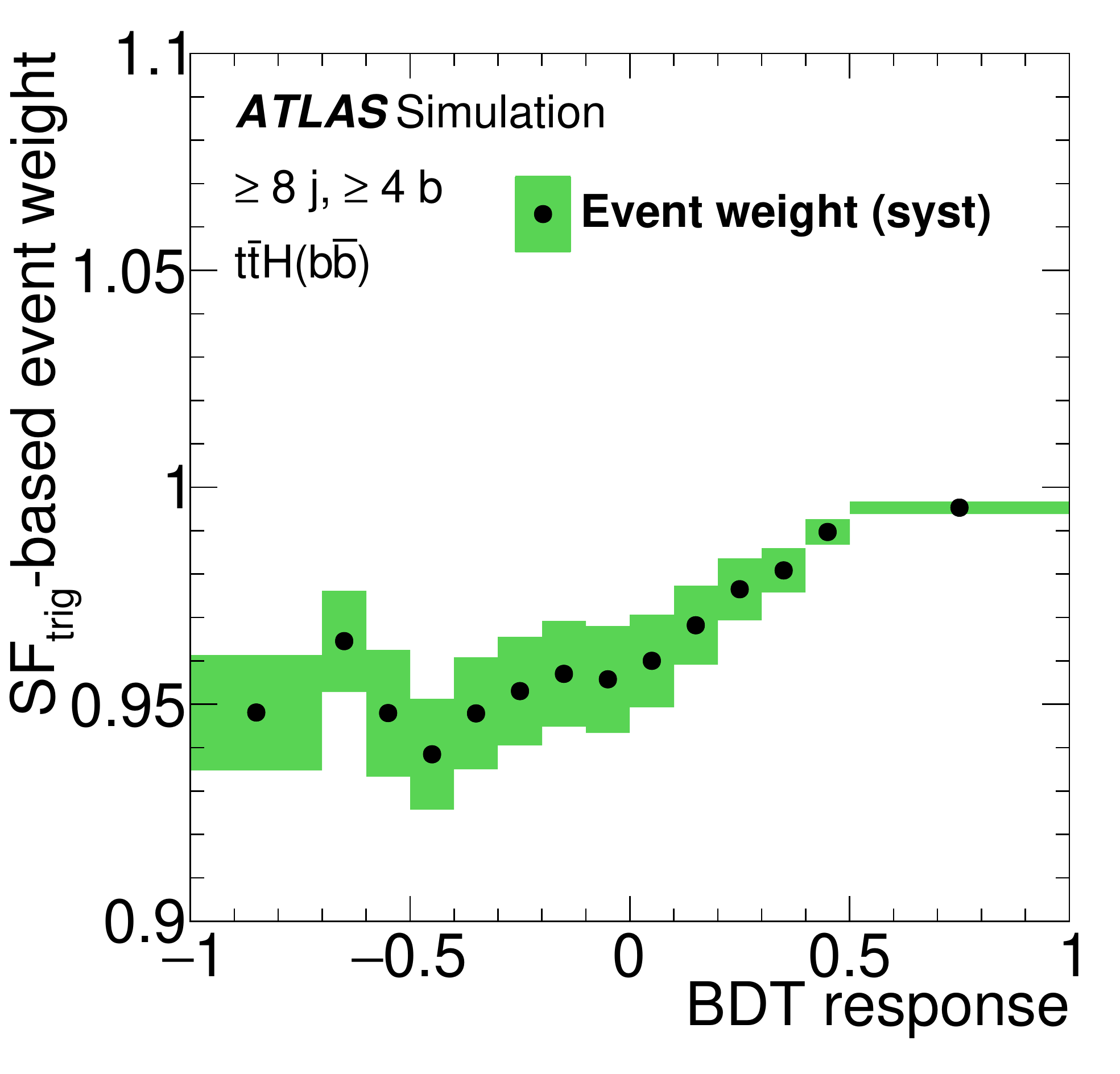}}\label{fig:syst_tr}
    \subfloat[]{\includegraphics[width=0.45\textwidth]{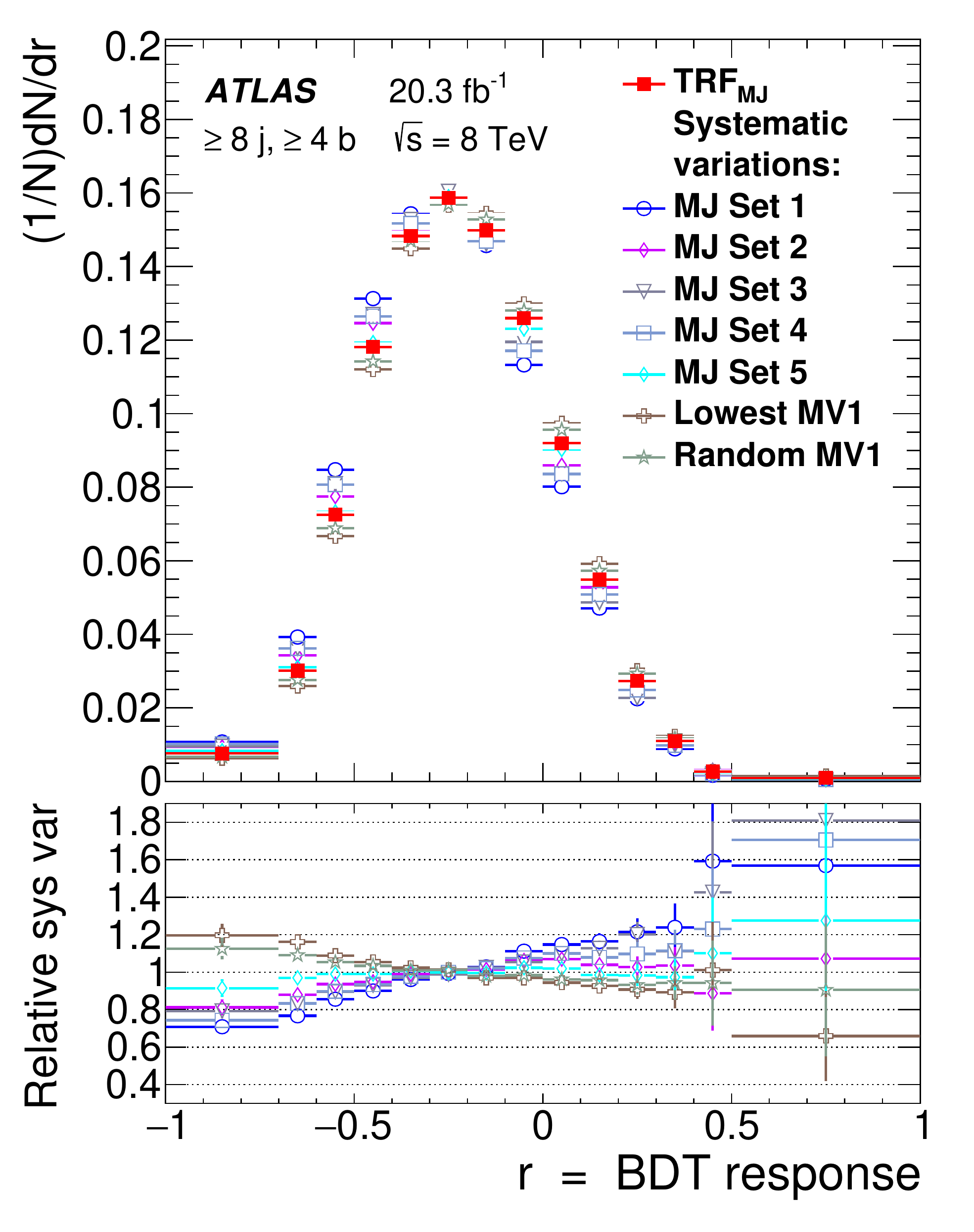}}\label{fig:syst_trfmj}
    \caption{(a) Per event trigger scale factor $\mathrm{SF}_{\mathrm{trig}}$
      (black dots) versus the BDT output of $t\bar{t}H$ events, shown with its corresponding
      systematic uncertainty (green band) for the ($\geq$8j,
          $\geq$4b) region.
      (b) Comparison of the BDT output of the multijet background predicted
      with different sets of $\text{TRF}_{\text{MJ}}$. The nominal $\text{TRF}_{\text{MJ}}$ is
      represented by the red points. The bottom panel shows the ratios
      of the alternative \trfmj predictions to the nominal set. 
    }
    \label{fig:TRFQCD_syst}
  \end{center}
\end{figure*}

The systematic uncertainty in the luminosity for the data sample is
  2.8\%.  It is derived following the same
methodology as that detailed in Ref.~\cite{lumi}.
The trigger uncertainty is determined from the difference between
$\epsilon_{\mathrm{trig}} $, estimated using \tth\ and dijet MC events. Each
jet in the event is weighted according to
$\mathrm{SF}_{\mathrm{trig}}(\pt,\eta)$, the uncertainty of which is propagated to the shape and normalisation of the BDT output distribution,
as shown in Figure~\ref{fig:TRFQCD_syst}(a). 
 
The uncertainties in physics objects are related to the reconstruction
and $b$-tagging of jets.
The jet energy resolution (JER) and  the jet
energy scale (JES) uncertainties are derived combining the information from
test-beam data and simulation~\cite{ATLASJetEnergyMeasurement}. The JES uncertainties are split
into 21 uncorrelated components. The largest of these uncertainties is
 due to the jet flavour composition.  
The JVF uncertainty is derived from $Z(\rightarrow
\ell^+\ell^-$)+ 1-jet events in data and simulation by varying the nominal
cut value by 0.1 up and down.

The uncertainty related to the $b$-tagging is modelled with six independent
parameters, while four parameters model the
$c$-tagging uncertainty~\cite{btagpaper}. These are eigenvalues obtained by diagonalising the 
matrix which parameterises the tagging efficiency as a function of $\pt$, taking into account bin-to-bin correlations. Twelve parameters, which depend on $\pt$ and $\eta$,
are used to parameterise the light-jet-tagging systematic
  uncertainties~\cite{ltagging}. The per-jet $b$-tagging uncertainties are 3\%--5\%, about 10\% for $c$-tagging and 20\% for light jet
tagging. An additional uncertainty is assigned to the $b$-tagging efficiency for jets with
$\pt> 300$~\GeV, which lacks statistics for an accurate calibration from data.  
 
A combined uncertainty of $\pm 6.0\%$ is assigned to the  $t\bar{t}$+jets production cross
section, including modelling components due to the value of $\alpha_{\rm s}$, the PDF used,
the process energy scale, and the top quark mass. Other systematic
uncertainties related to $t\bar{t}$+jets production are due to the
 modelling of parton showers and hadronisation.

The systematic uncertainties arising from the reweighting procedure to
improve $t\bar{t}$~background description by simulation
(Section~\ref{sec:bgmodel}), have been extensively studied in Ref.~\cite{Aad:2015gra} and
  adopted in this analysis. The largest uncertainties in the $t\bar{t}$ 
  background description arise from radiation modelling, the
  choice of generator to simulate $t\bar{t}$ production, the
  JES, JER, and flavour modelling. These systematic
uncertainties are applied to the $t\bar{t}$+light and $t\bar{t}+c\bar{c}$
components. Two additional systematic uncertainties, the full difference between applying and not applying
the reweightings of the $t\bar{t}$ system $\pt$ and top quark \pt, are assigned to the $t\bar{t}+c\bar{c}$
component.

Four additional systematic uncertainties in the $t\bar{t}+c\bar{c}$ estimate
are derived from the simultaneous variation of factorisation and
renormalisation scales in {\sc Madgraph}+{\sc Pythia}. For the $t\bar{t}+b\bar{b}$
background, three scale uncertainties are evaluated by varying the renormalisation
and resummation scales. The shower recoil model uncertainty and two
uncertainties due to the PDF choice in the {\sc sherpa+OpenLoops} NLO calculation are
also taken into account. 

The $t\bar{t}$+jets background is parameterised to allow a varying percentage of heavy
flavours $c$ and $b$ in the additional jets not originating from the top quark decay products. An uncertainty of $\pm50\%$ is
assigned to the $t\bar{t}+b\bar{b}$ and $t\bar{t}+c\bar{c}$ 
components of the $t\bar{t}+$jets cross section, which are treated as uncorrelated and are derived by
comparing {\sc Powheg}+{\sc Pythia} with a NLO result based on {\sc
sherpa+OpenLoops}. The uncertainty in the $t\bar{t}+b\bar{b}$ contribution
represents the dominant systematic effect in this analysis.  
An uncertainty of $\pm 30\%$ in the total cross section is assumed for $t\bar{t}+V$~\cite{ttbarVxs1,ttbarVxs2}.

The multijet background is estimated using data in regions with exactly two $b$-tagged jets after subtraction of contributions from other events using MC simulation. All systematic uncertainties
mentioned above are fully propagated to the data-driven multijet background
estimation and treated in a correlated manner.

To estimate the uncertainties associated with the multijet background, the values of \mjeps 
 are determined as a function of different sets of variables, listed in
the first part of Table~\ref{tab:QCDNP}, which are
sensitive to the amount and the mechanism of heavy-flavour production.   
Alternative variables used are $\Delta R_{(j,j)}^{\text{min}}$, the minimum $\Delta R$ between the probed
jet and any other jet in the event, $ \Delta
R_{(j,\text{hMV1})}^{\text{min}}$, the minimum $\Delta R$ between the probed
jet and the two jets with highest $b$-tag probability or $\langle
\Delta R_{(j,\text{hMV1})} \rangle$, its average
value, and $\Delta R_{\text{MV1}}$, the $\Delta R$ between the two jets with the highest $b$-tag probability. 
In addition, different choices of methods
to exclude $b$-tagged jets when determining \mjeps in the \trfmj method are
considered: the two $b$-tagged jets with the
lowest MV1 weight or a random choice of two jets
among all $b$-tagged jets in the event are chosen.
The different sets of variables used to define \mjeps affect the shape of the 
BDT distribution for the multijet background, as shown in
Figure~\ref{fig:TRFQCD_syst}(b). Each of these shape variations is
taken into account by a nuisance parameter in the fit. 
These parameterisations also affect the overall normalisation,
with a maximum variation of 18\% in the 3-$b$-tag regions and 38\% in the $\geq$4-$b$-tag
regions. 
Residual mismodelling of $H_\text{T}$ and $S_\text{T}$ from the extraction
region are also taken into account as systematic uncertainties.
The normalisation of the multijet background is evaluated independently in
each of the six analysis regions.

\begin{table*}  
\small
\begin{center}
\caption{\label{tab:QCDNP} Alternative predictions of the multijet
    background with the \trfmj method. Multijet sets 1 to 5 correspond to variations of the nominal set
of variables describing \mjeps. 
 The next two sets specify the variation in the nominal set based on the two
$b$-tagged jets which are used to compute \mjeps. The last two refer to
changes due to the residual mismodellings of $H_{\text{T}}$ and $S_{\text{T}}$.
Each of these variations of the multijet background shape is quantified by one
nuisance parameter in the fit. }   
\begin{tabular}{l|cc}
\hline
\trfmj predictions &  Parameterisation variables in the \trfmj method \\ \hline 
Nominal set &   \pt, $|\eta|, \; \langle \Delta R_{(j,\text{hMV1})} \rangle$ \\ \hline
Multijet set 1 &  \pt, $\Delta R_{\text{MV1}}$, $\Delta R_{(j,\text{hMV1})}^{\text{min}}$ \\
Multijet set 2 &  \pt, $\Delta R_{\text{MV1}}$, $\Delta R_{(j,j)}^{\text{min}}$ \\
Multijet set 3 &  \pt, $|\eta|$, $\Delta R_{(j,\text{hMV1})}^{\text{min}}$ \\
Multijet set 4 &  \pt, $|\eta|$, $\Delta R_{\text{MV1}}$, $\Delta R_{(j,\text{hMV1})}^{\text{min}}$ \\
Multijet set 5 &  \pt, $\Delta R_{\text{MV1}}$, $\langle\Delta R_{(j,\text{hMV1})}\rangle$ \\ \hline
Multijet lowest MV1 & Nominal set removing the two lowest MV1 jets from computation \\	 
Multijet random MV1 & Nominal set removing randomly two MV1 jets from computation\\ \hline
Multijet HT RW & Nominal set with $H_{\text{T}}$ reweighting \\
Multijet ST RW & Nominal set with $S_{\text{T}}$ reweighting \\ \hline% \hline
\end{tabular}
\end{center}  
\end{table*} 

For the signal MC modelling, the {\sc PowHel} factorisation and
renormalisation scales are varied independently by a factor
two and 0.5. The kinematics of the MC
simulated samples are then reweighted to reproduce the effects of these variations.   
The uncertainties related to the choice of PDFs are evaluated using the
recommendations of PDF4LHC~\cite{ref:pdf4lhc}.
The systematic uncertainties from the parton shower and fragmentation models are evaluated using  {\sc PowHel+Herwig} samples. The uncertainty
due to the choice of generator is evaluated by comparing {\sc
  PowHel}+{\sc Pythia8} with {\sc Madgraph5}\_{\sc aMC@NLO}+{\sc Herwig}++.  
%%%%%%%%%%%%%%

\section{Statistical methods}
\label{sec:statmeth}

The binned distributions of the BDT output
 discriminants for each of the six analysis} regions are combined 
as inputs to a test statistic to search for the presence of a signal.
The analysis uses a maximum-likelihood fit~\cite{cls_3}
to measure the compatibility of the observed data with
the background-only hypothesis, i.e., $\mu=0$, and to make statistical
inferences about $\mu$, such as upper limits, using the CL$_{\mathrm{s}}$
method~\cite{cls,cls_2} as implemented in the {\normalfont \scshape
  RooFit} package~\cite{RooFitManual}. 

A fit is performed
under the signal-plus-background hypothesis to obtain the value of the signal
strength, assuming a SM Higgs boson mass of $m_H$ = 125~\GeV.  
The value of $\mu$ is a free parameter in the fit.
The normalisation of each component of the
background and $\mu$ are determined simultaneously from the fit.
Contributions from $t\bar{t}$+jets, $t\bar{t}+V$ and single-top-quark backgrounds are constrained by the uncertainties of the respective
theoretical calculations, the uncertainty in the luminosity, and experimental data. The
multijet background normalisations are free parameters in the fit and are
  independent in each region.   
The performance of the fit is validated using simulated
events by injecting a signal with variable
strength and comparing the known strength to the fitted value. 

\section{Results}
\label{sec:results}

The yields in the different analysis regions considered in the analysis after the fit (post-fit)  are
summarised in Table~\ref{tab:postfityields}. In each region, the variation of
background and signal events with respect to the pre-fit values (cf. Table~\ref{tab:prefityields}) are modest and, in particular, the fitted
multijet background component is well constrained by the fit within an uncertainty of 8\%. 

\begin{table*}[!ht]
  \caption{\label{tab:postfityields} Event yields from simulated backgrounds and the signal as well as measured events in each
    of the analysis regions after the fit. The quoted 
uncertainties include statistical and systematical effects. The sum of
all contributions may slightly differ from the total value due to
  rounding. The $tH$ background is not shown as fewer than 1.5 events in each
region are predicted.}
  \small
  \begin{center}
  \begin{tabular}{l|ccccccc}
\hline
                      & 6j, 3b           &  6j, $\geq$4b  & 7j, 3b           & 7j, $\geq$4b  & $\geq$8j, 3b            &  $\geq$8j, $\geq$4b \\ \hline  
Multijet	      & \phantom{.}15940 $\pm$ 320\phantom{.} & \phantom{.}1423 $\pm$ 66\phantom{.} & \phantom{.}12060 $\pm$ 350\phantom{.} & \phantom{.}1233 $\pm$ 78\phantom{.} & \phantom{.}10020 $\pm$ 490\phantom{.} & \phantom{.}1280 $\pm$ 100\phantom{.} \\
$t\bar{t}$+light      &	\phantom{.0}1750 $\pm$ 270\phantom{.}  & \phantom{00.}55 $\pm$ 13\phantom{.}   & \phantom{0.}1650 $\pm$ 340\phantom{.}  & \phantom{00.}54 $\pm$ 15\phantom{.}   & \phantom{0.}1550 $\pm$ 450\phantom{.}  & \phantom{00.}54 $\pm$ 21\phantom{.0} \\
$t\bar{t}+c\bar{c}$   & \phantom{.00}350 $\pm$ 170\phantom{.}   & \phantom{00.}22 $\pm$ 11\phantom{.}   & \phantom{00.}490 $\pm$ 240\phantom{.}   & \phantom{00.}28 $\pm$ 14\phantom{.}   & \phantom{00.}750 $\pm$ 360\phantom{.}   & \phantom{00.}66 $\pm$ 33\phantom{.0} \\	 
$t\bar{t}+b\bar{b}$   & \phantom{.00}230 $\pm$ 120\phantom{.}   & \phantom{00.}31 $\pm$ 17\phantom{.}   & \phantom{00.}350 $\pm$ 190\phantom{.}   & \phantom{00.}63 $\pm$ 34\phantom{.}   & \phantom{00.}560 $\pm$ 320\phantom{.}   & \phantom{0.}139 $\pm$ 75\phantom{.0} \\
$t\bar{t}+V$	      & \phantom{00}15.0 $\pm$ 6.2\phantom{0}  & \phantom{00}1.9 $\pm$ 1.5 & \phantom{00}23.3 $\pm$ 8.9\phantom{0}  & \phantom{00}3.6 $\pm$ 2.2 & \phantom{000.}43 $\pm$ 15\phantom{.0}& \phantom{00}8.7 $\pm$ 4.2\phantom{0} \\
Single top	      & \phantom{.00}184 $\pm$ 59\phantom{.0}  & \phantom{00}6.7 $\pm$ 3.6 & \phantom{00.}153 $\pm$ 52\phantom{.0}    & \phantom{00}9.4 $\pm$ 4.4 & \phantom{0.}123 $\pm$ 48\phantom{.} & \phantom{0}11.8 $\pm$ 5.8\phantom{0} \\ 
\hline
Total background      & \phantom{.}18470 $\pm$ 320\phantom{.} & \phantom{.}1539 $\pm$ 58\phantom{.} & \phantom{.}14720 $\pm$ 320\phantom{.} & \phantom{.}1391 $\pm$ 69\phantom{.} & \phantom{.}13030 $\pm$ 340\phantom{.} & \phantom{.}1561 $\pm$ 63\phantom{.0} \\ \hline
$t\bar{t}H$ ($m_H$=125~\GeV)     & \phantom{00}23.4 $\pm$ 6.3\phantom{0}  & \phantom{00}5.6 $\pm$ 2.8 & \phantom{00}39.1 $\pm$ 8.9\phantom{0}  & \phantom{0}11.9 $\pm$ 4.5 & \phantom{00.}71 $\pm$ 15\phantom{.}     & \phantom{0}28.8 $\pm$ 8.5\phantom{0} \\ \hline
Data events                 & 18508           & 1545          & 14741           & 1402          & 13131           & 1587 \\ \hline

\end{tabular}
\end{center} 
\end{table*}
 
\begin{figure*}[!ht]
    \begin{center}
      \subfloat[]{\includegraphics[width=0.42\textwidth]{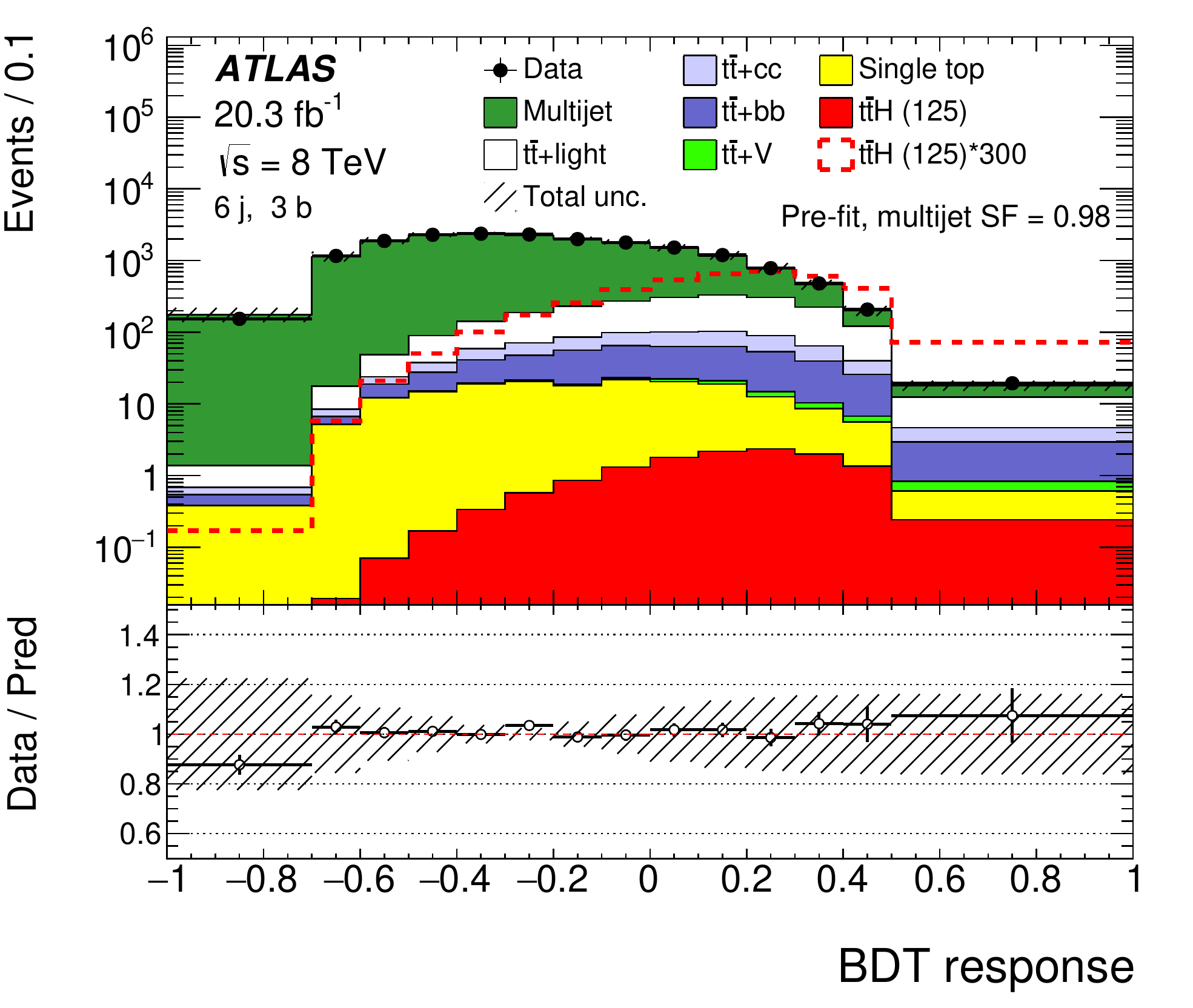}}\label{fig:prepost_3b_a}
      \subfloat[]{\includegraphics[width=0.42\textwidth]{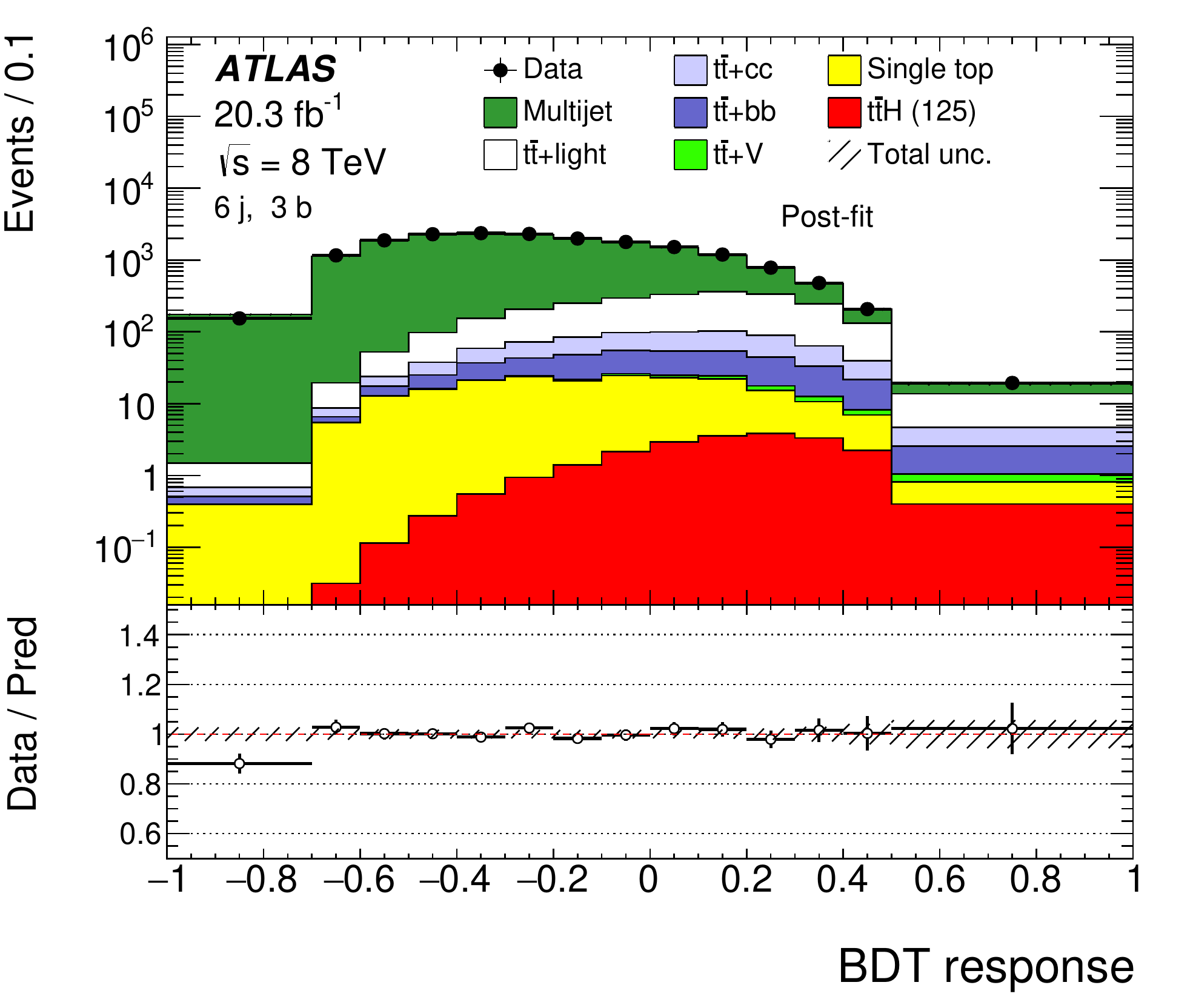}}\label{fig:prepost_3b_b}\\
      \subfloat[]{\includegraphics[width=0.42\textwidth]{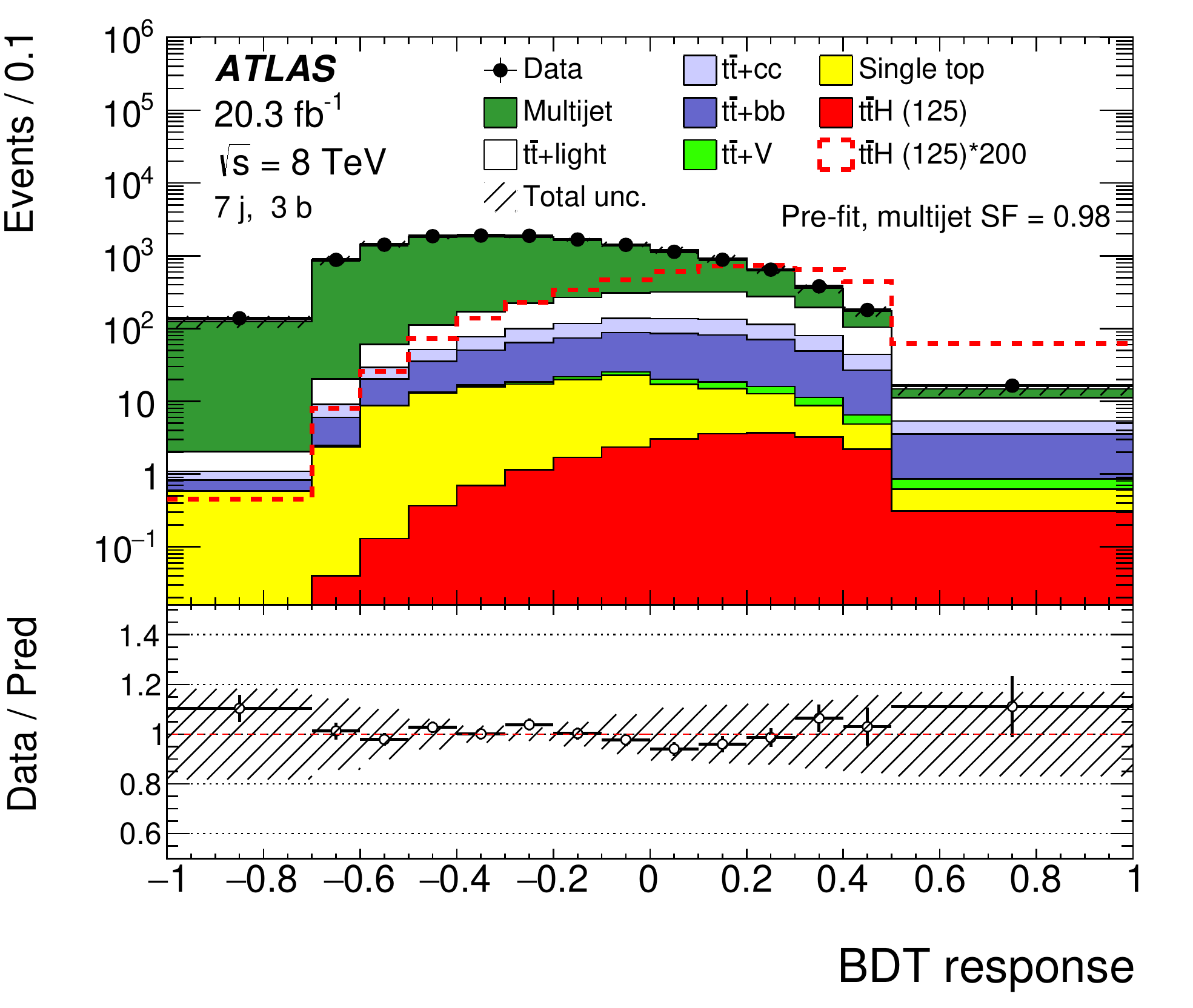}}\label{fig:prepost3b_c}
      \subfloat[]{\includegraphics[width=0.42\textwidth]{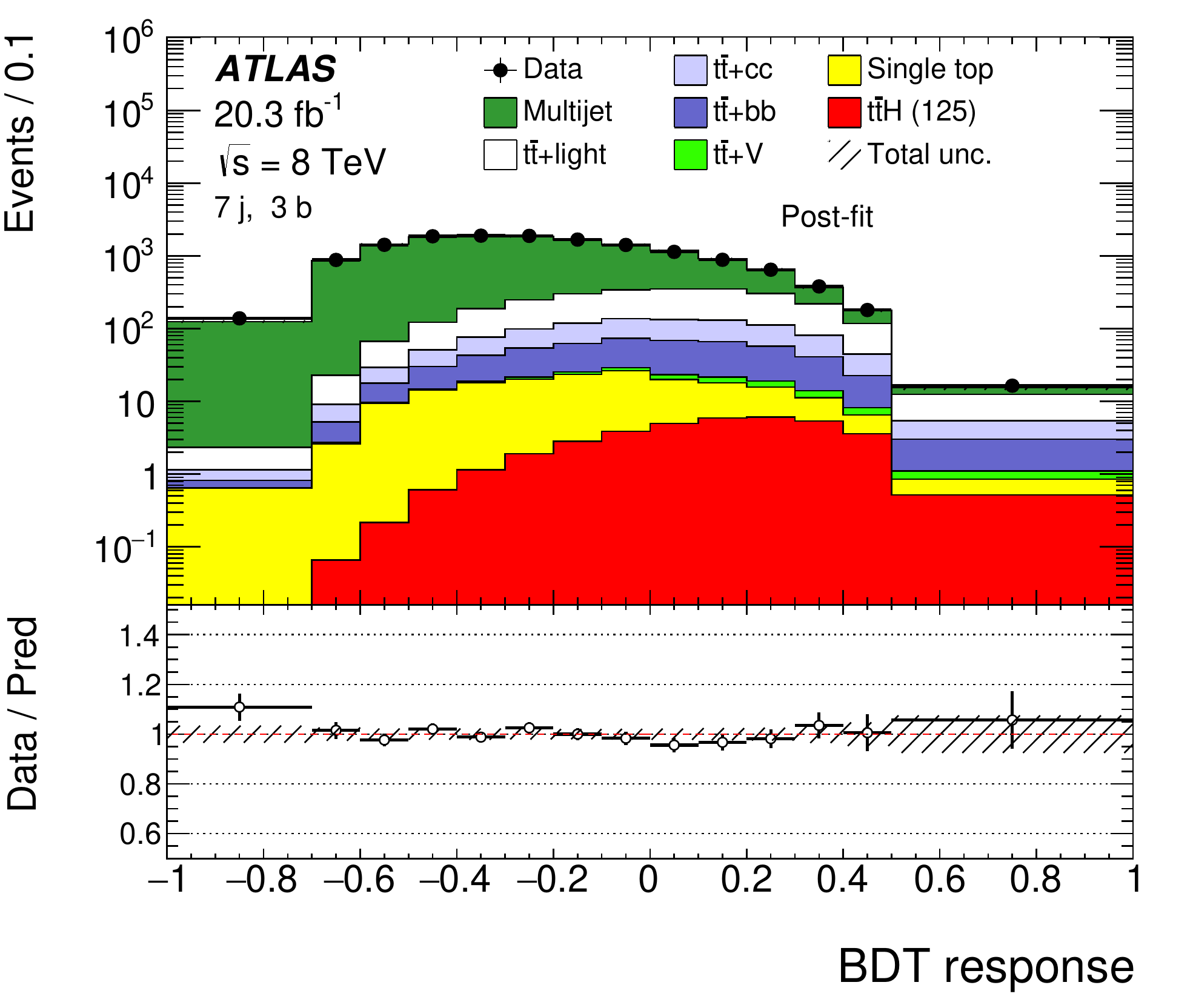}}\label{fig:prepost_3b_d}\\
      \subfloat[]{\includegraphics[width=0.42\textwidth]{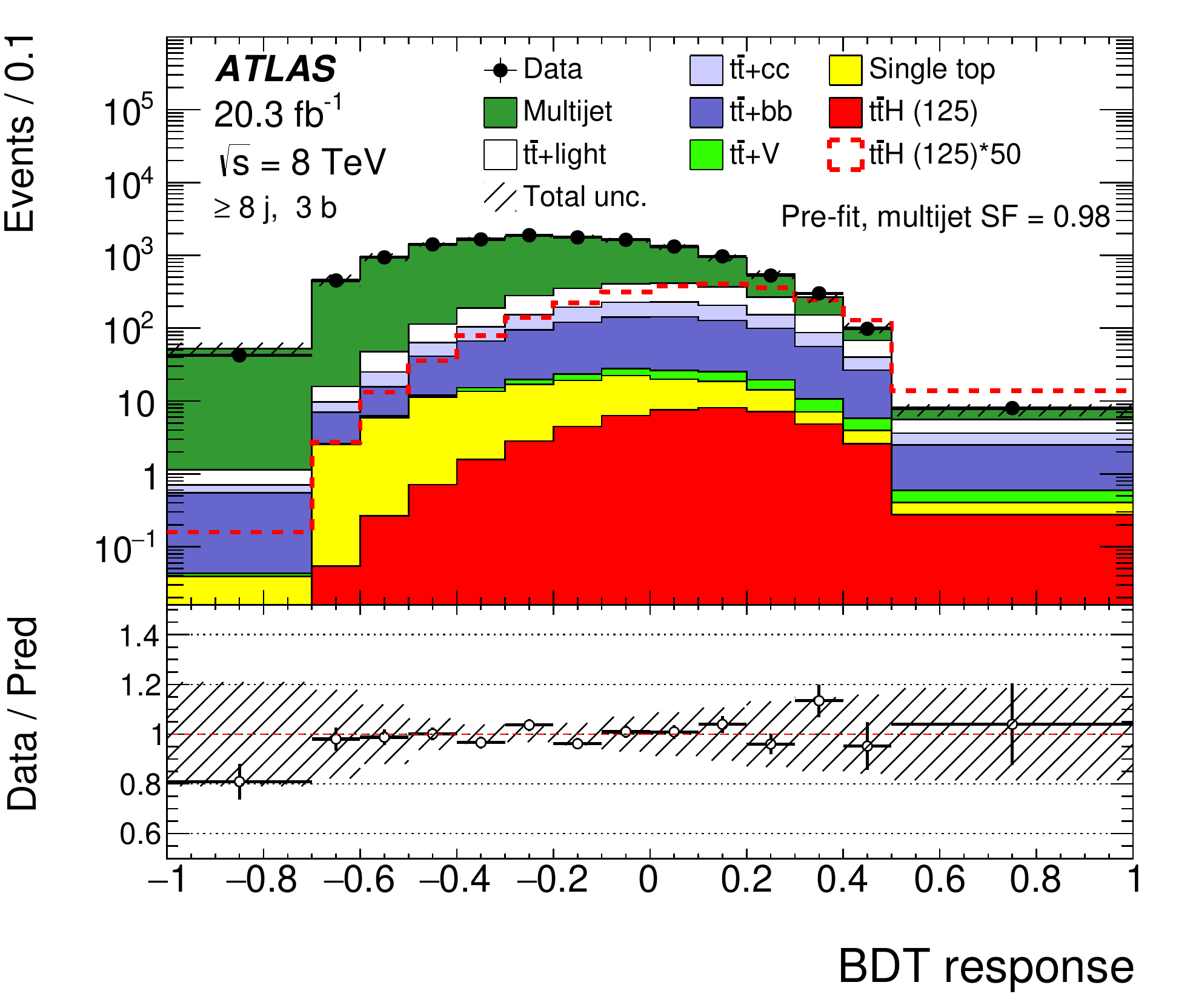}}\label{fig:prepost_3b_e}
      \subfloat[]{\includegraphics[width=0.42\textwidth]{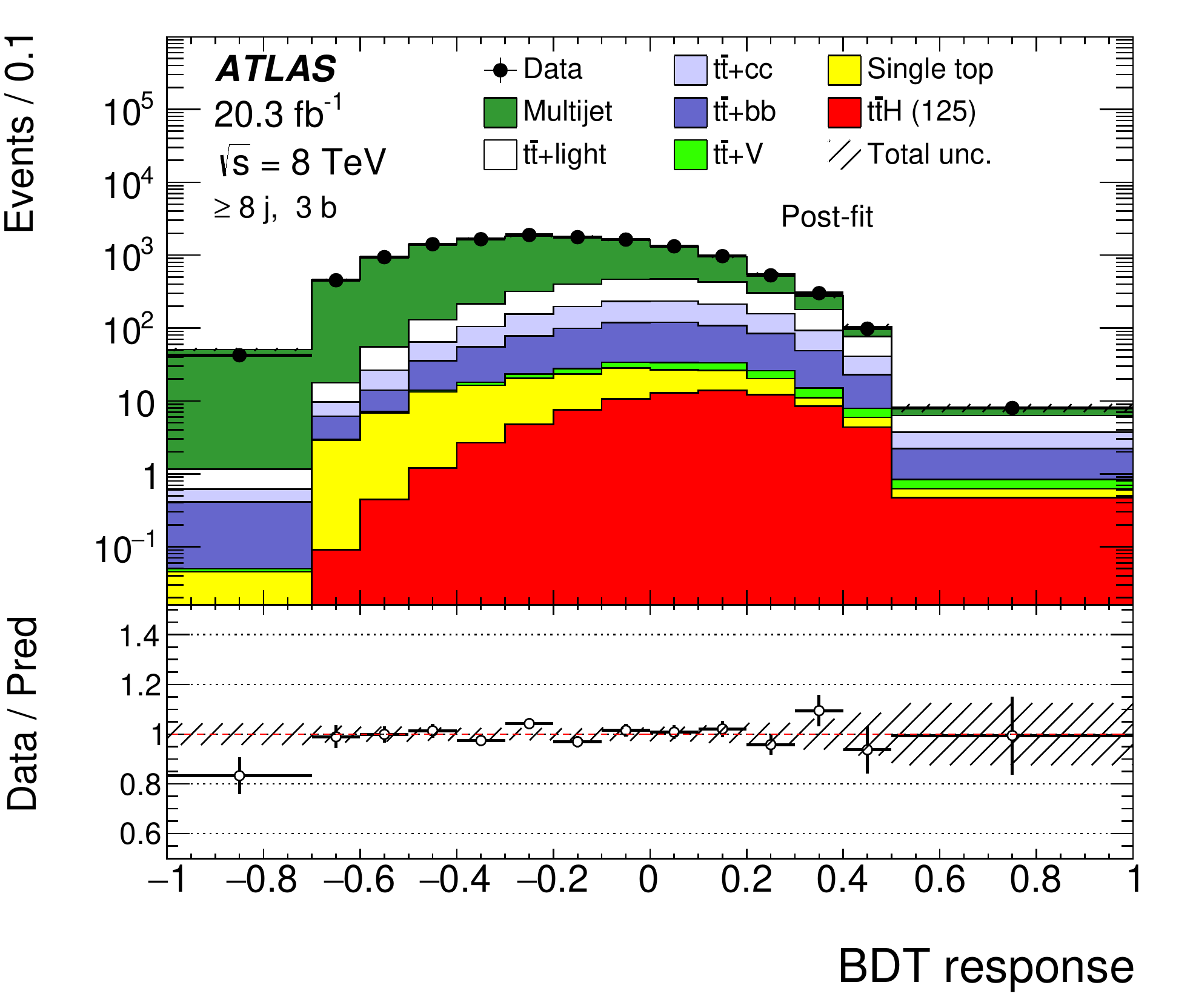}}\label{fig:prepost_3b_f}\\
      \caption{Comparison between data and prediction for the
        BDT discriminant in the, from top to bottom, (6--8j, 3b)
        regions before (left) and after (right) the fit. The fit is performed under the signal-plus-background
        hypothesis. Pre-fit plots show an overlay of the multijet distribution normalised to
        data for illustration purposes only.
        The bottom panels display the ratios of data to the total prediction.
        The hashed areas represent the total uncertainty in the
        background predictions. 
        The $t\bar{t}H$ signal yield (solid red) is scaled by a fixed factor
        before the fit.
      }
      \label{fig:prepost_3b} 
    \end{center}
\end{figure*}

\begin{figure*}[!ht]
  \begin{center}   
    \subfloat[]{\includegraphics[width=0.42\textwidth]{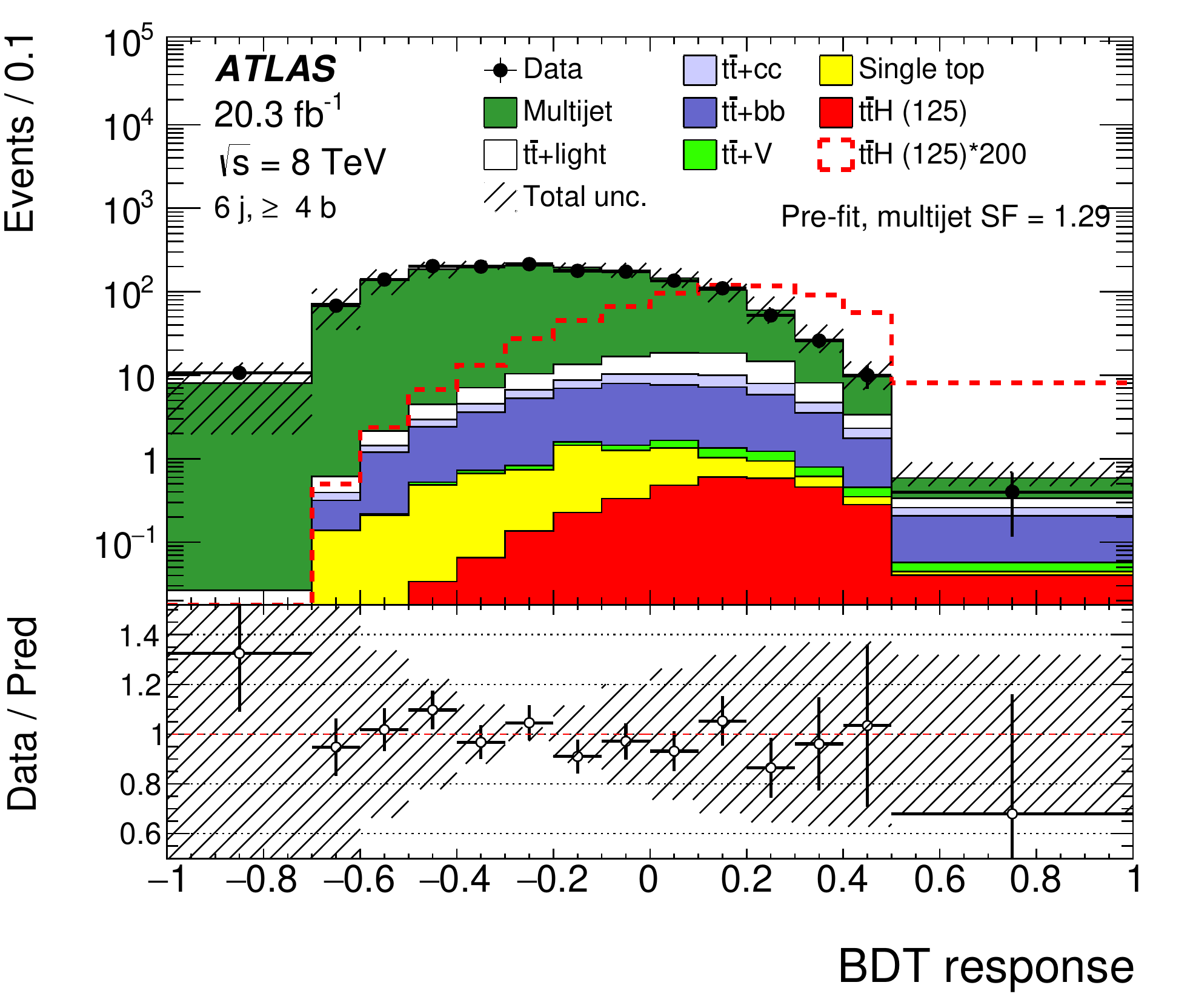}}\label{fig:prepost_4b_a}
    \subfloat[]{\includegraphics[width=0.42\textwidth]{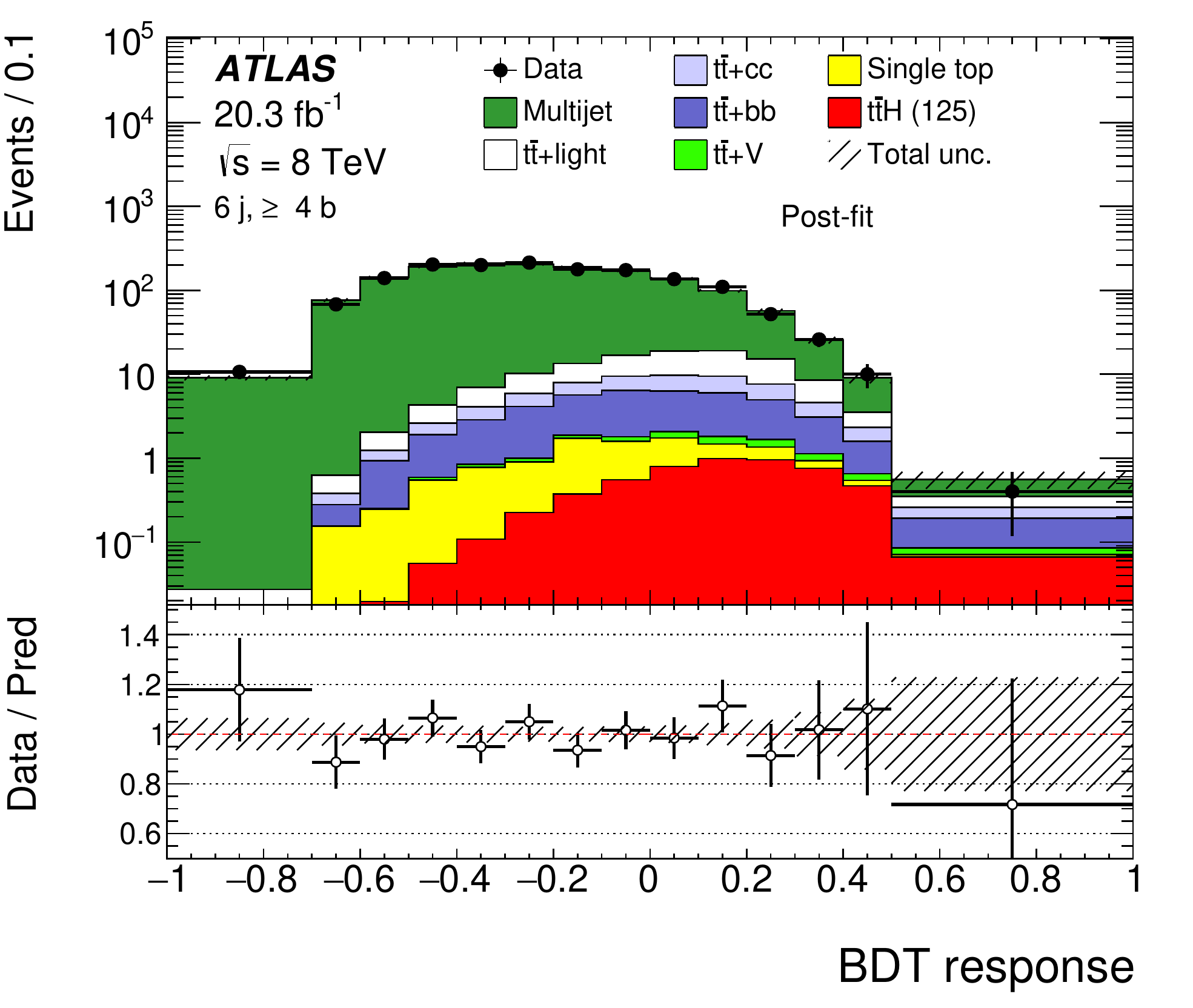}}\label{fig:prepost_4b_b}\\
    \subfloat[]{\includegraphics[width=0.42\textwidth]{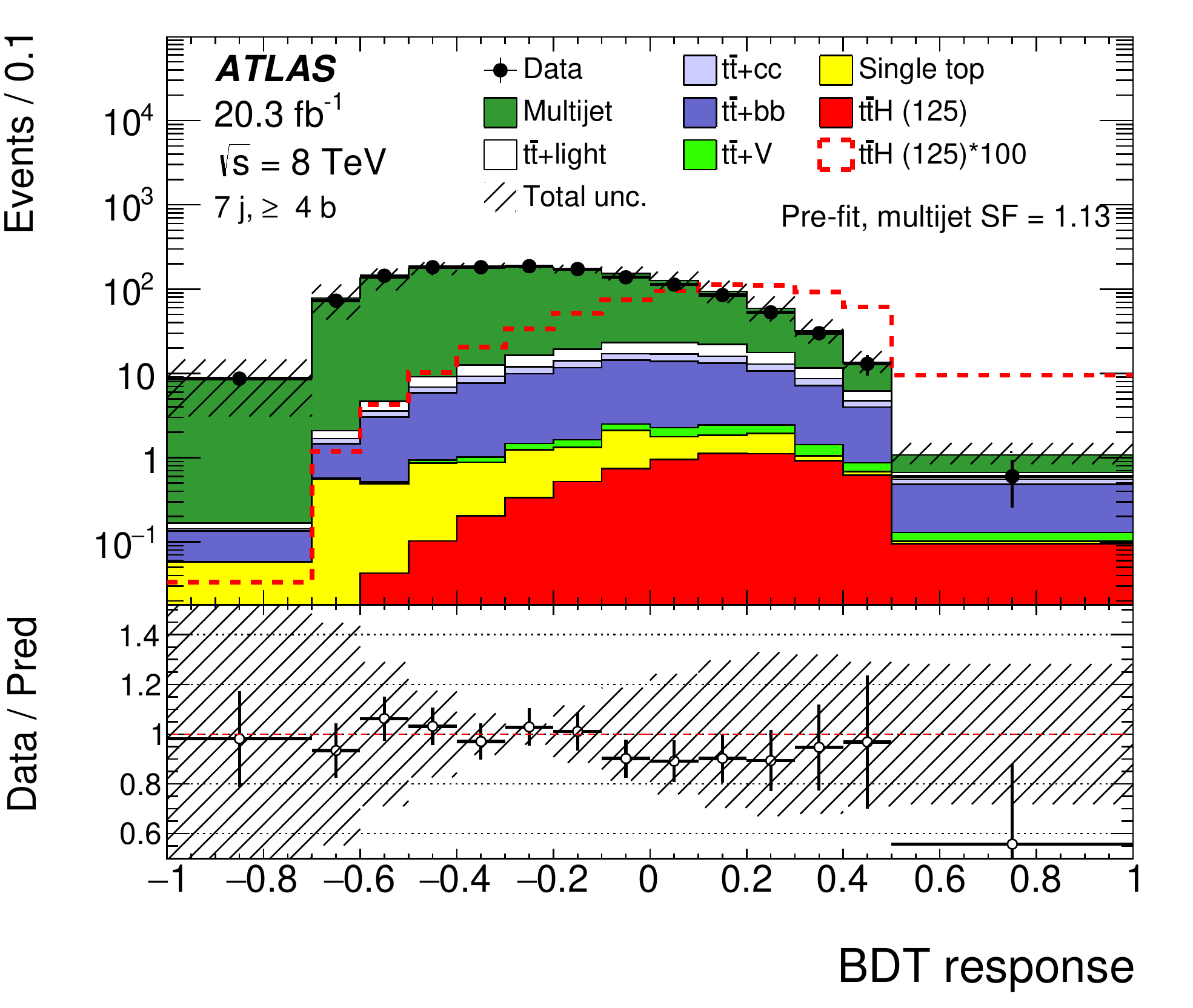}}\label{fig:prepost4b_c}
    \subfloat[]{\includegraphics[width=0.42\textwidth]{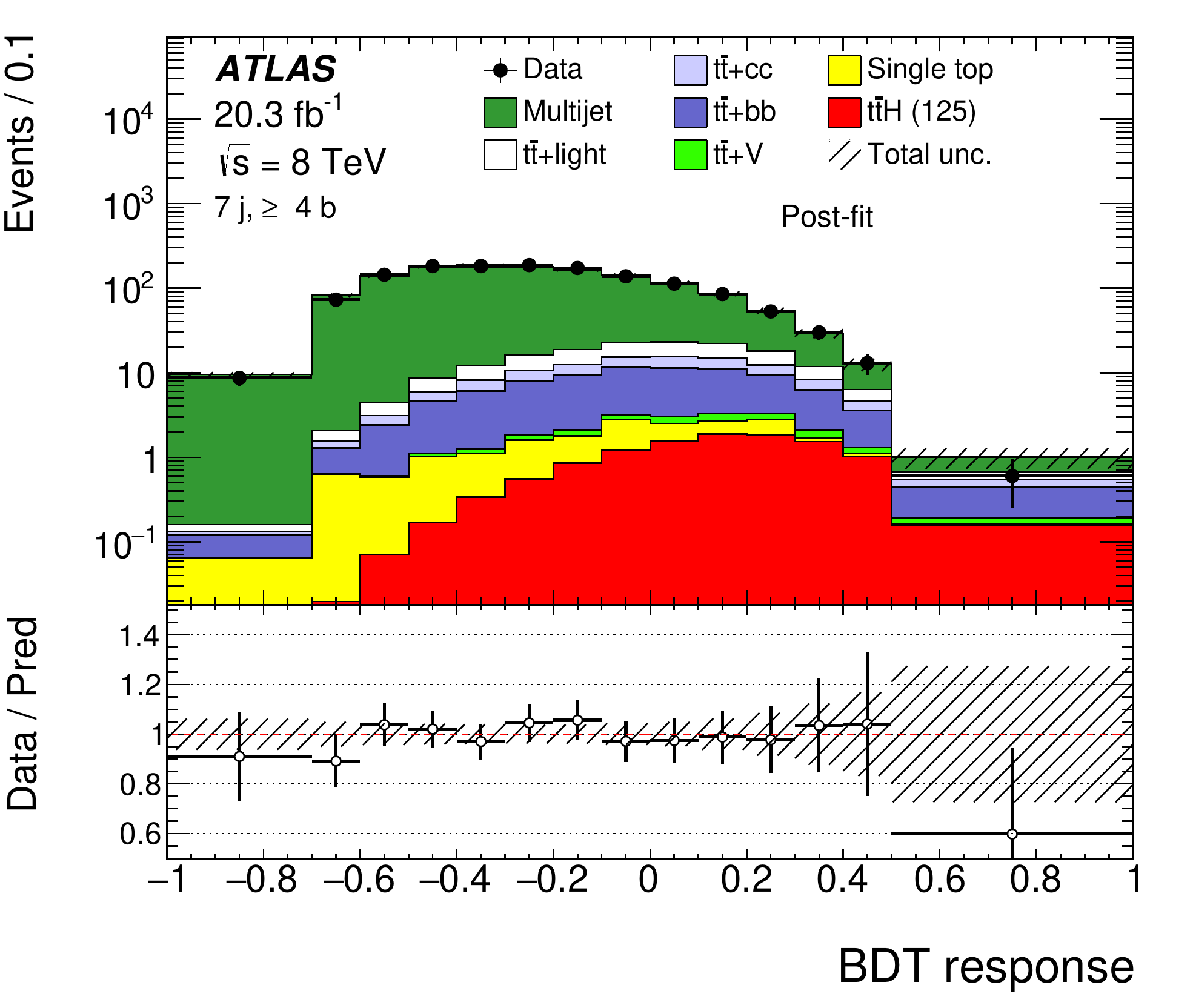}}\label{fig:prepost_4b_d}\\
    \subfloat[]{\includegraphics[width=0.42\textwidth]{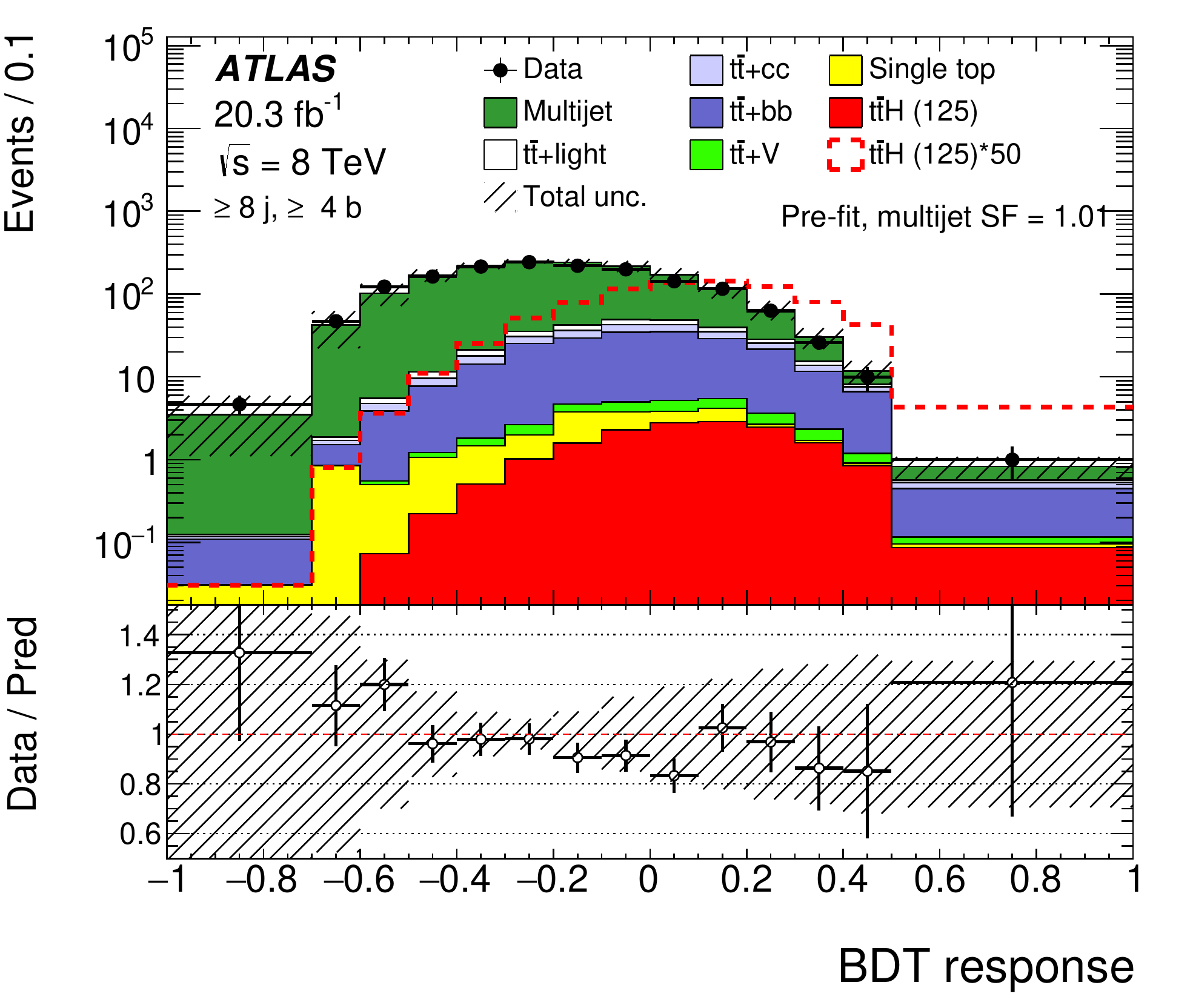}}\label{fig:prepost_4b_e}
    \subfloat[]{\includegraphics[width=0.42\textwidth]{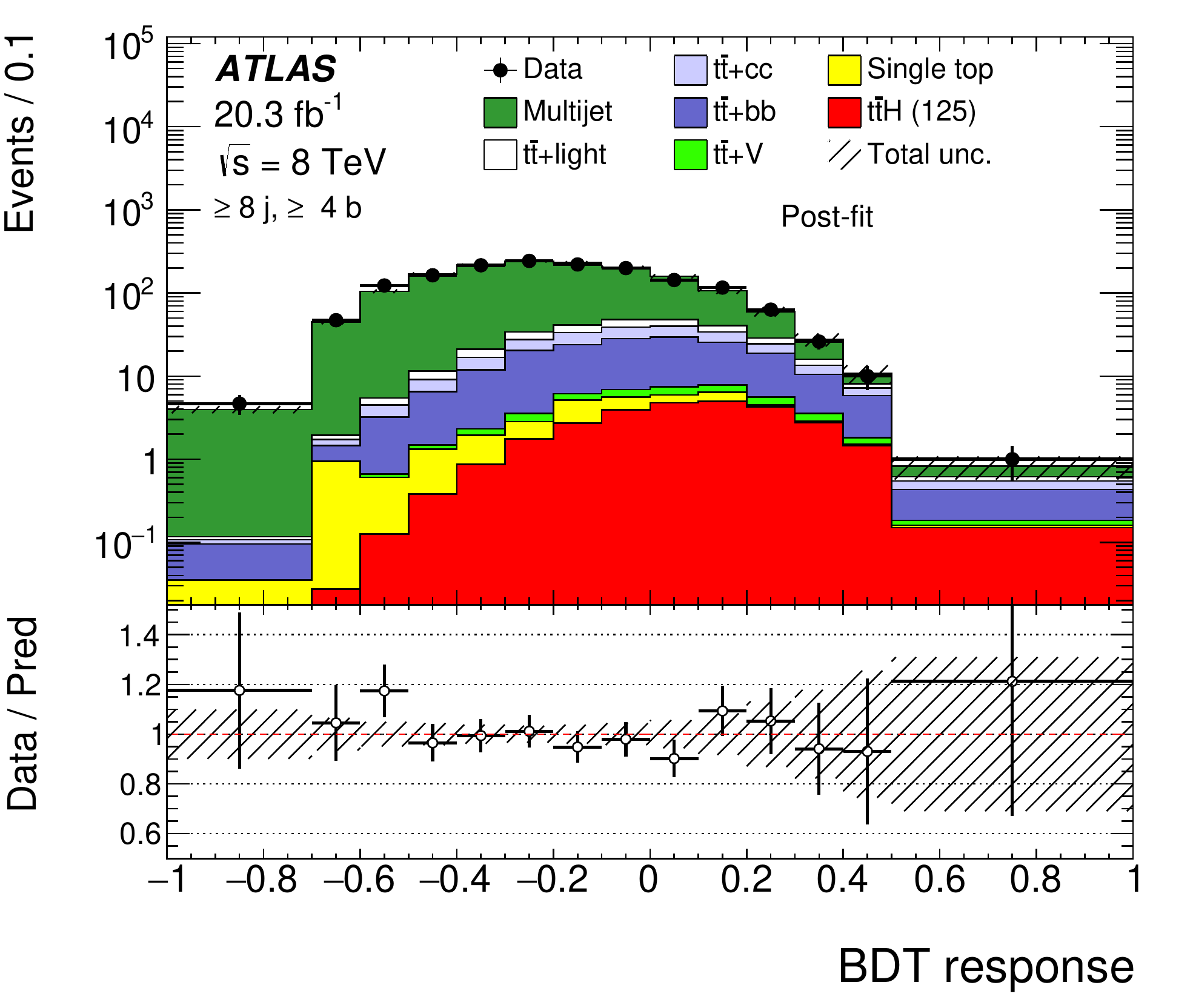}}\label{fig:prepost_4b_f}\\
    \caption{Comparison between data and prediction for the
        BDT discriminant in the, from top to bottom, (6--8j, $\geq$4b)
        regions before (left) and after (right) the fit. The fit is performed under the signal-plus-background 
        hypothesis. Pre-fit plots show an overlay of the multijet distribution normalised to
        data for illustration purposes only.
        The bottom panels display the ratios of data to the total prediction.
        The hashed areas represent the total uncertainty in the
        background predictions. 
        The $t\bar{t}H$ signal yield (solid red) is scaled by a fixed factor
        before the fit. 
      }
    \label{fig:prepost_4b} 
  \end{center}
\end{figure*}

Figures~\ref{fig:prepost_3b} and \ref{fig:prepost_4b} show the BDT output
distributions for data and the predictions in each analysis region, both
before (left panels) and after (right panels) the fit to data. The relative uncertainties decrease
significantly in all regions due to the constraints provided by the data,
exploiting the correlations between the uncertainties in the different analysis regions.

The signal strength in the all-hadronic $t\bar{t}H$ decay mode, for $m_H$ =
125~\GeV, is measured to be:
\begin{equation}
  \mu(m_H \! = \! 125~\GeV) = 1.6 \pm 2.6.
\end{equation}
The expected uncertainty in the signal strength ($\mu$ = 1)
is $\pm$2.8. The observed (expected) significance of the signal
is 0.6 (0.4) standard deviations, corresponding to
an observed (expected) $p$-value of 27\% (34\%), where the $p$-value is the probability to
obtain a result at least as signal-like as observed if no signal were present.

The observed and expected limits are summarised in
Table~\ref{tab:LimitVSAnalysis}. A $t\bar{t}H$ signal 6.4 times larger than
predicted by the SM is excluded at 95\% CL. A signal 5.4 times larger than the
signal of a SM Higgs boson is expected to be excluded for the background-only hypothesis.  

\begin{table*}[!ht]
    \caption{\label{tab:LimitVSAnalysis} Observed and expected
      upper limits at 95\% CL on $\sigma(t\bar{t}H)$ relative to the SM prediction assuming
      $m_H=125$~\GeV, for the background-only hypothesis. Confidence intervals around the
      expected limits under the background-only hypothesis are also provided,
      denoted by $\pm 1 \sigma$ and $\pm 2 \sigma$, respectively.  The
      expected (median) upper limit at 95\% CL assuming the SM prediction for
      $\sigma(t\bar{t}H)$ is shown in the last column. } 
  \small
  \begin{center}
    \begin{tabular}{l|c|ccccc|c}
      \hline
        \multirow{2}{*} & \multirow{2}{*}{Observed} &
        \multicolumn{5}{c|}{Expected if $\mu$ = 0} & Expected if $\mu$ = 1  \\ \cline{3-8}
           &  & $-2\sigma$  & $-1\sigma$  & Median    & $+1\sigma$  & $+2\sigma$ & Median \\
      
      \hline
      Upper limit on $\mu$ at 95\% &  6.4   &  2.9    & 3.9  &   5.4   & 7.5  & 10.1 & 6.4 \\
     \hline
    \end{tabular}
  \end{center}
\end{table*}

Figure~\ref{fig:elemer} summarises the post-fit event yields for data, total
background and signal expectations as a function of $\log_{10}(S/B)$. The signal is
normalised to the fitted value of the signal strength ($\mu$ = 1.6). A
  signal strength 6.4 times larger than predicted by the SM is also shown 
in Figure~\ref{fig:elemer}. 
 
Figure~\ref{fig:ranking} shows the effect of the major systematic
uncertainties on the fitted value of $\mu$ and the constraints
provided by the data. The ranking, from top to bottom, is determined by the post-fit impact on $\mu$. 
This effect is calculated by fixing the corresponding nuisance parameter at $\hat{\mathrm{\theta}} \pm \sigma_{\mathrm{\theta}}$
and performing the fit again. Here $\hat{\mathrm{\theta}}$ is the fitted value of the nuisance parameter and
$\sigma_{\mathrm{\theta}}$ is its post-fit uncertainty.
The difference between the default and the modified $\mu$, $\Delta\mu$,
represents the effect on $\mu$ of this particular systematic 
uncertainty. This is also shown in Table~\ref{tab:elemer}.

\begin{table*}[!ht]
    \caption{\label{tab:elemer}  Effect of the different
      sources of systematic uncertainties~on $\mu$ expressed in terms of percentage of the fitted value of $\mu$ sorted according to their post-fit effect.
      }
  \small
  \begin{center}
    \begin{tabular}{l|c}
      \hline
      Sources of systematic uncertainty & $\pm 1 \sigma$ post-fit impact on $\mu$ \\
      \hline
      $t\bar{t}$ normalisation  & 108\%  \\
      Multijet normalisation 		                         &  71\%  \\ 
      Multijet shape  						 & 60\%  \\
      Main contributions from $t\bar{t}$ modelling         &  34\%--41\% \\
      Flavour tagging 		                                 & 31\%  \\
      Jet energy scale                                              & 27\%  \\
      Signal modelling 	                                        & 22\%  \\
      Luminosity+trigger+JVF+JER 	                              & 18\%  \\
      \hline
    \end{tabular}
  \end{center}
\end{table*}

\begin{figure}[!ht]
  \begin{center}
    \includegraphics[width=0.48\textwidth]{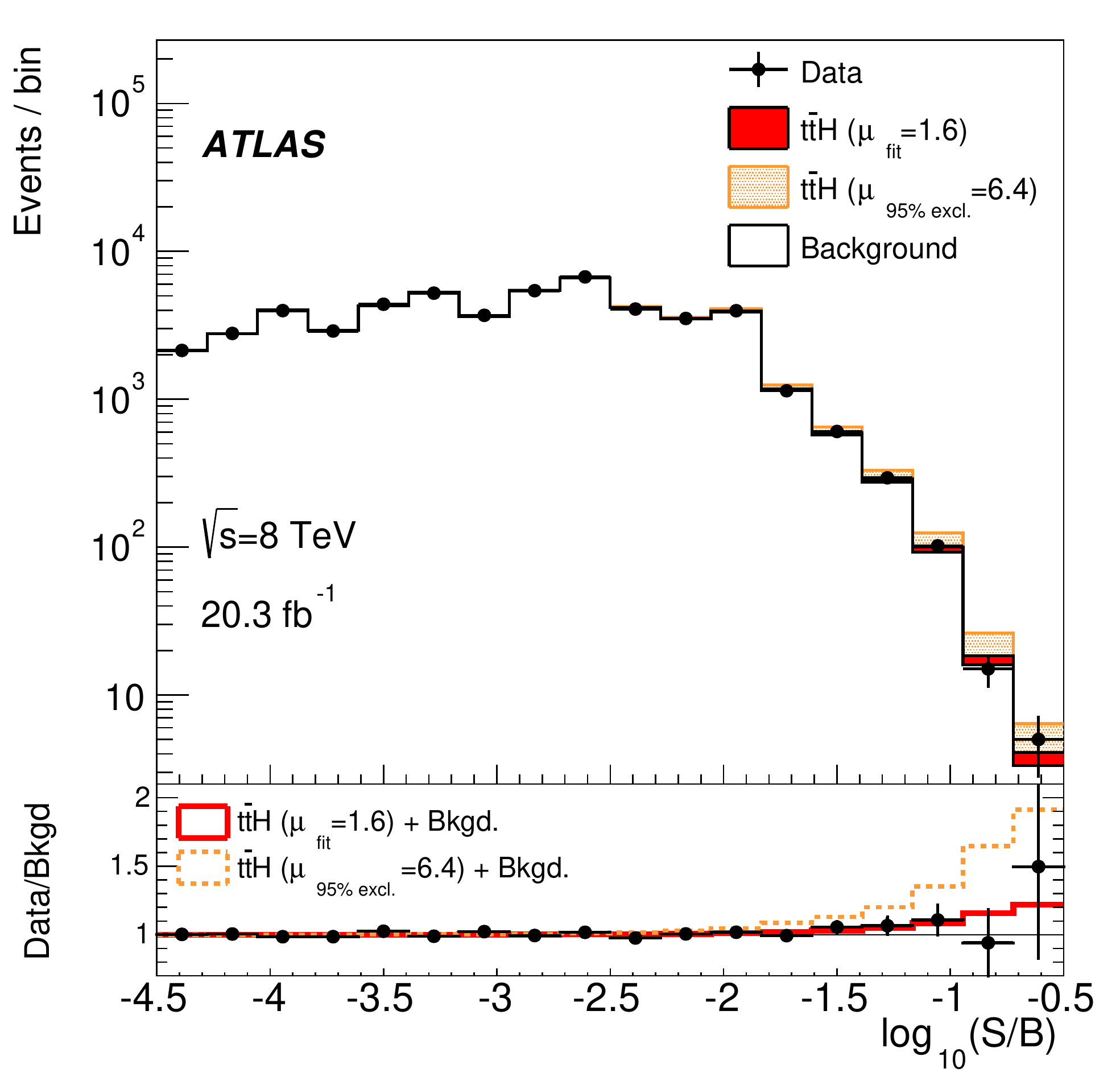}
    \caption{Event yields as a function of $\log_{10}(S/B)$, where $S$ 
      (expected signal yield) and $B$ (expected background 
      yield) are taken from the corresponding BDT discriminant bin. Events from all fitted regions are included. 
      The predicted background is obtained from the global signal-plus-background fit. 
      The $t\bar{t}H$ signal is shown both for the best-fit value ($\mu = 1.6$) and for the upper limit at 95\% CL ($\mu=6.4$).}
    \label{fig:elemer}
  \end{center}
\end{figure}

\begin{figure*}[!htb]
  \begin{center}
    \includegraphics[width=0.65\textwidth]{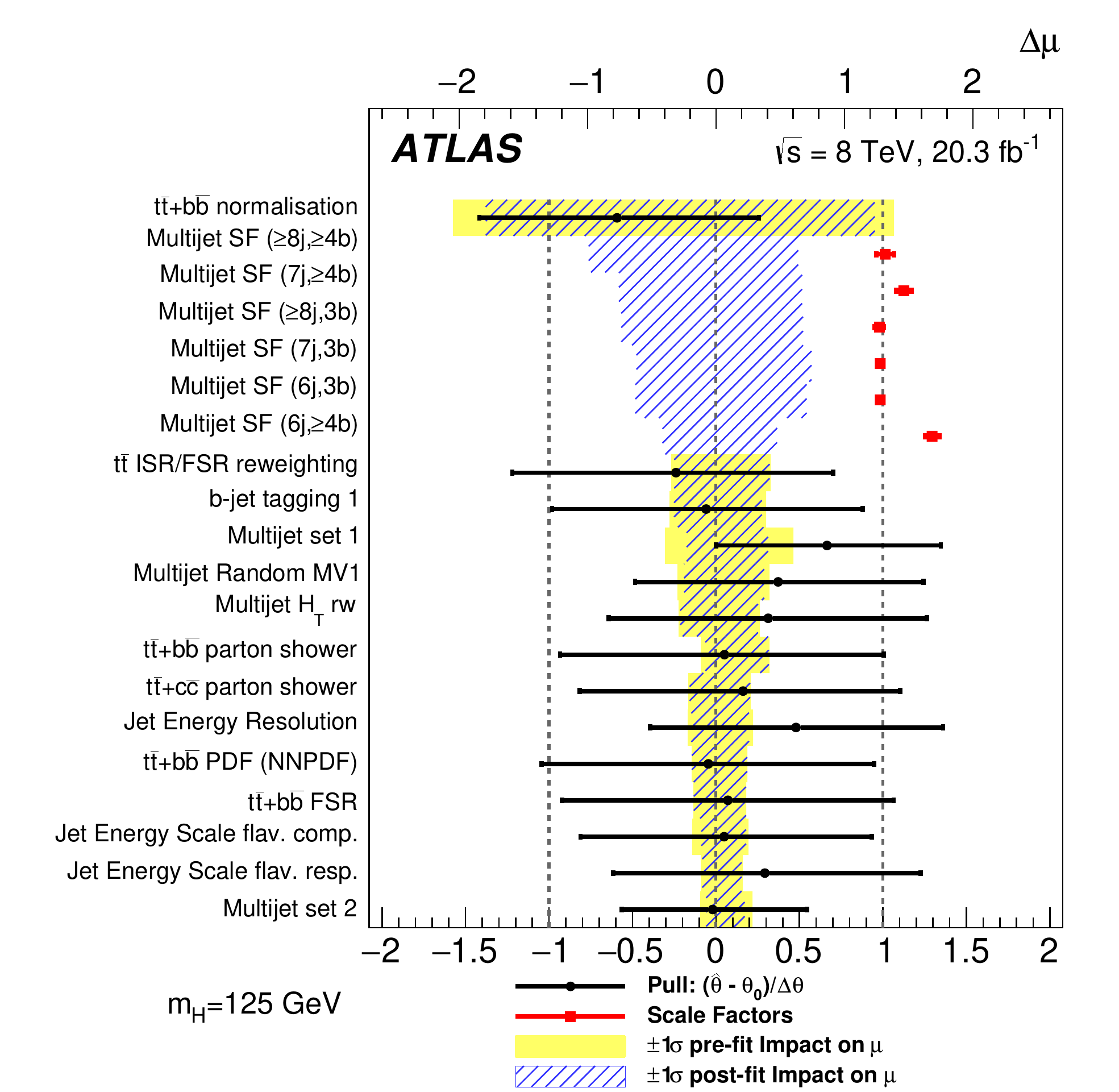}\hspace{0.15cm}
    \caption{The fitted values of the 20 nuisance parameters
      corresponding to the sources of systematic uncertainty with the
      largest impact on the fitted signal strength $\mu$. The
      points, which are drawn conforming to the scale of the bottom axis, show the
      deviation of each of the fitted nuisance parameters $\hat{\mathrm{\theta}}$ from
      $\mathrm{\theta_{0}}$, which is the nominal value of that nuisance parameter, in units
      of the pre-fit standard deviation $\Delta\theta$. The plain yellow area
      represents the pre-fit impact on $\mu$ and the hashed blue area its
      post-fit impact. The error bars show the
      post-fit uncertainties $\sigma_{\theta}$, which have size
        close to one if the data do not provide
      any further constraint on that uncertainty. Conversely, an error bar for
      $\sigma_{\theta}$ smaller than one indicates a 
      reduction with respect to the original uncertainty. 
      The nuisance parameters are sorted according to their post-fit impact
      $\Delta\theta$ (top horizontal scale). Multijet scale factors (SF) show
      the fitted values and uncertainties of the 
      normalisation parameters that are freely floating in the fit. These
      normalisation parameters have a pre-fit value of unity. 
    }
    \label{fig:ranking} 
  \end{center}
\end{figure*}

The largest effect arises from the uncertainty in the normalisation of the
irreducible $t\bar{t}+b\bar{b}$ background. The $t\bar{t}+b\bar{b}$
background normalisation is smaller by 30\% in the fit than the prediction, resulting in a decrease of the 
observed $t\bar{t}+b\bar{b}$ yield with respect to the {\sc Powheg}+{\sc Pythia} prediction. The second largest effect comes from the multijet background
normalisation. The data-driven method
focuses on  modelling the shape of the multijet background while the
  normalisation is constrained by the regions dominated by multijet background.
The uncertainty in the normalisation parameters amounts to few percent
and the values from each region are consistent with
the variations applied to these parameters to account for systematic
  uncertainties. Two of the multijet background shape uncertainties are ranked fourth and fifth,
and their pulls are slightly positive. 

Other important uncertainties include $b$-tagging and
JES. Uncertainties arising from jet energy resolution, jet vertex 
fraction, jet reconstruction and JES that affect primarily low-\pt jets, as
well as the $t\bar{t}$+light-jet background modelling
uncertainties, do not have a significant impact on the result.

\clearpage
\section{Combination of \tth\ results at $\sqrt{s}=7$ and 8~\TeV} 
\label{sec:tthcombi}

The sensitivity of the search for \tth\ production can be increased by statistically combining different Higgs boson decay channels. This combination is described in the following.

  %-------------------------------------------------------------------------------
  \subsection{Individual \ttH\ measurements and results}
  \label{sec:tthindiv}
  %-------------------------------------------------------------------------------
   
  The \ttH\ searches that are combined are:
  \begin{itemize}
	  \item \ttH(\Htobb ) in the single-lepton and opposite-charge dilepton~\ttbar\ decay channels using data at $\sqrt{s} = 8$~\TeV~\cite{Aad:2015gra},
	  \item \ttH(\Htobb ) in the all-hadronic \ttbar\ decay channel using data at $\sqrt{s} = 8$~\TeV\ as presented in this paper,% in the previous sections},
	  \item \ttH(\Htomulti ) with two same-charge leptons ($e$ or $\mu$), three leptons, four leptons, two hadronically decaying $\tau$ leptons plus one  lepton and one hadronically decaying $\tau$ lepton plus two leptons in the final state
	  using data at $\sqrt{s} = 8$~\TeV~\cite{Aad:2015iha},
	  \item \ttH(\Htoyy ) at $\sqrt{s} = 7$ and $8$~\TeV\ in both the hadronic and leptonic ($e$ or $\mu$) $t\bar{t}$ pair decay channels~\cite{ttHyy}.
  \end{itemize}

  First all \Htobb\ final states are combined, obtaining a signal strength for the \ttH(\Htobb) combination, and then the outcome is combined with the remaining (non-\Htobb) channels.
     
  %-------------------------------------------------------------------------------
  \subsubsection{\Htobb\ (single lepton and dilepton \ttbar\ decays)}
  \label{subsec:tthbb}
  %-------------------------------------------------------------------------------
  
  The search for \ttH\ production with \Htobb\ is performed in both the single-lepton and dilepton $t\bar{t}$ decay modes~\cite{Aad:2015gra}. 
The single-lepton analysis requires one charged lepton with at least four jets, of which at least two need to be $b$-tagged, while the dilepton analysis requires two opposite-charge leptons with at least two jets, of which at least two must be $b$-tagged.  The events are then categorised according to the jet and $b$-tagged jet multiplicity. The dominant background in the signal-enriched regions is from \ttbb\ events. In these regions, neural networks~\cite{Neurobayes} are built using kinematic information in order to separate the \ttH\ signal from \ttbar\ background.    Furthermore, in the single-lepton channel, a matrix-element discriminant is built in the most signal-enriched regions and is used as an input to the neural network.     
  
 %-------------------------------------------------------------------------------
 \subsubsection{\Htomulti}
 \label{subsec:multilepton}
 %-------------------------------------------------------------------------------
 
 The \ttH\ search with \Htomulti~\cite{Aad:2015iha} exploits several multilepton signatures resulting from Higgs boson decays to vector bosons and/or $\tau$ leptons.  Events are categorised based on the number of charged leptons and/or hadronically decaying  $\tau$ leptons in the final state.  The categorisation includes events with two same-charge leptons, three leptons, four leptons, one lepton and two hadronic  $\tau$ leptons, as well as two same-charge leptons with one hadronically decaying $\tau$ lepton. Backgrounds include events with electron charge misidentification, which are estimated using data-driven techniques, non-prompt leptons arising from semileptonic $b$-hadron decays, mostly from \ttbar\ events, again estimated from data-driven techniques, and production of \ttW\ and \ttZ, which are estimated using MC simulations. Signal and background event yields are obtained from a simultaneous fit to all channels. 
  
  %-------------------------------------------------------------------------------
  \subsubsection{\Htoyy}
  \label{subsec:tthyy}
  %-------------------------------------------------------------------------------
  
  The \ttH\ search in the \Htoyy\ channel~\cite{ttHyy} exploits the
  sharp peak in the diphoton mass distribution from the \Htoyy\ decay over the continuum background.  The analysis is split according to the decay mode of the \ttbar\ pair. A leptonic selection requires at least one lepton and at least one $b$-tagged jet, and missing transverse momentum if there is only one $b$-tagged jet, whereas a hadronic selection requires a combination of jets and $b$-tagged jets.
  % to improve the signal sensitivity.
   Contributions from peaking non-\ttH\ Higgs boson production modes are estimated from MC simulations.  The signal is extracted with a fit using the diphoton mass distribution as a discriminant.
    
  %-------------------------------------------------------------------------------
  \subsection{Correlations}
  \label{subsec:correlation}
  %-------------------------------------------------------------------------------
  
  Nuisance parameters corresponding to the same source of uncertainty in different analyses are generally considered to be correlated with each other, except for the following sets:
\begin{itemize}
\item Nuisance parameters related to $b$-tagging (also $c$-tagging and light mis-tagging) are considered to be independent among the analyses as different $b$-tagging working points are employed.  
	\item The electron identification uncertainty is considered to be uncorrelated between analyses due to different selections used. 
          
\end{itemize}	
  
  %-------------------------------------------------------------------------------
  \subsection{Results of the combination}
  \label{subsec:resultscombo}
  %-------------------------------------------------------------------------------
  
  \subsubsection{Signal strength}

  The result of the \ttH(\Htobb) combination for the signal strength is $\mu = 1.4 \pm 1.0$. 
 The observed signal strengths
 for the individual \ttH(\Htobb) channels and for their combination are summarised in Figure~\ref{fig:summary_mu}.
 The $t\bar{t}+b\bar{b}$ normalisation nuisance parameters obtained in the all-hadronic analysis ($-0.6 \pm 0.8$) and the leptonic analysis ($+0.8 \pm 0.4$)
 The expected significance increases from $1.0 \sigma$ for the leptonic final state of \ttH(\Htobb) to
 $1.1 \sigma$ for the combined \ttH(\Htobb). Because the combined
 \ttH(\Htobb) best-fit value of $\mu$ is lower than the leptonic-only value, the
 observed significance for the \ttH(\Htobb) combination is reduced from $1.4 \sigma$
 (leptonic \cite{Aad:2015gra}) to $1.35 \sigma$ (combined). 

 \begin{figure}[ht!]
 \begin{center}
 \includegraphics[width=0.65\textwidth]{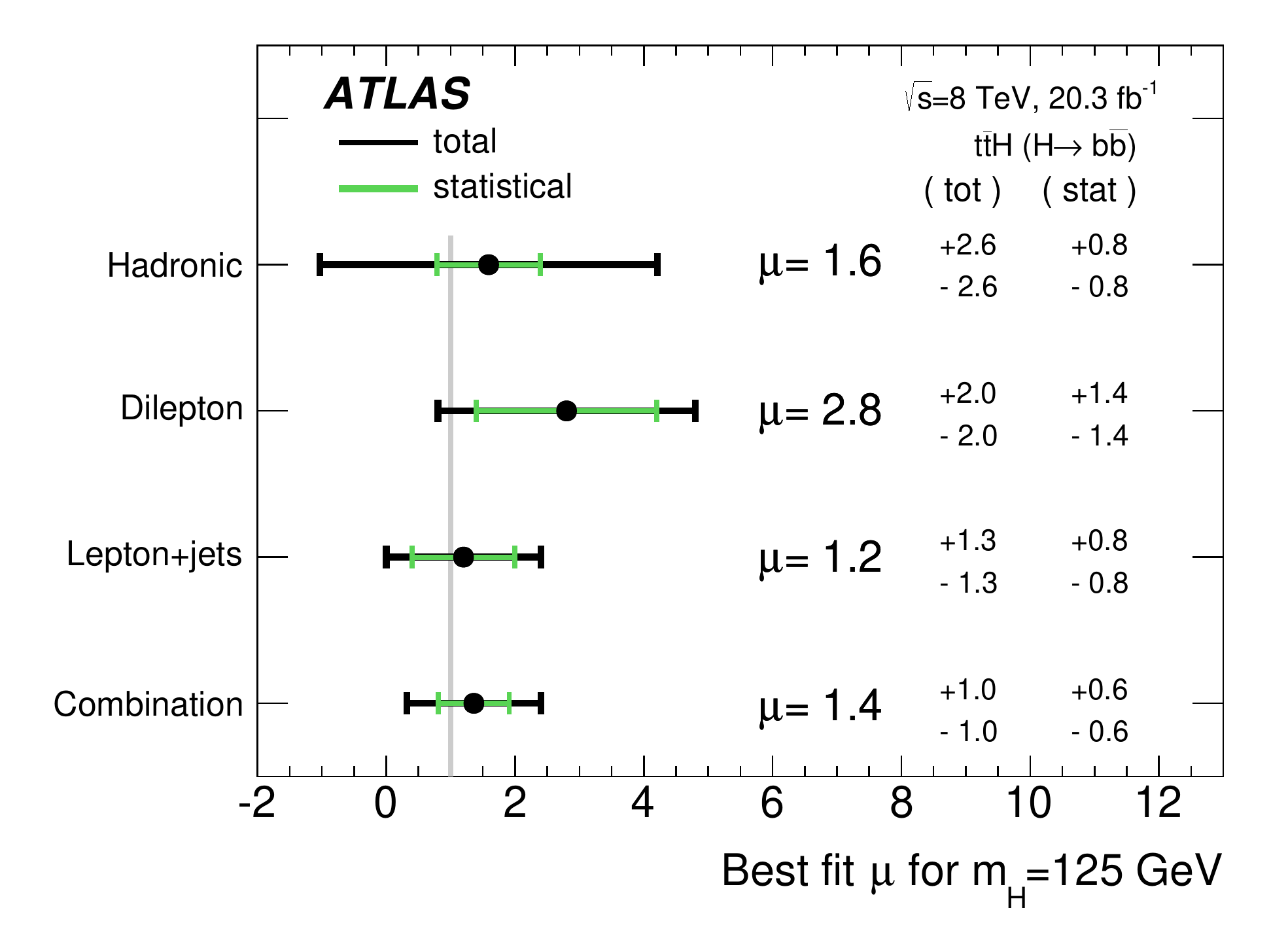} 
 \caption{Summary of the measurements of the signal strength $\mu$ for \ttH(\Htobb ) production for the individual \Htobb\ channels and for their combination, assuming $m_H = 125$~\GeV. The total (tot) and statistical (stat) uncertainties of $\mu$ are shown. The SM  $\mu=1$ expectation is shown as the grey line.
}
 \label{fig:summary_mu} 
 \end{center}
 \end{figure}
  
 Figure~\ref{fig:summary_muFULL} summarises the observed signal strength $\mu$\ of the individual \ttH\ channels (\Htobb, \Htoyy\ and \Htomulti) and the \ttH\ combination. 
The observed (expected) significance of the combined \ttH\ result is $2.33 \sigma$ ($1.53 \sigma$).  
  
  \begin{figure}[!h]
  \begin{center}
  \includegraphics[width=0.65\textwidth]{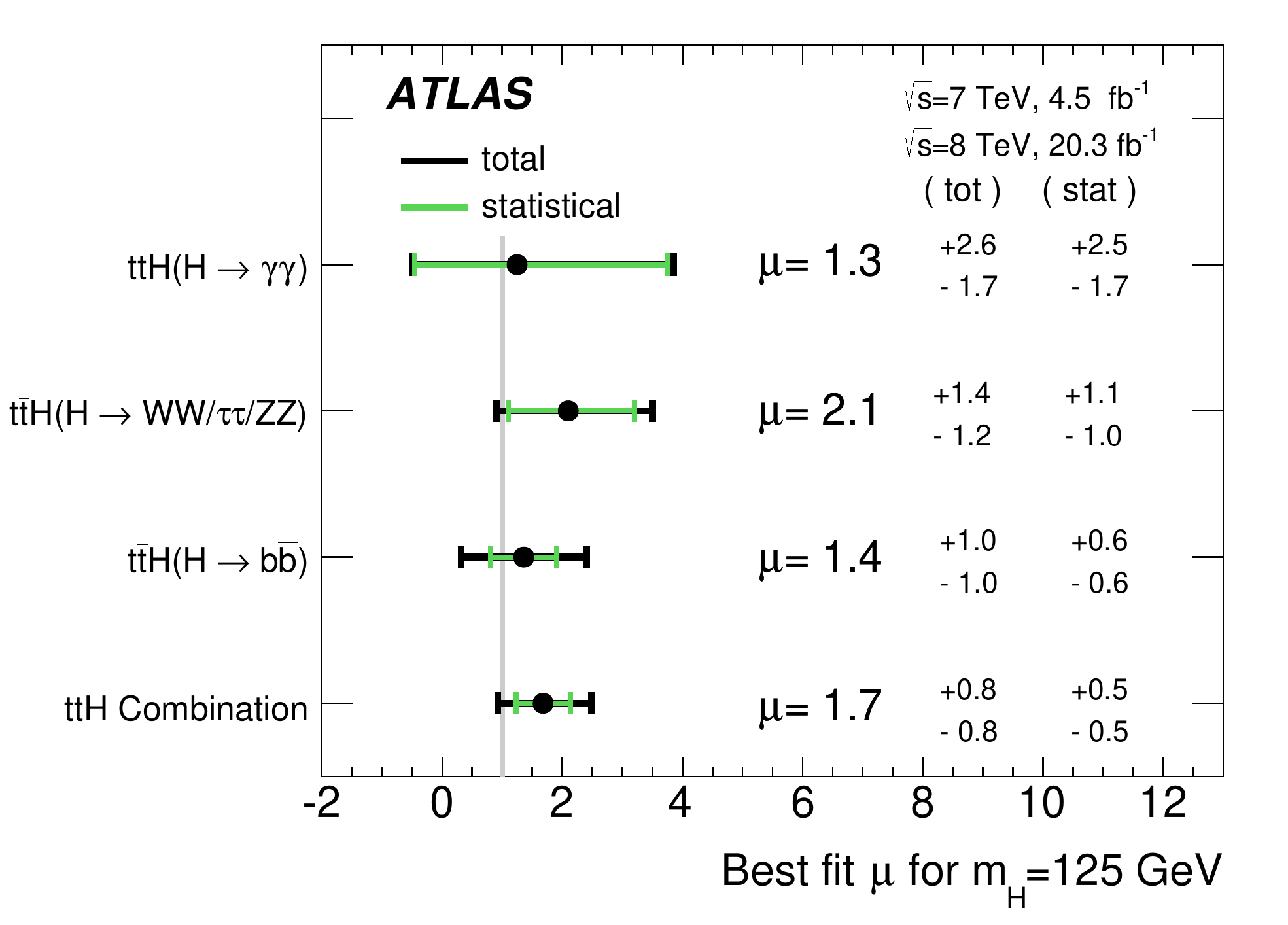} 
  \caption{Summary of the measurements of the signal strength $\mu$ for the individual channels and for their combination, assuming $m_H = 125$~\GeV. The total (tot) and statistical (stat) uncertainties of $\mu$ are shown. 
The SM  $\mu=1$ expectation is shown as the grey line.
}
  \label{fig:summary_muFULL} 
  \end{center}
  \end{figure}

The combination of all $t\bar{t}H$ analyses yields an observed (expected) 95\% CL upper limit of  3.1 (1.4) times  the SM cross section. The observed 95\% CL limits for the individual \tth\ channels and for the combination are shown in Figure~\ref{fig:limits} and in Table~\ref{tab:LimitsVSAnalysis}. 

  \begin{figure}[!ht]
  \begin{center}
  \includegraphics[width=0.65\textwidth]{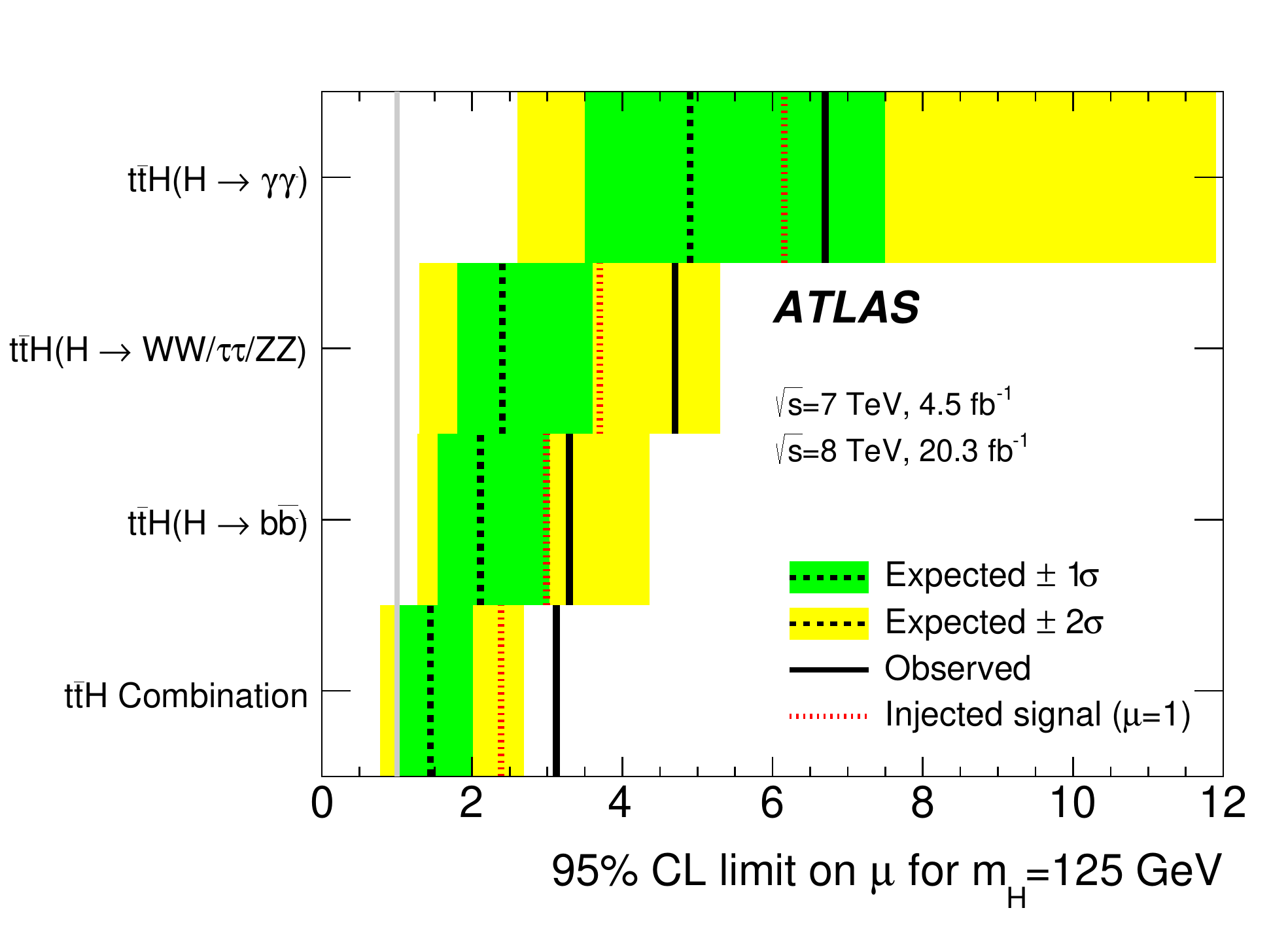}
  \caption{Upper limits on the signal strength $\mu$ for the individual channels as well as for their combination, at 95\% CL. The observed 
  limits (solid lines) are compared to the expected median limits under the background-only hypothesis (black dashed lines) and under the signal-plus-background hypothesis  
  assuming the SM prediction for $\sigma(t\bar{t}H)$ (red dotted
  lines). The surrounding green and yellow bands bands correspond to the $\pm 1 \sigma$ and $\pm 2 \sigma$ ranges around the
  expected limits under the background-only hypothesis.
  }
  \label{fig:limits} 
  \end{center}
  \end{figure}

\begin{table*}[!ht]
\caption{\label{tab:LimitsVSAnalysis} Observed and expected (median, for the
  background-only hypothesis) upper limits at 95\% CL on $\sigma(t\bar{t}H)$ relative to the SM prediction, for the individual channels as well as for their combination. The $\pm 1$$\sigma$ and $\pm 2$$\sigma$ ranges around the expected limit are also given. The expected median upper limits at 95\% CL assuming the SM prediction for $\sigma(t\bar{t}H)$ are shown in the last column. }
  \small
\begin{center}
\begin{tabular}{l|c|ccccc|c|c}
  \hline
  \multirow{3}{*}{Analysis} &  \multicolumn{7}{|c|}{95\% CL upper limit} & Signal strength \\ \cline{2-9}
   
  & \multirow{2}{*}{Observed}         & \multicolumn{6}{|c|}{Expected} & \multirow{2}{*}{$\mu$}  \\
                    &   & $-2\sigma$  & $-1\sigma$  & median    & $+1\sigma$  & $+2\sigma$  & median ($\mu = 1$) &  \\
  \hline  
  \ttH(\Htoyy)             &  6.7   &  2.6    & 3.5  &   4.9   & 7.5  & 11.9  & 6.2  & $1.2^{+2.6}_{-1.8}$\\[0.1cm]
  \ttH($H \rightarrow$ leptons)          &  4.7   &  1.3    & 1.8  &   2.4   & 3.6  & 5.3   & 3.7 & $2.1^{+1.4}_{-1.2}$ \\[0.1cm]   
  \ttH(\Htobb)             &  3.3   &  1.3    & 1.5  &   2.1   & 3.0  & 4.4   & 3.0 & 1.4 $\pm$ 1.0 \\ \hline
  \ttH~Combination        &  3.1  &  0.8   & 1.0 &   1.4  & 2.0 & 2.7  & 2.4 & 1.7 $\pm$ 0.8  \\
  \hline
\end{tabular}
\end{center}
\end{table*}

The result for the best-fit value is $\mu = 1.7 \pm 0.8$.

  \subsubsection{Couplings}
  
  Sensitivity to $t-H$ and $W-H$ couplings stems from several
  sources: from the \ttH\ production itself, from the Higgs boson decay branching
  fractions, from associated single top and Higgs boson production processes
  ($tHjb$ and $WtH$), where interference terms include both the $t\bar{t}H$
    and $WWH$ vertices, and from the $H \to \gamma\gamma$ branching
  fraction, where again interferences between loop contributions from the top quark and the $W$ boson
    are present. Different channels differ in their sensitivity to these components.    
  A two-parameter fit is performed, assuming that all boson couplings scale with the same modifier $\kappa_{V}$, while all fermion couplings scale with the same
  modifier $\kappa_{F}$.
  
  The parameterisation of the couplings for the \ttH\ and $tH$ production modes and for
  the different Higgs boson decay modes is taken from Refs.~\cite{Aad:2015gba,
    LHCHiggsXSWG}.
  Figure~\ref{fig:kappaV_kappaF_FULL} shows the log-likelihood
  contours of $\kappa_{F}$ versus $\kappa_{V}$ for the combined
  \ttH\ fit. The combination of all analysis channels slightly prefers positive $\kappa_F$. Additional studies, performed to determine the contribution of the individual analyses to the combined coupling measurement, indicate that the \ttH, \Htomulti\ analysis prefers somewhat enhanced $W-H$ coupling, which can only be compatible with the \ttH(\Htoyy) rate if the interference between $t\bar{t}H$
  and $WWH$ amplitudes is destructive, as expected in the SM.
  
  \begin{figure}[!h]
  \begin{center}
  \includegraphics[width=0.85\textwidth]{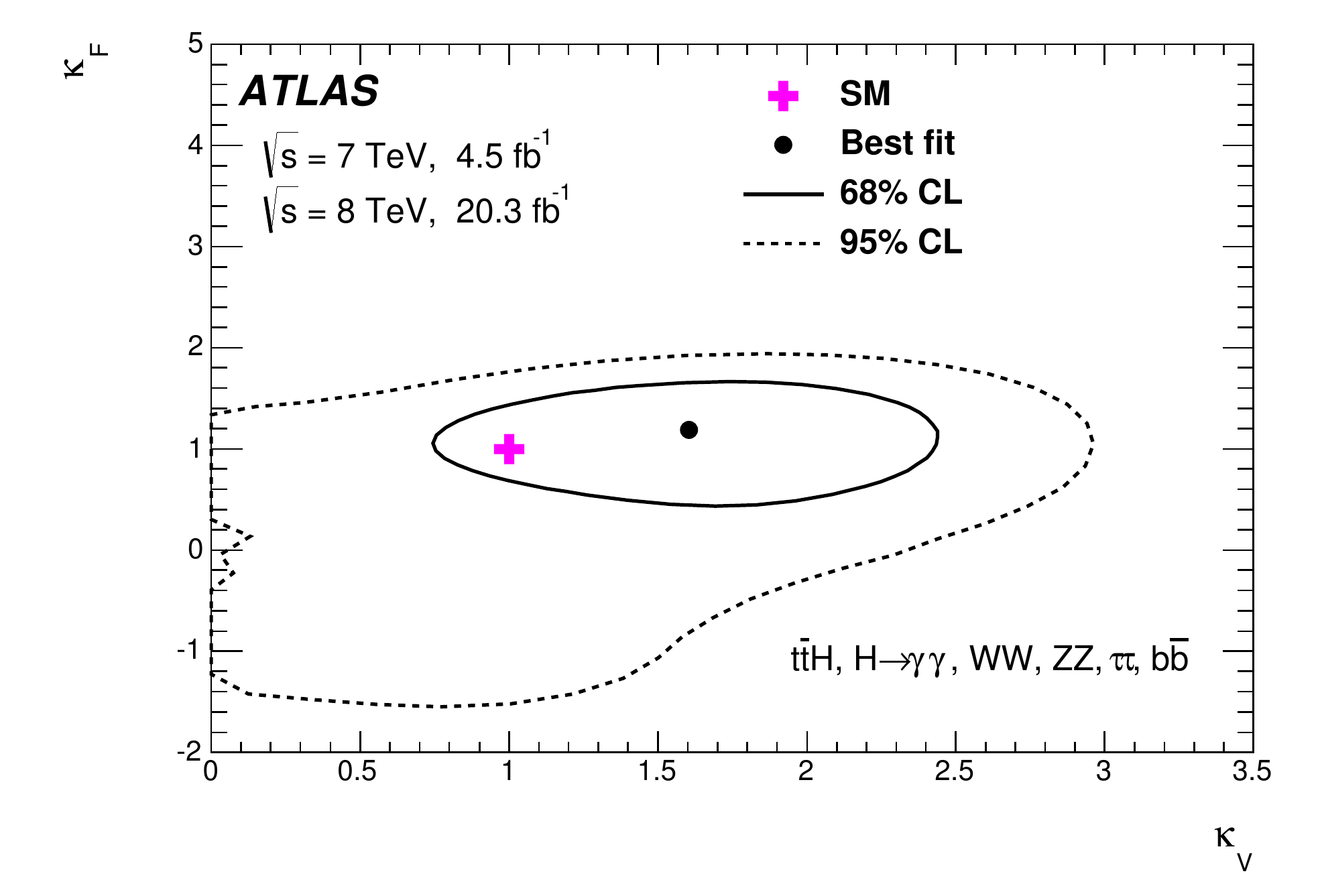} 
  \caption{Log-likelihood 
    for the combined \ttH\ fit. The fit
    agrees with the SM expectation within the 68\% CL
      contour. The physical boundary of $\kappa_{V} \geq 0$ is considered.} 
  \label{fig:kappaV_kappaF_FULL}  
  \end{center}   
  \end{figure}

\section{Conclusion}
\label{sec:conclusion}

A search for the SM Higgs boson produced in
association with a pair of top quarks ($t\bar{t}H$) has been carried out with the ATLAS detector at the Large Hadron Collider.
The search focuses on
\htobb\ decays with $t\bar{t}$ pairs decaying hadronically. The data
used correspond to an integrated luminosity of 20.3~\ifb~of
$pp$ collisions at $\sqrt{s}=8$~\TeV. 
The analysis is carried out in six different jet and
$b$-tagged jet multiplicity regions. Discrimination between signal and
background is obtained by employing a boosted decision tree multivariate classifier in all regions.  
No significant excess of events above the background expectation is
found for the SM  Higgs boson with a mass of 125~\GeV.
An observed (expected) 95\% CL upper limit of 6.4 (5.4) times
the SM  cross section is obtained. By performing a fit under the
signal-plus-background hypothesis,
the ratio of the measured signal strength
to the SM  expectation is found to be $\mu = 1.6 \pm 2.6$.

The statistical combination of all $t\bar{t}H$ analyses performed at  $\sqrt{s}=7$~\TeV\ and  $8$~\TeV\ yields an observed (expected) upper limit of 3.1 (1.4) times the SM cross section at 95\% CL.
The combined measured signal  strength is found to be $\mu = 1.7 \pm 0.8$.

\section*{Acknowledgements}

% Acknowledgements for papers with collision data
% Version 7-Feb-2016

% Standard acknowledgements start here
%----------------------------------------------
We thank CERN for the very successful operation of the LHC, as well as the
support staff from our institutions without whom ATLAS could not be
operated efficiently.

We acknowledge the support of ANPCyT, Argentina; YerPhI, Armenia; ARC, Australia; BMWFW and FWF, Austria; ANAS, Azerbaijan; SSTC, Belarus; CNPq and FAPESP, Brazil; NSERC, NRC and CFI, Canada; CERN; CONICYT, Chile; CAS, MOST and NSFC, China; COLCIENCIAS, Colombia; MSMT CR, MPO CR and VSC CR, Czech Republic; DNRF and DNSRC, Denmark; IN2P3-CNRS, CEA-DSM/IRFU, France; GNSF, Georgia; BMBF, HGF, and MPG, Germany; GSRT, Greece; RGC, Hong Kong SAR, China; ISF, I-CORE and Benoziyo Center, Israel; INFN, Italy; MEXT and JSPS, Japan; CNRST, Morocco; FOM and NWO, Netherlands; RCN, Norway; MNiSW and NCN, Poland; FCT, Portugal; MNE/IFA, Romania; MES of Russia and NRC KI, Russian Federation; JINR; MESTD, Serbia; MSSR, Slovakia; ARRS and MIZ\v{S}, Slovenia; DST/NRF, South Africa; MINECO, Spain; SRC and Wallenberg Foundation, Sweden; SERI, SNSF and Cantons of Bern and Geneva, Switzerland; MOST, Taiwan; TAEK, Turkey; STFC, United Kingdom; DOE and NSF, United States of America. In addition, individual groups and members have received support from BCKDF, the Canada Council, CANARIE, CRC, Compute Canada, FQRNT, and the Ontario Innovation Trust, Canada; EPLANET, ERC, FP7, Horizon 2020 and Marie Sk{\l}odowska-Curie Actions, European Union; Investissements d'Avenir Labex and Idex, ANR, R{\'e}gion Auvergne and Fondation Partager le Savoir, France; DFG and AvH Foundation, Germany; Herakleitos, Thales and Aristeia programmes co-financed by EU-ESF and the Greek NSRF; BSF, GIF and Minerva, Israel; BRF, Norway; the Royal Society and Leverhulme Trust, United Kingdom.

The crucial computing support from all WLCG partners is acknowledged
gratefully, in particular from CERN and the ATLAS Tier-1 facilities at
TRIUMF (Canada), NDGF (Denmark, Norway, Sweden), CC-IN2P3 (France),
KIT/GridKA (Germany), INFN-CNAF (Italy), NL-T1 (Netherlands), PIC (Spain),
ASGC (Taiwan), RAL (UK) and BNL (USA) and in the Tier-2 facilities
worldwide.
%----------------------------------------------

\clearpage
\printbibliography
\newpage 
% ATLAS Collaboration author list
% Data extracted on 12-Apr-2016 for paper reference HIGG-2015-05
\begin{flushleft}
{\Large The ATLAS Collaboration}

\bigskip

G.~Aad$^\textrm{\scriptsize 87}$,
B.~Abbott$^\textrm{\scriptsize 114}$,
J.~Abdallah$^\textrm{\scriptsize 65}$,
O.~Abdinov$^\textrm{\scriptsize 12}$,
B.~Abeloos$^\textrm{\scriptsize 118}$,
R.~Aben$^\textrm{\scriptsize 108}$,
M.~Abolins$^\textrm{\scriptsize 92}$,
O.S.~AbouZeid$^\textrm{\scriptsize 138}$,
N.L.~Abraham$^\textrm{\scriptsize 150}$,
H.~Abramowicz$^\textrm{\scriptsize 154}$
% ...
\bigskip
\\
% ...
$^{12}$ Institute of Physics, Azerbaijan Academy of Sciences, Baku, Azerbaijan\\
% ...
$^{65}$ University of Iowa, Iowa City IA, United States of America\\
% ...
$^{87}$ CPPM, Aix-Marseille Universit{\'e} and CNRS/IN2P3, Marseille, France\\
% ...
$^{92}$ Department of Physics and Astronomy, Michigan State University, East Lansing MI, United States of America\\
% ...
$^{108}$ Nikhef National Institute for Subatomic Physics and University of Amsterdam, Amsterdam, Netherlands\\
% ...
$^{114}$ Homer L. Dodge Department of Physics and Astronomy, University of Oklahoma, Norman OK, United States of America\\
% ...
$^{118}$ LAL, Univ. Paris-Sud, CNRS/IN2P3, Universit{\'e} Paris-Saclay, Orsay, France\\
% ...
$^{138}$ Santa Cruz Institute for Particle Physics, University of California Santa Cruz, Santa Cruz CA, United States of America\\
% ...
$^{150}$ Department of Physics and Astronomy, University of Sussex, Brighton, United Kingdom\\
% ...
$^{154}$ Raymond and Beverly Sackler School of Physics and Astronomy, Tel Aviv University, Tel Aviv, Israel\\
% ...
\end{flushleft}

% Created with xml2latex.py

\clearpage
\end{document}